\documentclass[12pt,amssymb]{article} 

\usepackage{mathrsfs}
\usepackage{amssymb}
\usepackage[dvips]{graphicx}

\newcommand{\newsection}[1]{
\addtocounter{section}{1} \setcounter{equation}{0}
\setcounter{subsection}{0} \addcontentsline{toc}{section}{\protect
\numberline{\arabic{section}}{{\rm #1}}} \vglue .6cm \pagebreak[3]
\noindent{ \bf  \thesection. #1}\nopagebreak[4]\par\vskip .3cm}
\newcommand{\newsubsection}[1]{
\addtocounter{subsection}{1}\setcounter{subsubsection}{0}
\addcontentsline{toc}{subsection}{\protect
\numberline{\arabic{section}.\arabic{subsection}}{#1}} \vglue .4cm
\pagebreak[3] \noindent{\it \thesubsection.
#1}\nopagebreak[4]\par\vskip .3cm}

\newcommand{\newsubsubsection}[1]{
\addtocounter{subsubsection}{1}
\addcontentsline{toc}{subsubsection}{\protect
\numberline{\arabic{section}.\arabic{subsection}.\arabic{subsubsection}}{
#1}} \vglue .4cm \pagebreak[3] \noindent{\it \thesubsubsection.
#1}\nopagebreak[4]\par\vskip .3cm}

\makeatletter
\newcommand{\seclabel}[1]{%
  \@bsphack
  \protected@write\@auxout{}%
     {\string\newlabel{#1}{{\thesection}{\thepage}}}
  \@esphack
  }
\newcommand{\subseclabel}[1]{%
  \@bsphack
  \protected@write\@auxout{}%
     {\string\newlabel{#1}{{\thesubsection}{\thepage}}}
  \@esphack
  }
\newcommand{\tablabel}[1]{%
  \@bsphack
  \protected@write\@auxout{}%
     {\string\newlabel{#1}{{\arabic{tabnum}}{\thepage}}}
  \@esphack
  }
\makeatother

\renewcommand{\theequation}{\thesection.\arabic{equation}}

\newlength{\extraspace}
\newlength{\extraspaces}
\newcounter{dummy}
\newcommand{\bc}{\begin{center}}
\newcommand{\ec}{\end{center}}
\newcommand{\be}{\begin{equation}
\addtolength{\abovedisplayskip}{\extraspaces}
\addtolength{\belowdisplayskip}{\extraspaces}
\addtolength{\abovedisplayshortskip}{\extraspace}
\addtolength{\belowdisplayshortskip}{\extraspace}}
\newcommand{\ee}{\end{equation}}

\newcommand{\ba}{\begin{eqnarray}
\addtolength{\abovedisplayskip}{\extraspaces}
\addtolength{\belowdisplayskip}{\extraspaces}
\addtolength{\abovedisplayshortskip}{\extraspace}
\addtolength{\belowdisplayshortskip}{\extraspace}}
\newcommand{\ea}{\end{eqnarray}}

\newcommand{\ban}{\begin{eqnarray*}
\addtolength{\abovedisplayskip}{\extraspaces}
\addtolength{\belowdisplayskip}{\extraspaces}
\addtolength{\abovedisplayshortskip}{\extraspace}
\addtolength{\belowdisplayshortskip}{\extraspace}}
\newcommand{\ean}{\end{eqnarray*}}
\newcommand{\baa}{
\addtocounter{equation}{1} \setcounter{dummy}{\value{equation}}
\setcounter{equation}{0}
\renewcommand{\theequation}{\thesection.\arabic{dummy}\alph{equation}}
\begin{eqnarray}
\addtolength{\abovedisplayskip}{\extraspaces}
\addtolength{\belowdisplayskip}{\extraspaces}
\addtolength{\abovedisplayshortskip}{\extraspace}
\addtolength{\belowdisplayshortskip}{\extraspace}}
\newcommand{\eaa}{
\end{eqnarray}
\setcounter{equation}{\value{dummy}}
\renewcommand{\theequation}{\thesection.\arabic{equation}}}

\input epsf

\newcounter{fignum}
\newcounter{tabel}

\newcounter{tabnum}
\newcommand{\vev}[1]{\left\langle #1\right\rangle}

\newcommand{\half}{\frac{1}{2}}
\newcommand{\del}{\partial}
\newcommand{\delb}{\bar{\del}}
\newcommand{\eol}{\nonumber \\}

\newcommand{\cO}{{\cal O}}

\newcommand{\bt}{{\bf 10}}
\newcommand{\bfv}{{\bf 5}}
\newcommand{\bfb}{{\overline{\bf 5 \!}\,}}
\newcommand{\btb}{{\overline{\bf 10 \!}\,}}

\begin{document}

%
%

\begin{flushright}
August 2008\\
AEI-2008-036
\end{flushright}
\vspace{2cm}

\thispagestyle{empty}

%
%

\begin{center}
{\Large\bf  Breaking GUT Groups in F-Theory
 \\[13mm] }

{\sc Ron Donagi}\\[2.5mm]
{\it Department of Mathematics, University of Pennsylvania \\
Philadelphia, PA 19104-6395, USA}\\[9mm]

{\sc Martijn Wijnholt}\\[2.5mm]
{\it Max Planck Institute (Albert Einstein Institute)\\
Am M\"uhlenberg 1 \\
D-14476 Potsdam-Golm, Germany }\\
[30mm]

 {\sc Abstract}

\end{center}

We consider the possibility of breaking the GUT group to the
Standard Model gauge group in $F$-theory compactifications by
turning on certain $U(1)$ fluxes. We show that the requirement of
massless hypercharge is equivalent to a topological constraint on
the UV completion of the local model. The possibility of this
mechanism is intrinsic to $F$-theory. We address some of the
phenomenological signatures of this scenario. We show that our
models predict monopoles as in conventional GUT models. We discuss
in detail the leading threshold corrections to the gauge kinetic
terms and their effect on unification. They turn out to be related
to Ray-Singer torsion. We also discuss the issue of proton decay in
$F$-theory models and explain how to engineer models which satisfy
current experimental bounds.

  \vfill

\newpage

\renewcommand{\Large}{\normalsize}

\tableofcontents

\newpage

\newsection{Introduction}

One of the most remarkable hints about physics at high energy scales
is the apparent unification of gauge couplings
\cite{Georgi:1974yf,Dimopoulos:1981yj}. This gives strong support
for the idea that the Standard Model originates from a single GUT
group at some high scale \cite{Georgi:1974sy}. Several other
properties of the Standard Model point to some kind of unification,
however there are also aspects that are hard to fit in a
conventional four-dimensional GUT model, particularly the
doublet-triplet splitting problem. As has been appreciated for some
time, the discrepancies have natural resolutions if the unification
takes place not in four but in higher dimensions
\cite{Hosotani:1983xw,Witten:1985xc}. Of course gauge theories in
higher dimensions do not make sense by themselves, but they have a
natural home in string theory, which provides their UV completion.

The first models of this type were explored in the heterotic string,
where unification takes place in ten dimensions
\cite{Candelas:1985en}. However the experimental fact of the
hierarchy between the TeV scale and the Planck scale, as well as
practical considerations, have led to the idea of local model
building
\cite{Antoniadis:2000ena,Aldazabal:2000sa,Berenstein:2001nk,Verlinde:2005jr}.
In this scenario, one considers models of particle physics where the
effects of four-dimensional gravity (and its attendant
complications) may be treated as a small perturbation. That is, a
local model is defined by the existence of a decoupling limit
\cite{Verlinde:2005jr}
\be g^2_{YM}(\mu)\  {\rm fixed}, \quad M_{Pl}/\mu \to \infty. \ee
In the context of GUT models, this means we would fix the coupling
at the GUT scale $\mu \sim M_{GUT}$, and we will have at least two
independent expansion parameters, $\alpha_{GUT}$ and
$M_{GUT}/M_{Pl}$.

Although perhaps not apparent at first sight, insisting on the twin
principles of unification and local model building cuts down the
landscape of possibilities considerably. In particular one might
have thought that models with $D$-branes provide a natural
realization of these principles. Unfortunately attempts to construct
GUTs in D-brane models ran into fundamental difficulties, due to the
absence of matter in the spinor representation of $SO(10)$, or the
perturbative vanishing of top quark Yukawa couplings for $SU(5)$
(since the epsilon tensor needed for these couplings cannot be
generated in perturbative open string theory). These problems can be
traced back to the absence of exceptional gauge symmetries on
perturbative $D$-branes. GUT groups can of course be naturally
incorporated in the $E_8 \times E_8$ heterotic string, however the
ideas about local model building and decoupling from the Planck
scale leads one to ask if exceptional Yang-Mills symmetries can also
be localized on branes with non-zero codimension.

\begin{figure}[t]
\addtocounter{tabnum}{1} \tablabel{ExcBranes}
\begin{center}
\renewcommand{\arraystretch}{1.5}
\begin{tabular}{|c|c|}
  \hline
  \ \  {\it dim} \qquad & \qquad  {\it stringy realization}\qquad \qquad \\
  \hline \hline
  10d \qquad & $E_8 \times E_8$ heterotic string \\
  9d & strongly coupled type I' \\
  8d & $F$-theory on ALE \\
  7d & $M$-theory on ALE \\
  6d & \qquad IIa on ALE/IIb with NS5 \qquad \quad \\
  \hline
\end{tabular}\\[5mm]
\parbox{10cm}
{\bf Table \arabic{tabnum}: \it Branes with exceptional gauge
symmetry in string theory.}
\renewcommand{\arraystretch}{1.0}
\end{center}
\end{figure}

Apart from the heterotic string, exceptional Yang-Mills theory can
in fact also appear on various branes in string theory (see table
\ref{ExcBranes}). However Yang-Mills theories realized on 5-branes
or branes of lower dimension do not lead to chiral matter, because
if the codimension is high enough then we can always separate the
branes. On the other hand, if the codimension is too small, one gets
into trouble with the requirement of decoupling. The situation in
type I' should be very similar to the situation in Horava-Witten
theory, to which it is related by compactification on a circle. For
a generic compactification in the Horava-Witten set-up, if the size
of the interval gets too large then the bare coupling of one of the
$E_8$s is pushed to infinity, which yields a bound of the form
\cite{Witten:1996mz}
\be G_N \gtrsim {\alpha_{GUT}^2 \over M_{GUT}^2 }\ee
This means that $M_{GUT}/M_{Pl}$ is never parametrically small and
the gravitational back-reaction may not be considered subleading.
Another issue for the strongly coupled type I' is that there are no
known constructive techniques. A similar issue affects the study of
$7d$ Yang-Mills theory in $M$-theory. This requires the study of
compactification on 7-manifolds of $G_2$ holonomy, a subject about
which so little is known that we cannot even write equations for
semi-realistic local models.\footnote{Since this paper first
appeared, the construction of local $M$-theory models has been
described in \cite{Pantev:2009de}.}

Thus if one is interested in a framework for local model building
with GUT groups, on second thought it turns out the options are
really rather limited. In $F$-theory, like in the heterotic string,
there are actually powerful constructive techniques available from
algebraic geometry \cite{Bershadsky:1996nh,Friedman:1997yq}. Up
until recently, the main issue hindering progress in this direction
was the lack of knowledge about engineering chiral matter, an issue
that was pointed out and solved in \cite{Donagi:2008ca} by turning
on fluxes (see also \cite{Beasley:2008dc}, and further refinement in
\cite{Hayashi:2008ba}). However although GUT groups and chiral
matter could be naturally obtained in this framework, there was not
yet a sound method for breaking the GUT group to the Standard Model
while preserving the main predictions from unification.

The main purpose this paper is to make such a proposal. As we
discuss in section \ref{GUTbreaking}, this can be achieved by
turning on suitable $U(1)$ fluxes. There are however a number of
subtleties in the implementation. In particular, analogous
mechanisms are not available in $M$-theory on $G_2$ or the heterotic
string. This is an additional motivation to study unification in
$F$-theory.

Having specified a GUT breaking mechanism, we can now start to
address some of the phenomenology of $F$-theoretic GUTs. There are
many issues one could discuss, and we focus on only a few of them.
This paper is a direct continuation of \cite{Donagi:2008ca}, so we
refer to that paper for basic concepts and notation.

In section \ref{Monopoles} we show that our models predict monopoles
carrying magnetic hypercharge. They correspond to certain
$D3$-branes ending on the GUT 7-brane.

In section \ref{KKThresholds} we give a detailed discussion of
threshold corrections to the gauge kinetic terms. It is well known
that the GUT prediction of $\alpha_3(M_Z)$ differs slightly from the
experimental value, and the natural question is if the heavy
threshold corrections in our scenario make the situation worse or
better. We will see that at least in our toy models, the corrections
come with varying signs and can make the discrepancy worse or
better. We will also see that the KK scale ends up parametrically
lower than the the GUT scale, although in practice this is a mild
effect.

In section \ref{ProtonDecay} we discuss proton decay. Dimension four
and five operators violating baryon number are generically present
and models have to be engineered to guarantee their absence. We find
that decay through dimension six operators is parametrically
enhanced as $\alpha_{GUT}\to 0$ compared to four-dimensional models.
A similar situation had earlier been discovered in other KK models
of unification \cite{Friedmann:2002ty,Klebanov:2003my}, however the
parametric dependence for the $p\to \pi^0 e^+_L$ channel is
different from what was seen in these papers. Moreover a second
effect, the lowering of the KK scale compared to the GUT scale, also
did not appear in $M$-theory.

{\it Note for revision:} The present version of this paper includes
significant clarifications and improvements over the original
version arXiv:0808.2223. Most of these were designed to bring the
paper more in line with the subsequent papers
\cite{Donagi:2009ra,Donagi:2011jy}. We also extended the discussion
of extra light $Z'$s, expanded the discussion of flux quantization
and exotic matter in appendix \ref{HetGfluxes}, and added appendix
\ref{E8roots}. As this paper was being prepared for submission, two
papers appeared (\cite{Beasley:2008kw,Tatar:2008zj}) that overlap
with some of our results.

\newpage

\newsection{Breaking the GUT group}
\seclabel{GUTbreaking}

\newsubsection{The basic proposal}

We start with a brief recap of the discussion in
\cite{Donagi:2008ca}. There are basically three ideas for breaking
the GUT group in string models. The first is to engineer an adjoint
Higgs or a Higgs in another large representation. Although this is
in principle possible in $F$-theory, in this way one ends up with a
traditional four-dimensional GUT model. Such models are beset by
phenomenological problems such as doublet/triplet splitting and a
tension between fast proton decay and unification. Therefore it is
more desirable to use the additional ingredients in our
$F$-theoretic constructions and to look for an intrinsically higher
dimensional mechanism which preserves the successes of unification
but evades the problems.

The second approach, which has been successfully implemented in the
heterotic string, is to use discrete Wilson lines. However in order
to avoid adjoint Higgses in $F$-theory, we would typically wrap our
7-brane on four-cycles with trivial fundamental group\footnote{One
could consider smooth models with discrete Wilson lines, based on
the Enriques surface or some surface of general type. However an
application of the Riemann-Roch formula shows that such models have
light lepto-quarks.}. Attempts to dualize heterotic models with
discrete Wilson lines lead to 7-branes with singularities on the
worldvolume. Although the presence of singularities is common in
recent phenomenological literature on extra-dimensional GUTs
\cite{Hall:2002ea,Dienes:1998vg}, in the present context we should
avoid them. There could very well be new degrees of freedom at such
singularities that spoil the running of the couplings or are not
mutually supersymmetric with the other ingredients in the
construction, or give rise to light particles with fractional
electric charges as happens in analogous heterotic orbifold
constructions. Given our current understanding of $F$-theory, it is
impossible to know. Until this point is clarified, this approach
does not appear promising in the context of $F$-theory.

The third idea is to turn on $U(1)$ fluxes on the worldvolume of the
7-brane in order to break the GUT group. For definiteness let us
consider the case of a 7-brane with a GUT group $G=SU(5)$ wrapping a
four-cycle $S$. We denote the elliptically fibered Calabi-Yau
four-fold by $Y_4$, the section by $\sigma(B_3)$ and the inclusion
of $S$ by $i: S \to \sigma(B_3)$. Then in order to break the GUT
group we could turn on a non-trivial flux on $S$ for the hypercharge
gauge field, which is embedded in $SU(5)$ as
\be Y = \left(
      \begin{array}{ccccc}
        -1/3 & 0 & 0 & 0 & 0 \\
        0 & -1/3 & 0 & 0 & 0 \\
        0 & 0 & -1/3 & 0 & 0 \\
        0 & 0 & 0 & 1/2 & 0 \\
        0 & 0 & 0 & 0 & 1/2 \\
      \end{array}
    \right)
\ee
This means we specify a line bundle $L$ on $S$, with $c_1(L)$ a
non-trivial class in $H^2(S)$ represented by a harmonic two-form,
and set the $SU(5)$ field strength to $\vev{F_{SU(5)}} = c_1(L)\,
Y$. More precisely, recall that abelian gauge fields descend from
harmonic two-forms in the lattice
\be \Lambda\ =\ \{\, \omega \in H^2(Y_4)\,|\, \omega \cdot \gamma =
0 \ {\rm when } \  \gamma \in  H_2(B_3)\ {\rm or}\ \gamma = [T^2]\,
\} \ee
Let us denote the harmonic two-form leading to the hypercharge gauge
field by $\omega_Y$:
\be C^{(3)}_{}\ =\ A_{Y} \wedge \omega^Y_{} \ee
Then let us turn on a $G$-flux\footnote{Recall that the $G$-flux
must be properly quantized, which means in the present case that the
flux of the line bundle $L$ must be quantized. This results in a
small puzzle regarding exotic matter, which we discuss and resolve
in the next subsection.}
\be {\sf G}/2\pi\ =\ c_1(L) \wedge \omega^Y \ee
More generally, we would consider a global Calabi-Yau four-fold with
a $G$-flux, which reproduces our local model with the above $G$-flux
in a degeneration limit. The two-form $\omega^Y$ is somewhat
singular, it has a delta-function piece localized on the $A_4$
singularity giving rise to the $SU(5)$ gauge group. (To make sense
of it mathematically, we should first resolve the $A_4$ singularity;
$F$-theory corresponds to the limit when the exceptional cycles are
taken to zero size).

Such a $U(1)$ flux will break the GUT group $G=SU(5)$ to the
subgroup which commutes with the $U(1)$, which is the Standard Model
gauge group $SU(3) \times SU(2) \times U(1)$. Further we assume
large volumes so that we can trust the predictions from the two
derivative eight-dimensional Yang-Mills action. Then with this
method of breaking the GUT group we recover the standard $SU(5)$
relations between the gauge couplings.

Unfortunately this argument is too quick, for two reasons. As we
will discuss in the next subsection, naively turning on such a flux
leads to lepto-quarks in the low energy spectrum, which is
unacceptable. Actually, as we will show later this problem is
naturally resolved by a more careful examination of the flux we
should turn on, so we will not worry about this here. The second
issue is that the Yang-Mills theory is coupled to additional closed
string modes which live in the bulk of space-time, which may lead the
abelian gauge fields to pick up a mass through the St\"uckelberg
mechanism. The $F$-theory effective action has a Chern-Simons term
of the form\footnote{The reader might wonder why we don't consider
couplings to other RR or NS potentials (we'd like to thank J.
Maldacena for asking the question). For instance, naively one should
also take couplings like $\int C^{(2)} \wedge {\rm Tr} (F \wedge F
\wedge F)$ into account. However due to the $Sl(2,Z)$ monodromies,
in a generic $F$-theory compactification there are no zero modes for
any of the RR or NS potentials with two indices on ${R^4}$, except
for $C^{(4)}_{RR}$. This is clear from the $M$-theory description.}
\be \mathscr{L}_{12}\ \sim\ -{i\over 4 \pi}\int C_4 \wedge {\sf G}
\wedge {\sf G} \ee
Now we expand the four-form as follows:
\be C_4\ =\ C_2^M \wedge {\beta_M} , \qquad \beta_M \in H^2(B_3) \ee
Note that these are bulk modes and know about all of $B_3$ (or more
precisely, the section $\sigma_{B_3}$). We can expand the $G$-flux
as
\be {\sf G}\ =\ F_Y \wedge \omega^Y + G_{\rm internal} \ee
where $\omega^Y \in \Lambda$ is the harmonic two-form giving rise to
hypercharge.

Then we find the following four-dimensional coupling:
\be\label{axioncoupling} \mathscr{L}_4\ \sim\ -{i\over
2\pi}\,\Pi_{M}^Y \int d^4x\ C_2^M \wedge F_Y \ee
where
\be 
\Pi_{M}^Y\ =\ \int_{Y_4} \beta_M \wedge \omega^Y \wedge G_{\rm
internal} \ee
As is well-known, the fields $C_2^M$ may be dualized to RR axions
$a_M$ in four dimensions, and (\ref{axioncoupling}) then leads to a
coupling of the form
\be\label{gaugemass}
\mathscr{L}_4 \sim \int d^4x\  G^{MN}g^{\mu\nu}( \Pi_{M} A_{Y,\mu}-  \del_\mu
a_M)( \Pi_{N} A_{Y,\nu}-  \del_\nu
a_N) \ee
where
\be G_{MN}\ =\ \int_{B_3} \beta_M \wedge  * \beta_N \ee
Next we plug in ${\sf G}_{\rm int}/2\pi = c_1(L) \wedge \omega^Y$.
In a local model (a $dP_8$ or $dP_9$  fibration over $S$), we may use
$\int_{dP_8} \omega_Y \wedge \omega_Y = {\rm Tr}(Y^2)$, leading to
\be \Pi_M^Y\ =\ -2\pi{\rm Tr}[Y^2]  \int_S c_1(L) \wedge i^*\beta_M
\ee
Therefore at first sight it seems that $\Pi_M^Y$ must be non-zero for
some $M$, and the hypercharge gauge field will pick up a mass
through (\ref{axioncoupling}) or (\ref{gaugemass}).

We would like to make a brief comment on our argument. We assumed
that $\omega^Y$ can be represented by a class that is localized at
the $SU(5)$ singularity, so that the overlap $\Pi_M^Y$ only depends
on the pull-back $i^*\beta_M$ of $\beta_M$ to $S$. This could hardly
be otherwise, because we would expect that for a $U(1)$ gauge symmetry
that is part of an unbroken non-abelian group, the abelian and non-abelian
components are localized in the same way, so it should be in the
singularity lattice of the $dP_9$
surface. We will see this explicitly for hypercharge below. For
abelian gauge symmetries that are not part of a non-abelian group,
there is no such localization and no guarantee that the
corresponding $\omega$ can be extended globally.

One way to avoid the conclusion that hypercharge is lifted,
suggested in \cite{Witten:1985bz}, is to turn on a flux in the same
cohomology class $[c_1(L)]$ in a hidden sector. The axion will then
couple to a linear combination of $A_\mu^Y$ and a $U(1)$ in the
hidden sector, and the orthogonal linear combination will remain
massless. But in this approach the kinetic term of the remaining
massless $U(1)$ is changed compared to our original hypercharge
generator, and we lose the relation $\sqrt{5/3}\,g_1 =g_2 =g_3$.

Here we would like to propose a different way out, which has the
virtue of preserving the GUT relations among the couplings at
leading order. The situation we discussed above is very analogous to
that of obtaining a massless hypercharge in models with branes at
singularities \cite{Buican:2006sn}. In fact in
\cite{Wijnholt:2007vn}, section 6, the mechanism of
\cite{Buican:2006sn} was already interpreted as a partial
unification scenario. The point is that because the brane has
non-zero codimension, the vanishing of $\Pi_M^Y$ for all $M$ does
{\it not} imply that $c_1(L)$ must vanish. Rather it is equivalent
to the following statement. If we embed $i: S \to B_3$ then
\be i^*:H^2(B_3) \to H^2(S) \ee
generally has non-trivial cokernel, and $\Pi_M^Y=0$ for all $M$
means that $c_1(L)$ is a generator of this cokernel. Thus the
requirement of a massless hypercharge gauge field would really
amount to a topological constraint on the UV completion of the local
model.

It is useful to restate this criterion using Poincar\'e duality.
Consider the two-cycle $\Xi \in H_2(S)$ dual to $c_1(L)$. Then the
requirement $\Pi_M^Y=0$ for all $M$ is equivalent to the statement
that $\Xi$ is in the kernel of the map $i: H_2(S) \to H_2(B_3)$. In
other words, even though $\Xi$ is a non-trivial two-cycle in $S$,
once we embed it in $B_3$ we can find a three-chain $\Gamma$ in
$B_3$ so that $\Xi = \del \Gamma$.

One quick observation is that the canonical divisor $K_S$ is always
a pull-back of a class in $H^2(B_3)$. This follows from the
adjunction formula, and the fact that $K_{B_3}$ and $\cO(S)$ yield
classes in $H^2(B_3)$. So in any local construction we should
certainly make sure that $c_1(L)$ is orthogonal to $c_1(K_S)$.
Generically we would expect there is one other class in the image of
$i^*$, namely the remaining linear combination of $c_1(K_{B_3})$ and
$c_1(\cO(S))$ (which may be proportional to $c_1(K_S)$ however), and
the remaining classes to be orthogonal to two-forms inherited from
$B_3$. This is because divisors in the local Picard group typically
get destroyed when higher order terms are added to the equations in
order to compactify the model \cite{Buican:2006sn}. However for
special UV completions the set of inherited classes may be larger.

\newsubsection{More refined version}
\subseclabel{RefinedGUTBreaking}

We now deal with the other problem briefly mentioned above. The
issue is that due to quantization constraints on the flux for $A_Y$,
naively it seems that turning on such a flux leads to massless
lepto-quarks, which is phenomenologically unacceptable. Here by
lepto-quarks we mean chiral fields with the same gauge quantum
numbers as the $X$ and $Y$ gauge bosons of conventional
$SU(5)_{GUT}$ models. To see this, consider the decomposition of the
$SU(5)$ adjoint under $SU(3) \times SU(2) \times U(1)_Y$:
\ba {\bf 24} &=& ({\bf 3},{\bf 1})_0 \oplus ({\bf 1},{\bf 8})_0 \oplus
({\bf 1},{\bf 1})_0 \oplus({\bf 2},{\bf 3})_{\bf -5/6} \oplus  ({\bf
2},\overline{\bf 3\!}\,)_{\bf 5/6} \eol[2mm] &\equiv& {\bf R}_0\
\oplus \ {\bf R}_{-5/6}\ \oplus\ {\bf R}_{5/6} \ea
In order to get the MSSM at low energies, we cannot have any massless modes
in the representation ${\bf R}_{5/6}$. This means in particular that
\be
\chi(S, L^{5/6})=0
\ee
Since ${\rm deg}(L)=c_1(L)\cdot K_S =0$ as noted above, and since we
take $S$ to be a Del Pezzo surface in order to avoid massless
adjoint valued fields, we have
\be\label{HRRSimplified}
\chi(S,L^{5/6})= 1 + \half\, c_1(L^{5/6})^2\ee
from the Hirzebruch-Riemann-Roch formula. So the absence of
lepto-quarks requires that $c_1(L^{5/6})^2 = -2$. On the other hand,
since
\ba
{\bf 10} &=& ({\bf 2},{\bf 3})_{\bf 1/6} + ({\bf 1},{\bf 1})_{\bf 1} + ({\bf
1},\overline{\bf 3\!}\,)_{\bf -2/3}  \eol
\overline{\bf 5\!}\, &=& ({\bf 2},{\bf 1})_{\bf -1/2} + ({\bf
1},\overline{\bf 3\!}\,)_{\bf 1/3}
\ea
it may naively seem that we also need $c_1(L^{1/6})$ and
$c_1(L^{1/3})$ to be integer classes, so that
$L^{5/6},L^{1/6},L^{1/3}$ are all honest line bundles on $S$. But
this requirement is inconsistent with $c_1(L^{5/6})^2 = -2$, so it
seems we would not get any sensible models without extra
lepto-quarks this way.

This issue arose because we tried to apply the quantization
constraints the hypercharge flux and the flux for generating chiral
matter separately. However the correct thing to do is to apply the
constraints only to the total flux. We will now analyze the model in
more detail and see that quantization only requires $L^{5/6}$ to be
an honest line bundle, and is therefore perfectly consistent with
the absence of massless lepto-quarks.

\begin{figure}[t]
\begin{center}
            \scalebox{.45}{
               \includegraphics[width=\textwidth]{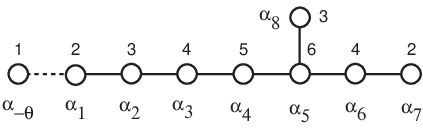}
               }
\end{center}
\vspace{-.5cm} \caption{ \it The extended $E_8$ Dynkin diagram and
Dynkin indices.}\label{E8Dynkin1}
\end{figure}

Consider a local model for an $SU(5)$ GUT obtained from unfolding an
$E_8$ singularity. That is we consider an $E_8$ ALE fibered over
$S_2$, with the vanishing cycles labelled by the roots of $E_8$ (see
figure \ref{E8Dynkin1}). We take $\{\alpha_{-\theta}, \alpha_1,
\alpha_2, \alpha_3, \alpha_4 \}$ to be of finite volume
and varying over $S_2$, and we take the monodromy for $\{\alpha_5,
\ldots, \alpha_8\}$ to be trivial. This breaks the gauge group from
$E_8$ to $SU(5)$. The first group of cycles may undergo monodromies
corresponding to Weyl reflections generated by $\{\alpha_{-\theta},
\alpha_1, \alpha_2, \alpha_3 \}$. These Weyl reflections
leave the second group of cycles invariant. We denote the
corresponding Weyl group as $\mathscr{W}_{A_4}$.

When the two-cycle $\alpha_4$ in the ALE (or some
$\mathscr{W}_{A_4}$ image) shrinks to zero size, the singularity
increases to $SO(10)$, so this corresponds to the matter curve
$\Sigma_{\bf 10}$. Similarly when the two-cycle $\alpha_3 +
\alpha_4$ (or some $\mathscr{W}_{A_4}$ image) shrinks to zero size,
we get the matter curve $\Sigma_\bfv$.

Let us define a canonical basis of Cartan generators $\omega_i$
satisfying
\be [\omega_i,\alpha_j] = \delta_{ij}\ \ee
Let us write $\omega_Y$ in this basis, taking the roots
$\{\alpha_6,\alpha_7\}$ and $\{\alpha_8\}$ in the Dynkin diagram to
correspond to the roots of $SU(3)_c \times SU(2)_w$ respectively.
Naively one might think that the hypercharge generator should be
identified with $\omega_Y = -5/6 \,\omega_5$. However this generator
assigns a non-zero $U(1)$ charge to $\alpha_{-\theta}$, hence it
does not commute with the $\mathscr{W}_{A_4}$ monodromy group and
receives a mass of order the KK scale. Furthermore it gives
incorrect hypercharge assignments to the hypermultiplets, so we
should really not call this generator hypercharge. Since we know the
hypercharge assignments of the hypermultiplets on the matter curves
and of the $SU(5)$ gauge bosons, we see that the correct hypercharge
generator $\omega_Y$ must satisfy
\be [\omega_Y,\alpha_5]\ =\ -{5\over 6}, \qquad [\omega_Y,\alpha_4]\
=\ 1 \ee
and all the remaining commutators zero. Note that these equations
are consistent in that the equation $[\omega_Y,\alpha_{-\theta}]= 0
$ actually follows from the definition of $\alpha_{\theta}$ and the
remaining commutators. Hence we deduce
\be \omega_Y\ =\ -{5\over 6}\omega_5 + \omega_4 \ee
Dually we may also represent this by a linear combination of
algebraic cycles on the $dP_9$ (fibered over the base), and we find
\be\label{HypDualExpression} \omega_Y \ \simeq \ {1\over 6}(2 \alpha_7 + 4 \alpha_6 + 6
\alpha_5 + 3 \alpha_8) \ee
This is indeed a linear combination of effective cycles in the
singularity lattice, as expected for a $U(1)$ which is part of an
unbroken non-abelian gauge symmetry. Turning on a flux for this
$U(1)_Y$ breaks the $SU(5)$ group to $SU(3)\times SU(2) \times
U(1)$. Furthermore this generator actually commutes with the
${\mathscr{W}}_{A_4}$ Weyl group that breaks the $E_8$ to
$SU(5)_{GUT}$; in particular it is defined globally over $S$ and
defines an unbroken $U(1)$ gauge symmetry.

As discussed in \cite{Donagi:2008ca}, in $F$-theory the Cartan
generators $\omega_i$ correspond to two-forms on the ALE which we
denote by the same name, and the $\alpha_j$ correspond to two-cycles
of the ALE. The integral of the $\omega_i$ yield the $U(1)$ charges
of of BPS membranes (or rather $(p,q)$-strings) wrapped on the
corresponding two-cycle, in particular with our definition we have
$\int_{\alpha_j}\omega_i = \delta_{ij}$ so the periods of the
$\omega_i$ are all integral. Now turning on a flux for the $U(1)$
gauge field associated to $\omega_Y$, corresponds to turning on an
additional ${\sf G}$- flux of the form
\be\label{roughGzeta} {{\sf G}_Y\over 2\pi} \ = \  c_1(L) \wedge
\omega_Y\ =\ - c_1(L^{5/6}) \wedge \omega_5 + c_1(L) \wedge \omega_4
\ee
where $L$ is a (possibly fractional) line bundle on $S$. The last
piece looks problematic for satisfying the quantization conditions.
However since the last piece commutes with the generators of the
$SU(5)_{GUT}$ group, it can be cancelled by including a similar
fractionally quantized piece with opposite sign in the flux we turn
on to generate chiral matter. Hence for the total flux to be
properly quantized, only $c_1(L^{5/6})$ and its multiples need to be
integer quantized, but not $c_1(L)$ or $c_1(L^{1/6})$. As we already
noted, this is perfectly compatible with the requirement of the
absence of massless lepto-quarks.

In this way of stating it, it also becomes clear that the same
freedom should exist in spectral cover constructions for the
heterotic string. What we have discussed is the $F$-theory dual of
heterotic models with structure groups of the form $S(U(n) \times
U(1)) \subset SU(n+1)$
\cite{Andreas:2004ja,Blumenhagen:2005ga,Blumenhagen:2006ux,Blumenhagen:2006wj},
where for us the relevant case is $n=5$. See \cite{Donagi:2008ca}
for the relevant aspects of heterotic/$F$-theory duality.

In appendix \ref{HetGfluxes} we discuss how to write down the fluxes
in the heterotic language, and their dual $G$-flux constructed using
the projected cylinder map, which gives a precise algebro-geometric
definition of (\ref{roughGzeta}). We also show in more detail there
that the quantization constraint is compatible with the absence of
light exotic matter, taking into account the issue of ramification
(which we did not treat carefully above). To compare with our
description here, recall that the sheets of the spectral cover are
labelled by generators of the adjoint representation of $E_8$. For
$S(U(5)\times U(1))\subset SU(6)$, the most important piece is a
six-fold cover whose sheets are labelled by the six roots
\be\label{SU6Cover}
\begin{array}{lcl}
  \sigma_1=\alpha_5 & \qquad & \sigma_4=\alpha_5+\alpha_4 + \alpha_3 + \alpha_2 \\
  \sigma_2=\alpha_5+\alpha_4 &  & \sigma_5=\alpha_5+\alpha_4 + \alpha_3 + \alpha_2 + \alpha_1 \\
 \sigma_3= \alpha_5+\alpha_4 +
\alpha_3 &  & \sigma_6=\alpha_5+\alpha_4 + \alpha_3 + \alpha_2 +
\alpha_1 + \alpha_{-\theta}
\end{array} \ee
This spectral cover is naturally reducible under the action of
$\mathscr{W}_{A_4}$ -- it splits into a copy of the zero section,
labelled by $\{\alpha_5\}$, and a five-fold spectral cover whose
sheets are labelled by $\{ \sigma_2, \ldots, \sigma_6 \}$.
 Since $\omega_Y$ evaluates to $1/6$
on each of the sheets of the five-fold cover, the $G$-flux
$c_1(L)\wedge \omega_Y$ maps to a line bundle on this five-fold
spectral cover which is morally $L^{1/6}=(L^{5/6})^{1/5}$ on each
sheet, i.e. it is a $U(5)$ spectral cover rather than an $SU(5)$
spectral cover. Similarly this $G$-flux yields a line bundle
$L^{-5/6}$ on $\sigma_1$, the zero-section proportional to the
hypercharge generator of $SU(5)_{GUT}$. One should keep in mind
though that we only want part of a heterotic model (the observable
sector, i.e. the local neighbourhood of the GUT brane in $F$-theory
language) while we would like to replace the hidden sector by
something else. Indeed as we will see in the next subsection,
$F$-theory models which implement our GUT breaking mechanism cannot
have global heterotic duals.

The hypercharge $G$-flux defined through the cylinder map depends on
the choice of the line bundle $L^{5/6}$ on $S_2$, which we
henceforth call $\zeta$. In order to eliminate the axion couplings, $c_1(\zeta)$
must be orthogonal to any class inherited from $B_3$, in particular
$c_1(K_S)$ and $c_1(N_S)$. For primitiveness, $\zeta$ must be
orthogonal to the pull-back to $S$ of the K\"ahler class of $B_3$.
This is also clearly guaranteed by the previous condition. Finally,
in order to eliminate the light $({\bf 2},{\bf 3})_{\bf -5/6}$ exotics
from the spectrum, from (\ref{HRRSimplified}) we see that $c_1(\zeta)$
must be one of the `simple roots', i.e. a
cohomology class with $c_1(\zeta)^2=-2$ and $c_1(\zeta)\cdot K_S=0$.
It is not hard to make toy models with three generations and no
exotic matter; in appendix \ref{HetGfluxes} we explain how to do
this using the cylinder map.

\newsubsection{Other comments}

\newsubsubsection{Doublet-triplet splitting}

It may be useful to point out that our GUT breaking mechanism does
not lead to a doublet-triplet splitting problem. The zero modes for
the doublets and triplets come from different Dolbeault cohomology
groups, so we can have light doublets without having light triplets.
This story is essentially the same as for GUT breaking with discrete
Wilson lines: because there is never a four-dimensional $SU(5)$
gauge group at any scale, there is no need for four-dimensional
triplet partners of the Higgs doublets of the Minimal Supersymmetric 
Standard Model. Since there is still an
eight-dimensional GUT group, there will of course still be
Kaluza-Klein modes with the same quantum numbers as the triplets,
with masses of order the radius of curvature of the surface $S$, and
they can be an issue for proton decay and precision unification. We
discuss this in more detail in sections \ref{KKThresholds} and
\ref{ProtonDecay}.

\newsubsubsection{No heterotic or $M$-theory duals}

If $B_3$ is fibered over $S$, then the map $i^*:H^2(B_3) \to H^2(S)$
is automatically surjective.  This is easy to see because any class
$c_1(\zeta) \in H^2(S)$ may be pulled back to $B_3$. For such
compactifications, our GUT breaking mechanism can therefore not be
implemented. In particular, it does not work for duals of
perturbative heterotic models, which correspond to ${\bf P}^1$
fibrations over $S$. In fact it cannot be implemented in the
perturbative heterotic string at all, whether or not it has an
$F$-theory dual, because the zero modes of the axions and gauge
fields are both supported on all of the heterotic Calabi-Yau
three-fold. One may still use the trick of getting a massless
hypercharge as a linear combination of two different sectors, as
studied in
\cite{Andreas:2004ja,Blumenhagen:2005ga,Blumenhagen:2006ux,Blumenhagen:2006wj},
though at the cost of giving up unification at leading order.

One might also consider implementing this mechanism in $M$-theory on
manifolds of $G_2$ holonomy. Local models in this context consist of
ALE fibrations over a compact three-manifold $Q$. In order to avoid
adjoint Higgses one needs $b_1(Q)=0$. But this implies that
$b_2(Q)=0$ also and so we cannot turn on any fluxes.

Thus the possibility of breaking the GUT group by fluxes is
intrinsic to $F$-theory and should contribute to distinctive
phenomenological signatures.

\newpage

\newsection{GUT monopoles}
\seclabel{Monopoles}

In GUT models we expect to find solitons, particularly monopoles,
which will be generated in the early universe through the Kibble
mechanism (for a review, see \cite{Preskill:1984gd}). This is true
even for the higher dimensional models here; in a sense the role of
the Higgs field is played by the internal components of
eight-dimensional $U(1)$ gauge fields breaking the GUT group. For
the analgous situation in the heterotic string, see
\cite{Wen:1985qj}.

Let us review quickly some aspects of four-dimensional GUT
monopoles. If the GUT group $G$ is broken to a subgroup $H$,
monopoles are classified by $\pi_2(G/H)$. From the long exact
sequence
\be
 \ldots \to \pi_2(G) \to \pi_2(G/H) \to \pi_1(H) \to \pi_1(G) \to \ldots
 \ee
and the fact that for any compact simple Lie group we have
$\pi_2(G)=0$, we find that $\pi_2(G/H) = \pi_1(H)$. In the case of
adjoint breaking of $G= SU(5)$ to the SM, we have
\be H = [SU(3) \times SU(2) \times U(1)]/Z_6 \ee
The action of $Z_6$ on $SU(3) \times SU(2)$ is identified with the
center of $SU(3) \times SU(2)$. Thus there exists a monopole with
charge $1/2g_1$. The monopole also carries $Z_3$ colour and $Z_2$
$SU(2)$ magnetic charge.

Now let us try to find the monopoles in our $F$-theory set-up. The
important feature which allowed us to break the GUT group without
mixing with other fields was the
existence of an extra two-cycle on the worldvolume of the 7-brane,
which becomes the boundary of a three-cycle when embedded in
space-time. So we expect the existence of hypercharged monopoles
must be related to this feature.
 Let us denote
by $\Xi \in H_2(S)$ the Poincar\'e dual of $c_1(\zeta)$ in $S$.
Since $c_1(\zeta)$ was automatically orthogonal to the K\"ahler
class, then if $S$ is non-singular $\Xi$ can not be represented by a
single geometric two-cycle in $S$, only by a linear combination of
them. For definiteness let us say that $\Xi = A_1 - A_2$ where $A_1$
and $A_2$ are geometric cycles (in particular they do not become
boundaries when embedded in $B_3$).

Since our mechanism requires that $PD(\alpha)$ becomes a boundary
when embedded in $B_3$, there exists a three-chain $\Gamma$ in $B_3$
such that
\be A_1 - A_2 = \del \Gamma \ee
Now we can wrap a $D3$-brane on $\Gamma$. There must be a
representative of $\Gamma$ with minimal volume greater than zero,
because if we restrict to a sufficiently small neighbourhood of the
brane we know that $\Gamma$ does not exist. This is similar to a
$D3$-ending-on-$D5$ system and so should correspond to the
sought-after monopole.

Let us consider a small `six-sphere' surrounding the end of a
stack of $D3$-branes. The integral of $dF^{(5)}_{RR}$ over any such six-sphere
must vanish, but we get a contribution $+1$ for every open $D3$-brane in the stack,
as it intersects the sphere precisely once. However due to the Chern-Simons terms
in the action, we also get contributions to
$dF^{(5)}_{RR}$ from other fluxes.
Thus we can relate the flux through the sphere to the number
of open $D3$-branes.

Let $D$ be a divisor in $S_2$ such that $D \cdot [A_1 - A_2] = 1$.
Such a divisor is guaranteed to exist by Poincar\'e duality. In
the present case, we can take our `six-sphere' to correspond to a
unit three-ball in $R^3$ times $D$ times a unit circle in the
normal bundle to $S_2$ in $B_3$, minus a unit two-sphere in $R^3$
times $D$ times the unit disk in the normal bundle to $S_2$ in
$B_3$. We have to take a linear combination, because otherwise the
boundary is non-empty and the integral of $dF^{(5)}_{RR}$ need not
vanish. Our `six-sphere' intersects precisely once with each
$D3$-brane wrapped on $\Gamma$. Then the total flux through this
`six-sphere' must be equal to the number $n$ of $D3$ branes
wrapped on $\Gamma$. More precisely, we consider the
eight-manifold $T$ which consists of the `six-sphere' with the
elliptic fibration on top of it. Then we must have
\be N_{D3}(\Gamma) = {\chi(T)\over 24} - {1\over 8\pi^2}\int_{T}
{\sf G} \wedge {\sf G} \ee
We will assume ${\chi(T)}=0$, which seems reasonable but which we
have not derived. The four-form $\omega^Y \wedge \omega^Y$ is a
delta-function on $Y_4$ localized on our $7$-brane. The six-sphere
intersects our 7-brane in a four-cycle, which is given by $S^2
\times D$ with $[A_1-A_2] \cdot D = 1$ and $S^2$ the unit
two-sphere in $R^3$ surrounding the origin. The flux integral can
be written as
\be n = {1\over 2\pi} \int_{S^2} F_Y \cdot {1\over
2\pi}\int_{D} {F_Y} \ 2 \half {\rm Tr}(Y^2) \ee
Since
${1\over 2\pi}\int_D F_Y = D \cdot [A_1 - A_2]=1$, we see that the
integral $\int_{S^2} F_Y$ is non-zero, and the
particle obtained by wrapping a $D3$ brane on $\Gamma$ carries
magnetic hypercharge. It might be interesting to do the topological
analysis carefully and see if the colour $Z_3$ charge and the
$SU(2)$ $Z_2$ charge can be recovered also.

Our monopoles are also similar to those of
\cite{Greene:1996dh,Verlinde:2006bc}. The fact that the monopole is
not supported on the GUT brane may seem somewhat puzzling, since one
may expect that it can be understood as a bound state of the fields
in the eight-dimensional gauge theory. A similar situation appears
in the context of the $D3$-ending-on-$D5$ system, to which our
configuration is similar. In that context, a field configuration on
the $D5$-brane satisfying the Nahm equations can be understood as a
`fuzzy funnel', a spike growing out of the $D5$-brane. The two points
of view have a slightly different range of validity.

In addition to monopoles, $F$-theory vacua have several other
interesting types of solitons. There are cosmic strings from
$D3$-branes wrapped on two-cycles, and domain walls interpolating
between vacua with different values of the $G$-flux. It should be
interesting to examine their effect on early universe cosmology.

\newpage

\newsection{ Precision of unification: Threshold corrections}
\seclabel{KKThresholds}

\newsubsection{Leading corrections to gauge kinetic terms}

Even though we have engineered gauge coupling unification at leading
order, we cannot expect unification to be precise, because there are
many massive states charged under the Standard Model gauge group.
The corrections to the leading terms at the GUT scale are organized
in a power series in $\alpha_{GUT}$, and we have
\be {16\pi^2\over g_a^2(M_{GUT})} = {4\pi k_a \over \alpha_{GUT}} +
\delta_a^{(0)} + \delta_a^{(1/2)}\, \alpha_{GUT}^{1/2} +
\delta_a^{(1)}\, \alpha_{GUT} + \ldots \ee
Such corrections are expected in any GUT model, for example in conventional
$4d$ GUTs one will get corrections due to the Higgs sector responsible for
breaking the GUT group to the Standard Model gauge group.
In this section we would like to discuss in detail the leading
threshold corrections, here denoted by $\delta_a^{(0)}$.

Actually there are two dimensionless parameters in our set-up,
$\alpha_{GUT}$ and $M_{GUT}/M_{Pl,4}$. Since we assume the existence
of a decoupling limit, we can make an expansion of the threshold
corrections to the gauge kinetic terms:
\be \delta \sim \delta_{local} + {M_{GUT}\over M_{Pl}} \delta' +
\ldots \ee
The $M_{pl}$ suppressed contributions can be regarded as
sub-subleading. Thus we may focus on the charged states that remain
dynamical in the local geometry.

The assumption of decoupling can fail because some corrections may
become divergent if certain tadpoles are not cancelled in the local
model, but only globally. We will see how this can happen explicitly
below. In fact it could happen that the tadpole
is already cancelled in a non-compact model which is slightly larger
than the first infinitesimal neighbourhood of the gauge brane, as
was seen explicitly in \cite{Buican:2006sn}. In other words, it
could be cancelled already even when the four-dimensional Planck
scale is infinite. To account for the general case, we will have to
introduce a cut-off $\Lambda$ for the loop integrals which in a
concrete global model could be anywhere between the local KK scale
and $M_{Pl}$ a priori.

The charged states in the local geometry can be separated into two
types. First there are the ground states of open strings solitons
which give us our $SU(5)$ gauge fields and ${\bf 10}$ and
${\overline{\bf 5\!}\,}$ matter. Even though the string scale is not
parametrically separated from the $10d$ Planck scale, these modes
still have a weakly coupled description in terms of a higher
dimensional gauge theory with enlarged gauge group containing
$SU(5)$. They give a one-loop correction proportional to
$\alpha_{GUT}^0$ which may be expanded in KK harmonics, and they
will be discussed at length in subsection \ref{OneLoopKK}. The
upshot will be that they are related to Ray-Singer torsion and can
in principle be computed in any given model, even without knowing
the Calabi-Yau metric. Second, there are the massive excitations of
these same open strings solitons, the open strings which gave us the
eight-dimensional gauge fields and ${\bf 10}$ and ${\overline{\bf
5\!}\,}$ matter. For finite $g_s$ they have masses of order the
ten-dimensional Planck scale and were not included in our gauge
theory. Integrating out such modes yields local higher derivative
terms to the eight-dimensional gauge theory action, localized on
$R^4 \times S$ and $R^4 \times \Sigma$. Since we don't have a weakly
coupled description of these modes in $F$-theory, we cannot
calculate the coefficients of such corrections, so we had better
hope that they do not affect unification at order $\alpha_{GUT}^0$.
We will argue that this is indeed the case.

The leading higher derivative terms in eight dimensions that may
affect unification are the ${\rm Tr}(F^4)$ and ${\rm Tr}(F^2)^2$
terms. Since they are local in eight dimensions and BPS protected,
we may use $F$-theory/hetetotic duality to view these same
corrections from the heterotic side, where they are well-studied.
Let us first consider the heterotic string on $T^2$, with Wilson
lines turned on so that the gauge group is broken to $SU(5)$. It is
believed that the ${\rm Tr}(F^4)$ and ${\rm Tr}(F^2)^2$  amplitudes
are BPS saturated at one loop and receive no contributions beyond
one-loop, up to non-perturbative corrections by $NS5$-instantons.
Now heterotic on $T^2$ is dual to $F$-theory on $K3$, and we may
take a low energy limit such that the heterotic string scale goes to
infinity but the mass of the $SU(5)$ $W$-bosons remains finite. In
this limit, the $K3$-surface degenerates to an $A_4$ ALE space. The
only surviving BPS states in short multiplets are the $W$-bosons of
$SU(5)$.

On the heterotic side, there are also tree level contributions to
the ${\rm Tr}(F^2)^2$ terms. They come from ten dimensions as the
gauge fields on the heterotic side live in ten dimensions. Upon
compactifying to eight dimensions and mapping to $F$-theory, tree
level $F^4$ terms are proportional to $V_{{\bf P}^1}^{-2}$ and one
loop $F^4$ terms are independent of $V_{{\bf P}^1}$. The gauge
kinetic terms are also independent of $V_{{\bf P}^1}$. Here $V_{{\bf
P}^1}$ denotes the volume of the base of the $K3$ surface. Therefore
in the local limit considered here, the heterotic tree level $F^4$
terms decouple. We still get contributions at one loop, but only by
short BPS multiplets. In our low energy limit, the only surviving
BPS states are the $SU(5)$ $W$-bosons, and so the amplitude
degenerates to a straightforward one-loop amplitude in the $SU(5)$
gauge theory. (See eg. \cite{Aharony:2003vk} for a review of the
analogous statements in six dimensions). In this limit, we find the
following divergent contribution from the eight-dimensional
supersymmetric Yang-Mills theory
\cite{Green:1982sw,Metsaev:1987ju}
\be \log (\Lambda^2/\mu^2)\,[t_8 +  \epsilon_8]\,{\rm Tr}_{adj}
(F^4) \ee
It is useful to express the trace in the adjoint in terms of
fundamental traces using the identity (for $SU(n)$ gauge group)
\be {\rm Tr}_{adj}(F^4) = 2 n\, {\rm Tr}_f(F^4) + 6\, ({\rm
Tr}_f(F^2))^2 \ee
Heterotic/type II duality predicts this to be the exact answer for
the $F^4$ amplitude in the limit we are considering. (One can probably
argue this more directly from the superspace structure of the $F^4$ terms).

From the perturbative type II perspective this may look somewhat
strange, since we would expect a tree level $F^4$ contribution from
the Born-Infeld action, as well as loop corrections and
non-perturbative corrections due
to $D_{-1}$-instantons. This is not a contradiction, as this
calculation is performed in a different weakly coupled frame. Since
the amplitude is protected one should get the same answer either way
after proper extrapolation, although clearly the full type II
calculation (including $D$-instantons) is much more involved.

The above picture for the relation between the $F$-theory and IIb computations
is confirmed in a number of instances where a comparison
between heterotic and perturbative type II could be done explicitly.
It was found that the heterotic one-loop result actually already
contained and agreed with the type II result based on the
DBI/Chern-Simons action in the weakly coupled type II limit
\cite{Tseytlin:1995fy,Bachas:1996bp,Kiritsis:2000zi,Lerche:1999de},
and sums up further $D_{-1}$ instanton corrections to it. We expect
that here too the heterotic result reproduces the full type II
contribution in the IIb weak coupling limit. This is not completely
unexpected, because the classical $F$-theory geometry sums up the
$D_{-1}$ instanton corrections and so a one-loop calculation in an
$F$-theory background should know about the instantons.
It would be interesting to check this in more detail.%
\footnote{In light of the recent work \cite{CDW} which shows that the Sen IIb limit
is a generalization of the heterotic $SO(32)$ limit,
old results about threshold corrections
in heterotic/type I duality such as \cite{Bachas:1997mc} are particularly relevant. The $D_1$-instantons
in \cite{Bachas:1997mc} correspond to $D_{-1}$-instantons in type IIb. Localization techniques for
summing $D_{-1}$-instanton corrections have recently been discussed in \cite{Fucito:2009rs}.} %
Since the
heterotic calculation is well-established, it is reasonable to
assume the heterotic answer (really a one-loop field theory
calculation) is the reliable result.\footnote{This revises a claim
in v1 of this paper, where we added the DBI $F^4$ correction to the
above one-loop amplitude. Since this DBI correction worsens a
discrepancy with experimentally measured deviations from
unification, this means that precision unification is in better
shape than claimed in v1.}

Once we compactify and turn on a background flux, the above one-loop
computation in the eight-dimensional gauge theory can be expanded in
a sum over KK modes. This computation will be discussed in the next
subsection, although we will use a more four-dimensional point of
view there.

There could be further corrections to the effective action in eight
dimensions at higher order in derivatives. However they will be
suppressed by additional powers of $\alpha_{GUT}$ and so are not of
interest to us. Moreover, they should not affect the
four-dimensional gauge kinetic terms at any rate because that would
change the tension of a BPS $D3$ instanton configuration in the
7-brane, but not its charge.

So far we have discussed contributions from the strings whose ground
states yield the $SU(5)$ gauge fields in eight dimensions. We
further have to include the modes of the open strings stretching
between different 7-branes, whose ground states give rise to charged
matter in the ${\bf 10}$ and $\overline{\bf 5\!}\,$.  The heterotic
perspective seems less useful in this case, so we will argue more
directly that the massive stringy tower cannot affect unification to
the order we are considering.

Strings stretching between intersecting 7-branes could give rise to
local higher derivative corrections of the form
\be\label{typeII}  \int_{R^4 \times \Sigma}\sqrt{ g}\,  F^3 \ee
The terms that can appear are constrained by $5+1$ dimensional
Lorentz invariance. There could also be terms of type
\be \int_{R^4 \times \Sigma} \sqrt{g}\, F^2\, R, \ee
however upon reduction they would give an $SU(5)$ symmetric
contribution to the gauge kinetic terms that can be absorbed in the
bare coupling. Upon reduction of (\ref{typeII}) to four dimensions,
if non-vanishing the four-dimensional gauge kinetic terms can only
be proportional to $\int_\Sigma F_Y$. But we have already seen that
for GUT breaking with fluxes to be viable, we must have $\int_\Sigma
F_Y=0$. More precisely, this holds for the total matter curves
$\Sigma_\bfv$ or $\Sigma_\bt$. In practice these matter curves may
be reducible, and $\int_{\Sigma_{i}} F_Y\not=0$ for each irreducible
piece. However the contribution to a given gauge kinetic term is
proportional only to the total flux through $\Sigma_\bfv$ and
$\Sigma_\bt$:
\be \sum_{i} \int_{\Sigma_{i}} F_Y = \int_{\Sigma_{\bfv,\bt}} F_Y =
0 \ee
so this does not affect the argument that such corrections cannot
affect unification. Therefore we have to go at least to the $F^4$
corrections on $R^4 \times \Sigma$ to get a contribution that is not
$SU(5)$ symmetric. However such higher derivative contributions will
be proportional to at least $\alpha'/R_{KK}^2 \sim
\alpha_{GUT}^{1/2}$ (i.e. they are suppressed by at least
$\alpha_{GUT}^{3/2}$ compared to the leading terms ), so they are
not relevant for our discussion as promised.

We have argued that in local models the leading corrections to
unification come only from one-loop amplitudes of the gauge theory
itself, i.e. from the eight-dimensional super-Yang-Mills multiplet
and from the six-dimensional hypermultiplet matter. Thus it remains
to expand these one-loop corrections as a sum over KK modes. This is
the subject of the following subsections.

\newsubsection{One-loop KK thresholds}
\subseclabel{OneLoopKK}

We would now like to discuss loop corrections to unification from
integrating out Kaluza-Klein modes of the gauge theory. We will see
that they can be determined without knowing the Calabi-Yau metric,
and discuss the role of an anomaly for determining their moduli
dependence.

\newsubsubsection{Generalities}

We briefly recall some relevant formulae. The one-loop running for
the gauge couplings in four dimensions is given by
\be\label{oneloop} {16 \pi^2\over g_a^2(\mu)} \ =\  {16 \pi^2 \,k_a \over g^2} +
b_a \log(\Lambda^2/\mu^2) + S_a \ee
with $(k_1,k_2,k_3) = (5/3,1,1)$. The one-loop beta-function
coefficients are given by
\be\label{beta} b_a = 2\, {\rm Tr}_{m=0}\,(Q_a^2\,(-1)^F({1\over 12}
- \chi^2)) \ee
where $Q_a$ is normalized so that ${\rm Tr}(Q_a^2) = k_a/2$. In the
MSSM we have $(b_1,b_2,b_3) = (11,1,-3)$. Finally the threshold
corrections are given by
\be\label{thresholddef} S_a = 2\, {\rm Tr}_{m\not =0}\, Q_a^2\, (-1)^F({1\over 12}-\chi^2) \
\log(\Lambda^2/m^2) . \ee
In our case this will be a sum over $N=1$ vector multiplets and
chiral multiplets. The ${\rm Tr}\, Q_a^2$ piece in
(\ref{beta}),(\ref{thresholddef}) factors out for each
supermultiplet, and we have
\be {\rm Tr}(-1)^F({1\over 12}-\chi^2) = \left\{
\begin{array}{cl}
  -3/2 \quad & ({\rm vector\ multiplet}) \\
  1/2
 \quad & ({\rm chiral\
multiplet})
\end{array} \right.
 \ee
For later purposes we would like to expand
(\ref{oneloop}) a little. If all matter would come in complete
$SU(5)$ multiplets, then we would have $(b_1,b_2,b_3) \propto
(k_1,k_2,k_3)$. However the gauge multiplets and the electro-weak
Higgses do not come in complete multiplets.  Then
we can decompose
\be b_a = b_a^{(c)} + b_a^{(g)} + b_a^{(h)} \ee
where $b^{(g)}_a$ is the contribution from the $SU(3) \times SU(2)
\times U(1)$ gauge fields, $b^{(h)}_a$ is the contribution from the
Higgs doublets, and $b^{(c)}_a$ are the contributions from the
remaining complete multiplets (the three generations of ${\bf 10}$
and $\overline{\bf 5\!}\,$). In other words we have
\be b_a^{(g)}=(0,-6,-9), \qquad b^{(h)}_a = (1,1,0), \qquad
b_a^{(c)} = (10,6,6) \propto (k_1,k_2,k_3) \ee

To compute the one-loop threshold corrections, we need the massive
spectrum of the Dirac operator of the eight-dimensional gauge
theory. As explained in \cite{Donagi:2008ca}, after compactification
this Dirac operator decomposes into various Dolbeault operators,
coupled to holomorphic bundles. Let us recall some generalities.
Consider a holomorphic bundle $V$ on $S$, and its associated
Dolbeault operator
\be \Omega_S^{0,0}\otimes R(V)\quad
\mathop{\longrightarrow}^{\delb}\quad \Omega_S^{0,1}\otimes R(V)
\quad \mathop{\longrightarrow}^{\delb} \quad \Omega_S^{0,2}\otimes
R(V) \ee
where $R$ is some representation of the structure group of $V$. Let
us denote by%
\footnote{Unfortunately this is {\it not} exactly the standard
Laplacian, which differs by a factor of a two (i.e. $\Delta_{d} =2
 \Delta_{\delb}$). Since the squared masses of KK modes are
conventionally defined to be the eigenvalues of $\Delta_{d}$, we
should really replace the radius $R$ by $R/\sqrt{2}$ in all the
formulae below, and similarly replace $M_{KK}$ by
$\sqrt{2}M_{KK}$.}
\be \Delta_{n,R(V)} =
(\delb + \delb^\dagger)^2 = \delb \delb^\dagger + \delb^\dagger
\delb \ee
the Laplacian acting on $\Omega_S^{0,n}\otimes R(V)$, where the
adjoint is defined with respect to the hermitian metric. Then we are
interested in the spectrum of $\delb + \delb^\dagger$, or
equivalently the spectrum of $\Delta_{n,R(V)}$ which gives the
squares of the eigenvalues. More precisely, we are interested in the
combination of eigenvalues appearing in (\ref{thresholddef}). This
is the logarithm of the determinant of $\Delta_{n,R(V)}$. Naively
the sum over eigenvalues is not well defined. Traditionally it is
defined using zeta function regularization \cite{RaySinger}. That is
we consider
\be \zeta_\Delta(s) = \sum_{\lambda_i>0} {1\over \lambda_i^s} \ee
For elliptic operators one may show that $\zeta_\Delta(s)$ extends
to a meromorphic function on the complex plane. Then we define
\be \log {\rm det}' \Delta = -\zeta'_\Delta(0) \ \sim \  \log \prod
\lambda_i \ee
Here we explicitly excluded the zero modes, because in
(\ref{thresholddef}) we want only the massive modes. This is
indicated by the prime on the determinant.

We need to say a few words about the zeta-function regularization
scheme used here.  There are other regularization schemes for the
sum over KK modes, see eg. \cite{Ghilencea:2003xy} for a discussion
of some different schemes and their relations. The zeta-function
regularization used in the definition of the holomorphic torsion is
special in that it throws out power law divergences, and keeps only
the logarithmic divergences. Power law divergences are however quite
natural in a KK theory; in the present case they signal that the
gauge coupling becomes dimensionful above the KK scale and has power
law running. Thus the regularization provided by the microscopic
theory underlying $F$-theory will certainly not be zeta-function
regularization.\footnote{We would like to thank S.~Raby for a
conversation on this point.}

However we have essentially already argued that the power law
divergences cannot affect unification. We can also argue this
directly; a term with power law divergence that could affect the
relations between the $4d$ gauge couplings must be of the schematic
form
\be \Lambda^2 F^3 \ee
on $S$. Since $g_{i\bar j}F_Y^{i\bar j}=0$ for the internal flux,
such a term would have no effect on the $4d$ gauge couplings.

\newsubsubsection{KK modes of $SU(5)$ gauge fields}

Now we are ready to write expressions for the one-loop contributions
of Kaluza-Klein modes to the gauge couplings. As discussed in
\cite{Donagi:2008ca} there are several scenarios we can consider.
Here we will take the case of an $SU(5)$ gauge field propagating in
the bulk and all chiral matter localized on 7-brane intersections.
It is not hard to write down analogous formal expressions for other
scenarios, with one or more chiral fields and a larger gauge group
supported in the bulk of the 7-brane (though of course evaluating
them in explicit examples is another matter). Let us first consider
the eight-dimensional $SU(5)$ gauge fields. Turning on a $U(1)$ flux
proportional to hypercharge breaks $SU(5)$ to $SU(3)\times
SU(2)\times U(1)$, and we decompose the adjoint representation of
$SU(5)$ as
\ba {\bf 24} &=& ({\bf 3},{\bf 1})_0 \oplus ({\bf 1},{\bf 8})_0 \oplus
({\bf 1},{\bf 1})_0 \oplus({\bf 2},{\bf 3})_{\bf -5/6} \oplus  ({\bf
2},\overline{\bf 3\!}\,)_{\bf 5/6} \eol[2mm] &\equiv& {\bf R}_0\
\oplus \ {\bf R}_{-5/6}\ \oplus\ {\bf R}_{5/6} \ea
We would like to sum up the contributions of the Kaluza-Klein modes
of the $SU(5)$ gauge fields. For brevity we denote by $\Delta_{n,Y}$
the Laplacian acting on $\Omega^{0,n}(L^Y)$. Then for each
eigen-vector of $\Delta_{0,Y}$ we get a vector multiplet in four
dimensions in the representation ${\bf R}_Y$. Similarly for each
eigenvector of $\Delta_{1,Y}$ we get a chiral multiplet, and for
each eigenvector of $\Delta_{2,Y}$ we get an anti-chiral multiplet
in the representation ${\bf R}_Y$, or equivalently a chiral multiplet in
${\bf R}_Y^*$. Note however that ${\rm Tr}_{{\bf R}_Y}Q_a^2 = {\rm Tr}_{{\bf R}_Y^*}
Q_a^2$. Thus we consider the linear combination
\be\label{KYdef} {\bf K}_Y =
 {3\over 2}\log {\rm det}' \ {\Delta_{0,R(V)}\over \Lambda^2}
- {1 \over 2} \log {\rm det}'\ {\Delta_{1,R(V)}\over \Lambda^2}
-{1\over 2}\log {\rm det}'\ {\Delta_{2,R(V)}\over \Lambda^2} \ee
We also need
\be {\rm Tr}_{{\bf R}_0} (Q_a^2) = (0,2,3), \qquad {\rm Tr}_{{\bf
R}_{5/6}} (Q_a^2) = (25/6,3/2,1) \ee
Altogether then we find
\be S_a^g = \sum_{Y=0,\pm 5/6}2\, {\rm Tr}_{{\bf R}_Y} (Q_a^2)\
{\bf K}_Y \eol \ee
or more explicitly
\be\label{gaugethreshold} S_1^g = {25\over 3}\,({\bf K}_{5/6}+{\bf
K}_{-5/6}), \qquad S_2^g = 4\, {\bf K}_0 + 3\,({\bf K}_{5/6}+{\bf
K}_{-5/6}), \qquad S_3^g = 6\,{\bf K}_0 + 2\,({\bf K}_{5/6}+{\bf
K}_{-5/6}) \ee
where $g$ indicates that these are the contributions from the gauge
fields. Note incidentally that since $c_1(L^{5/6})$ corresponds to a
simple root, the Del Pezzo has a diffeomorphism symmetry (namely the
Weyl reflection generated by $c_1(L^{5/6})$ itself) which takes
$L^{5/6} \to L^{-5/6}$, so we expect ${\bf K}_{5/6}$ and ${\bf
K}_{-5/6}$ to be identical.

The expression we have written can be related to holomorphic
Ray-Singer torsion. The appearance of Ray-Singer torsion is not
unexpected; in the heterotic string \cite{Bershadsky:1993cx} and
in $M$-theory on manifolds of $G_2$ holonomy
\cite{Friedmann:2002ty}, the threshold corrections can also be
related to Ray-Singer torsion. To see this, let us first note that
the above vector and chiral multiplets pair up in a natural way
into massive multiplets. We expand the linear fluctuations of the
$8d$ gauge field in terms of a complete set of eigenmodes of the
Laplacian. In fact, it is convenient to look at the $(0,1)$ part
of the gauge field with respect to a complex structure on ${\bf
R}^4 \times S_2$, so that the relevant Laplacian is given by
$\Delta_{\delb} = \not \! D^2$:
\be A_{8d} \ = \ A_{cl} + \sum_n a_{(n)\mu}dx^\mu \wedge
\psi_{(n)}^{0} + \sum_m \phi_{(m)} \psi_{(m)}^{1} \ee
where the zero-forms $\psi_{(n)}^{0}$ and the $(0,1)$-forms
$\psi_{(m)}^{1}$ are eigenmodes. The $(1,0)$ part of the $8d$
gauge field is recovered by hermitian conjugation. In our
application, $A_{cl}$ will be the hypercharge gauge field $A^Y$.
Similarly, we expand the Higgs field
\be
\Phi_{8d} \ = \ \Phi_{cl} + \sum_m\chi_{(m)} \psi_{(m)}^2
\ee
where $\psi_{(m)}^2$ are $(0,2)$ forms which are eigenmodes of
$\Delta_{\delb}$. We will take $\Phi_{cl}=0$. In addition, the set
of $(0,p)$ forms with values in $R(V)$ admits an orthogonal
decomposition known as the Hodge decomposition:
\be
\Omega^{0,p}(R(V)) \ = \ {\bf H} \oplus im(\delb) \oplus im(\delb^\dagger)
\ee
where ${\bf H}$ are the zero modes (harmonic forms) of
$\Delta_{\delb}$. Since the Laplacian commutes with $\delb$ and
$\delb^\dagger$, this decomposition is compatible with the
decomposition in terms of eigenmodes of $\Delta_{\delb}$.

Now, given an eigenmode $\psi_{(n)}^{0}$ of the Laplacian with
eigenvalue $M_{(n)}^2$, we get another eigenmode with the same
eigenvalue:
\be \not \! D \psi_{(n)}^{0}\ =\ \psi_{(n)}^{1} \ee
Similarly, given such an exact one-form $\psi_{(n)}^1$, by
applying $\not \! D$ we recover the zero-form $\psi_{(n)}$.
Furthermore, $\delb\delb^\dagger$ annihilates $im(\delb^\dagger)$.
Therefore the massive spectrum of $\delb^\dagger \delb$ acting on
zero-forms is the same as the massive spectrum of
$\delb\delb^\dagger$ acting on $(0,1)$ forms.

This has a simple physical interpretation. Let us consider the
effect of an eight-dimensional gauge transformation with parameter
$\lambda$. Expanding $\delta A_{8d} = (d\lambda)^{0,1}$ in KK
modes, we see that the $4d$ fields have the following
transformation laws:
\be \delta a^{(n)} \ = \ \delb \lambda^{(n)}, \qquad \delta
\phi^{(n)} \ = \ \lambda^{(n)} \ee
In particular, we find that $\phi^{(n)}$ describes the
longitudinal part of a massive $4d$ gauge field, whose $\pm 1$
polarizations are given by $a_{(n)}$. Thus the fact that the
massive spectrum of $\delb^\dagger \delb$ acting on zero-forms is
the same as the massive spectrum of $\delb\delb^\dagger$ acting on
$(0,1)$ forms corresponds physically to the fact that a vector
superfield eats a chiral field in order to obtain a mass.
Similarly, the massive spectrum of $\delb^\dagger \delb$ acting on
$(0,1)$-forms is the same as the massive spectrum of $\delb
\delb^\dagger$ acing on $(0,2)$-forms. This corresponds to the
fact that two chiral fields with opposite charges need to pair up
in order to gain a mass.

Thus we may rewrite (\ref{KYdef}) as
\be\label{KYdeftwo} {\bf K}_Y = 2\, \log {\rm det}' \
{\Delta_{0,R(V)}\over \Lambda^2} -  \log {\rm det}'\
{\Delta_{1,R(V)}\over \Lambda^2}\ee
Now we define the holomorphic Ray-Singer torsion as
\be\label{RSDefinition} {\bf  T}_{R(V)} = \half \sum_{k=0}^2
(-1)^{k+1} k \log {\rm det}' \ \Delta_{k,R(V)} \ee
The determinants are defined as before using zeta-function
regularization. Note that the conjugate linear $\bar{*}$-operator
commutes with the Laplacian, and maps the eigenvectors and
eigenvalues of $\Delta_{n,R(V)}$ to those of
$\Delta_{2-n,R^*(V)\otimes K_S}$. Therefore we have
\ba {\bf  T}_{R(V)} &=& \half (-1)^{n+1}\sum_{k=0}^n (-1)^{n-k+1}
(n-k) \log {\rm det}' \ \Delta_{n-k,K \otimes R(V)^*} \eol & =&
(-1)^{n+1} {\bf T}_{R(V)^* \otimes K } \ea
and
 \ba 2 \,{\bf T}_{R^*(V)\otimes K_S} &= &
-2\log {\rm det}'\ \Delta_{0,R(V)} +  \log {\rm det}'\
\Delta_{1,R(V)}  \eol \ea
is the the same as ${\bf K}_Y$, modulo the extra dependence on
$\Lambda^2$. The relation between them can be found using
\be\label{scaledeigenvalues}
 \zeta_{\Delta/\Lambda^2}(s)= \Lambda^{2s}\zeta_\Delta(s)\qquad
\Rightarrow \qquad -\zeta'_{\Delta/\Lambda^2}(0) = -\zeta_\Delta(0)
\log \Lambda^2 - \zeta_\Delta'(0) \ee
Hence we may write
\be\label{KYtorsion} {\bf K}_Y = {\bf K}_{-Y} = 2\,\left. {\bf
T}(S_2,L^Y) \right|_{R \to R \Lambda}\ee
where by $R=V_{S_2}^{1/4}$ we mean the scale of $S_2$.

\newsubsubsection{Anomaly and dependence on $\Lambda$}

In order to understand the dependence on the cut-off, we need to
know the scaling dependence of the holomorphic torsion. Computing
the torsion explicitly can quickly get somewhat complicated. But
fortunately the metric dependence of the holomorphic torsion is
determined by an anomaly (i.e. the torsion is a section of a
non-trivial line bundle over the configuration space) and can be
obtained rather easily. The anomalous variation of the torsion is
described by the metric anomaly formula derived by Bismut, Gillet
and Soul\'e \cite{BGS3}.

In appendix \ref{MetricAnomaly} we have worked out the dependence of
the torsion on a rescaling of the metric, and hence the dependence
on $\Lambda$. We find that
\ba\label{BGS} 2\, {\bf T}(X,V) &\sim& \left[ \sum (-1)^{q}(2-q)
h^q(X,V) \right. \eol & & \qquad \qquad\left.
 - \int_S (ch_2(V) +  {5\over 12} c_1(V) c_1(T) +
 {1\over 24}c_1(T)^2 +{1\over 12} c_2(T))\right] \log
 R^2\Lambda^2
\eol \ea
Let us further rewrite the zero mode dependence as
\be [2h^0(V)-h^1(V)]\log(R^2\Lambda^2) = \left[\half \chi(V)
-\left(-{3\over 2}h^0(V)+\half h^1(V) + \half h^2(V)\right)\right]
\log(R^2\Lambda^2) \ee
We also have the Riemann-Roch formula
\be \chi(V) = \int_S {\rm ch}_2(V) + \half c_1(V)c_1(T) + {1\over
12}c_1(T)^2 + {1\over 12}c_2(T) \ee
where we may also recall that $\int_S c_2(T) = \chi(S)$. Then we may
rewrite the contribution of the zero modes and the KK modes of the
eight-dimensional gauge fields to the running as
\ba\label{KKgaugerunning}
 {16\pi^2\over g_a^2(\mu)} =
 {16\pi^2k_a \over g^2}\!\!\!\!\!\! & \!\!\! &\!\!\!
 + 2 \sum_{Y=0,\pm {5\over 6}}\!\!{\rm Tr}_{R_Y}(Q_a^2)\left(-{3\over
2}h^0(V_Y)+\half h^1(V_Y) + \half
h^2(V_Y)\right)\log(M_{KK}^2/\mu^2) \eol & &\!\!\! + 2 \sum_{Y=0,\pm
{5\over 6}}\!\!{\rm Tr}_{R_Y}(Q_a^2)\, 2\, {\bf T}(S,V_Y)|_{R\to 1}
\eol & &\!\!\!  + 2 \sum_{Y=0,\pm {5\over 6}}\!\!{\rm
Tr}_{R_Y}(Q_a^2)\!\left[\int_S\! -{1\over 2} {\rm ch}_2(V_Y)-{1\over
6}c_1(V_Y) c_1(T) -{1\over 24}c_2(T)\right] \log(R^2 \Lambda^2)\eol
\ea
with $M_{KK} = 1/R_S$. Here we have collected the finite pieces in
the first two lines and the divergent pieces in the third line. The
first line describes the running of the massless modes up to the KK
scale. The second line describes the effect of integrating out the
KK modes at the KK scale; this term depends only on complex
structure moduli. Finally the third line describes the running above
the KK scale. We see that the divergent piece is entirely
proportional to a local density on $S$, as required by the Lorentz
invariance of the underlying eight-dimensional $SU(5)$ gauge theory.

The important thing to notice about (\ref{KKgaugerunning}) is that
the divergent contributions in the third line are not necessarily
proportional to $k_a$, and hence cannot necessarily be absorbed in
the bare gauge coupling $g$. Indeed, for the case of interest with
GUT breaking by hypercharge flux, for $Y=0$ we must take $V_Y \to
\cO$ where $\cO$ is the trivial line bundle on $S$, and for $Y=5/6$
we must take $V_Y \to \zeta = L^{5/6}$ which satisfies
$c_1(\zeta)\cdot c_1(T)=0$ and $c_1(\zeta)^2 = -2$. Then clearly we
find a divergence which is not proportional to $k_a
\log(\Lambda^2)$.

Of course we already predicted such a divergence earlier. In the
$8d$ Yang-Mills theory with 16 supersymmetries, the ${\rm Tr}(F^2)$
terms are not corrected at one loop, but as we discussed there is a
one loop correction of the form $\log (\Lambda^2)t_8{\rm
Tr}_{adj}(F^4)$. After compactifying on a background with GUT
breaking flux, this gives a divergent contribution to the
four-dimensional kinetic terms of the form
\be \log \Lambda^2\, n_a\, {\rm Tr} F_a^2 \ee
Note in particular that for the $U(1)_Y$ gauge field the ${\rm
Tr}(F^2){\rm Tr}(F^2)$ piece gives a contribution where the two
four-dimensional gauge fields come from different traces, whereas
such contributions for the non-abelian gauge fields are always zero.
Now one may check using our earlier expressions that ${\rm Tr}_{{\bf
R}_0}(Q_a^2)$ (or equivalently ${\rm Tr}_{{\bf R}_{5/6}}(Q_a^2)$) is
in fact a linear combination of $k_a$ and $n_a$. Thus the
divergences can be understood as one-loop contributions to the local
${\rm Tr}(F^2){\rm Tr}(R^2)$ and ${\rm Tr}_{adj}(F^4)$ terms in
eight dimensions, which is precisely what we find in the third line
of (\ref{KKgaugerunning}).

There are additional divergent contributions coming from the modes
living on the matter curves. Indeed anomaly cancellation associated
to the ${\rm Tr}(F^4)$ terms gives the relation
\be [\Sigma_\bfv] - 3 [\Sigma_\bt]-5c_1(T) = 0 \ee
so it would in fact be inconsistent not to include these additional
modes. However we will analyze the contributions to these modes
later and see that they cannot cancel the non-$SU(5)$ symmetric
divergences in (\ref{KKgaugerunning}) proportional to ${\rm
ch}_2(V_Y)$.

 Thus the divergence is
unavoidable in the local model, and we have to ask how it is
naturally cut off in an $F$-theory compactification. Naively one may
have thought that the running is naturally cut off when we include
modes in the local geometry with masses of order the ten-dimensional
Planck scale, which we denote $m_{10}$. However the $\log \Lambda^2$
dependence of the one-loop amplitude in the eight-dimensional gauge
theory has a simple physical interpretation: it indicates that the
back-reaction grows far from the 7-brane. In fact in a IIb weak
coupling limit it can be interpreted as the Green's function of
closed string field like the RR axion, i.e. the Green's function of
the Laplacian in the two dimensions transverse to the 7-brane. Thus
this divergence generally is not cut off by Planck scale modes in
the local geometry. In fact as we already mentioned, due to the BPS
properties of the ${\rm Tr}(F^4)$ term, such massive modes can not
contribute at all at one loop \cite{Douglas:1996yp,Bachas:1996bp},
and the running is unmodified at $m_{10}$.

Instead the divergence is cancelled by including suitable 7-branes
with opposite charges at infinity, i.e. it corresponds to a 7-brane
tadpole cancellation condition in a compact model. This introduces
new open string soliton sectors whose ground states are BPS, and so
can contribute to the amplitude. If the tadpole is cancelled
globally (i.e. only for finite $M_{Pl}$), a natural cut-off would
generally be the mass of an open string soliton stretching across
$B_3$, which is of order $\Lambda \sim m_{10}^2 V_{B_3}^{1/6}$.
However if it is cancelled in a higher order neighbourhood of the
GUT brane, as in \cite{Buican:2006sn}, then it could be much
smaller. Hence we introduce a fudge parameter $\lambda$, which is
model dependent, and write our cut-off as
\be \Lambda\ =\ \lambda \cdot M_{KK} \ee
where
\be m_{10}/M_{KK}\ <\ \lambda\ \leq\ R_{B_3}m_{10}^2/M_{KK}. \ee
Here $M_{KK}= 1/R_{S_2}$ is the scale where the Kaluza-Klein modes
of the eight-dimensional gauge theory become important.\footnote{In
v1 of this paper we took the cut-off $\Lambda$ at the lower end. It
was emphasized in \cite{Conlon:2009qa} that $\Lambda$ could be much
larger, although the divergence in \cite{Conlon:2009qa} is due to a
$D5$-tadpole and thus different from the divergence encountered
here. We thank J.~Conlon for conversations about this issue.}

After some rearranging , we can express the running incorporating
only the contributions from the $SU(5)$ gauge fields as
\be
 {16 \pi^2\over g_a^2(\mu)} \ =\  {16 \pi^2 \,k_a \over \tilde g^2} +
b^g_a\, \log(\lambda^{2/3} M_{KK}^2/\mu^2)  + \delta_a^g \ee
where
\be \delta_a^g = 2\, b_a^{5/6}\,\left.({\bf T}_{5/6}-{\bf
T}_0)\right|_{R\to 1} \ee
and we used our results on the scaling dependence of the holomorphic
torsion to put the $R$ and $\Lambda$ dependence in the second term
on the right hand side, and we absorbed several universal pieces by
shifting $g \to \tilde g$.
We have $b_a^{5/6}= (50/3,6,4)$, and as before $M_{KK} = 1/R_S =
V_{S_2}^{-1/4}$. The logic behind writing the formula this way is
that formally if we let $L^{5/6} \to \cO$, then ${\bf T}_{5/6} \to
{\bf T}_\cO$ plus a contribution for the extra zero modes in the
representation ${\bf R}_{5/6}$.

Assuming generic values for the complex structure moduli, the
correction $\delta_a^{(g)}$ can only give a small numerical
correction to the relation between the GUT scale and the KK scale.
However $\lambda$ may depend on the ratio $R_{S_2}R_{B_3}/l_{10}^2$
and give a parametrically large correction, with the relation
\be M_{GUT} \sim \lambda^{1/3} M_{KK} \ee
Since we regard the GUT scale as fixed by measurements of couplings
at low energies, this means that the KK scale may end up
parametrically lower than $10^{16}\, GeV$, although we will see that
in practice the ratio is presumably not very large.

\newsubsubsection{KK modes of matter fields}

Similarly we may write expressions for the Kaluza-Klein towers of
the modes localized on the matter curves, $\Sigma_{\overline{\bf
5\!}\,}$ and $\Sigma_{{\bf 10}}$. Actually the matter curve
$\Sigma_{\overline{\bf 5\!}\,}$  is usually singular and it turns
out that charged matter naturally propagates on the normalization of
the matter curve \cite{Hayashi:2008ba}, so in the following we shall
actually use the notation $\Sigma_{\overline{\bf 5\!}\,}$ to denote
this normalization. We decompose the $SU(5)$ multiplets as
\ba
{\bf 10} &=& ({\bf 2},{\bf 3})_{\bf 1/6} + ({\bf 1},{\bf 1})_{\bf 1} + ({\bf
1},\overline{\bf 3\!}\,)_{\bf -2/3}  \eol
\overline{\bf 5\!}\, &=& ({\bf 2},{\bf 1})_{\bf -1/2} + ({\bf
1},\overline{\bf 3\!}\,)_{\bf 1/3}
\ea
Again it is a convenient shorthand to omit the $SU(3)\times SU(2)$
transformation properties and denote these representations by their
hypercharge, ${\bf R}_Y$. Let us concentrate on $\Sigma_{\bf 10}$.
The total flux on this curve consists of the flux of a line bundle
$M_{\bf 10}$ which we may think of as coming from a non-compact
7-brane, and the flux of the hypercharge gauge field (corresponding
to a line bundle $L^Y$). We consider the Laplacian $\Delta_{n,Y} =
(\delb + \delb^\dagger)^2$ acting on \be \Omega^{0,n}_\Sigma \otimes
M_{\bf 10}|_\Sigma \otimes L^Y|_\Sigma \otimes K_\Sigma^{1/2}. \ee
There should be no confusion with the Laplacians $\Delta_{n,Y}$ on
$S$ introduced earlier, because they act on representations with
different values of $Y$. For each eigenvector of $\Delta_{n,Y}$ we
get a chiral or anti-chiral field (depending on whether $n=0$ or
$n=1$) in the representation ${\bf R}_Y$. Thus we will need the
following combination of determinants:
\be {\bf K}_{\Sigma_{\bf 10},Y} = -\half \log {\rm det}'\
{\Delta_{0,Y}\over \Lambda^2} - \half \log {\rm det}' \
{\Delta_{1,Y}\over \Lambda^2} \ee
In fact it is not hard to see that the massive spectra of
$\Delta_{0,Y}$ and $\Delta_{1,Y}$ are identical. Intuitively, this
is the statement that we need a chiral and an anti-chiral fermion in
order to write down a mass term. Hence modulo the $\Lambda$
dependence we are actually interested in twice the holomorphic
Ray-Singer torsion on $\Sigma_{\bf 10}$:
\be {\bf K}_{\Sigma_{\bf 10},Y} = -2\, {\bf T}_{M_{\bf 10} \otimes
L^Y \otimes K_\Sigma^{1/2}} + \zeta(0) \log \Lambda^2 \ee
The resulting contributions to the threshold corrections are
summarized in table \ref{KKThresTable}. These expressions really
apply to the total matter curves, so it includes possible matter
curves where the Higgses or additional messenger fields for
supersymmetry breaking are localized, if these are different from
the curves where quarks and lepton are localized.

\begin{figure}[t]
\addtocounter{tabnum}{1} \tablabel{KKThresTable}
\begin{center}
\renewcommand{\arraystretch}{1.5}
\begin{tabular}{|c|c|c|c|}
  \hline
         & $S_a^g$ & $S_a^{\bf 10}$ & $ S_a^{\overline{\bf 5\!}\,}$ \\
  \hline \hline
  $U(1)_Y$ \qquad & ${50\over 3}\,{\bf K}_{5/6}$ & ${1\over 3}\, {\bf K}_{\Sigma_{\bf 10},1/6} +
2\,{\bf K}_{\Sigma_{\bf 10},1} + {8\over 3}\, {\bf K}_{\Sigma_{\bf
10},-2/3}$ &  ${\bf K}_{\Sigma_{\overline{\bf 5\!}\,},-1/2} +
{2\over 3}\,{\bf K}_{\Sigma_{\overline{\bf
5\!}\,},1/3} $\\
  $SU(2)$ & $ 4\, {\bf K}_0 + 6\,{\bf K}_{5/6}$ & $3 \,{\bf K}_{\Sigma_{\bf 10},1/6}$ & ${\bf K}_{\Sigma_{\overline{\bf
5\!}\,},-1/2}$ \\
  $SU(3)$ & $6\,{\bf K}_0 + 4\,{\bf K}_{5/6} $ & $2\, {\bf K}_{\Sigma_{\bf 10},1/6} + {\bf
K}_{\Sigma_{\bf 10},-2/3}$ & ${\bf
K}_{\Sigma_{\overline{\bf 5\!}\,},1/3}$ \\
  \hline
\end{tabular}\\[5mm]
\parbox{10cm}
{\bf Table \arabic{tabnum}: \it KK Threshold corrections.}
\renewcommand{\arraystretch}{1.0}
\end{center}
\end{figure}

Again it is convenient to separate out the divergent pieces. Using
the result from appendix \ref{MetricAnomaly}, we have
\be K_{\Sigma,Y} = -2 \, {\bf T}(\Sigma,V)_{R\to 1} - \left[ h^0(V)
- \int_\Sigma \half c_1(V) + {1\over 6}c_1(T)\right] \log
(R^2\Lambda^2) \ee
Further we may rewrite $h^0 = \half \chi(V) + \half h^0 + \half h^1$
and use Riemann-Roch:
\be \chi(V) = \int_\Sigma c_1(V) + \half c_1(T) \ee
Then we can write the contribution of the massless and massive modes
on $\Sigma$ to the running as
\ba {16\pi^2\over g_a^2(\mu)} =
 {16\pi^2k_a \over g^2}\!\!\!\!\!\! & \!\!\! &
 + \sum_Y 2 {\rm Tr}_{R_Y}(Q_a^2)\left(\half h^0(V_Y) + \half
h^1(V_Y)\right) \log({M_{KK}^2/ \mu^2}) \eol
& & -\sum_Y 2 {\rm Tr}_{R_Y}(Q_a^2)\, 2\, {\bf
T}(\Sigma,V_Y)|_{R\to 1} \eol
& & -\sum_Y 2 {\rm Tr}_{R_Y}(Q_a^2)\, {1\over 12}\chi(\Sigma)\, \log
(R_\Sigma^2\Lambda^2) \ea
where we used $\int_\Sigma c_1(T) = \chi(\Sigma)$. Also $M_{KK}$ is
strictly given by $1/R_\Sigma$, but normally we would expect that
$R_\Sigma \sim R_S$. From the third line we see that the divergent
terms are $SU(5)$ symmetric and can be absorbed in the bare
coupling, as we promised earlier.

Putting all the one-loop KK corrections together, we get
\be {16 \pi^2\over g_a^2(\mu)} \ =\  {16 \pi^2 \,k_a \over g^2} +
b_a \log(\Lambda^2/\mu^2) + S^g_a + S_a^{\bf 10} +S_a^{\overline{\bf
5\!}\,} \ee
where $S^g_a, S_a^{\bf 10}$ and $S_a^{\overline{\bf 5\!}\,}$ are
given in table \ref{KKThresTable}. According to our earlier
results, we may write this as
\ba\label{TotalOneLoopV1} {16 \pi^2\over g_a^2(\mu)}  =  {16 \pi^2
\,k_a \over \tilde g^2}\!\!\!\!\!\! & &\!\!\! +\, (b^{(g)}_a +
b_a^{(c)}+b^{(h)}_a) \log{M_{KK}^2\over \mu^2} +\delta_a^g +
\delta_a^{\bf 10} + \delta_a^{\overline{\bf 5\!}\,} \eol & & \qquad
\qquad \qquad \quad- 2\, {\rm Tr}_{R_{5/6}}(Q_a^2)\int_S {\rm
ch}_2(L^{5/6}) \log( \Lambda^2/M_{KK}^2) \ea
By $\delta_a^{\bf 10},\delta_a^{\overline{\bf 5\!}\,}$ we mean the
the same expressions as in table \ref{KKThresTable}, except with the
scale dependent $\log \Lambda^2 R_{KK}^2$ terms removed. (In other
words, we take ${\bf K}_{\Sigma,Y} = -2\, {\bf T}_{\Sigma,Y}$ with
the volume of the matter curves set to one, $R_{KK}=1$). Also $b_a =
b_a^g + b_a^{(c)} + b_a^{(h)} =(11,1,-3)$ are the beta function
coefficients of the MSSM, and
\be \int_S {\rm ch}_2(L^{5/6}) = \half c_1(L^{5/6})^2 = -1 \ee
We would like to further rewrite the formula so that we can compare
it to running in a conventional four-dimensional model. This can be
done by absorbing various further pieces in the bare coupling. Note
that
\be 2 {\rm Tr}_{R_{5/6}}(Q_a^2)\log(\Lambda^2/M_{KK}^2) = -{3\over
2} \, 2\, {\rm Tr}_{R_{0}}(Q_a^2)\log(\lambda^{2/3}) \ {\rm mod}\,
k_a \ee
The chiral fields come in complete $SU(5)$ multiplets, so
\be b_a^{(c)} \log(\lambda^{2/3}) = 0 \ {\rm mod}\, k_a \ee
The higgses do not come in a complete multiplet, but we may write
\be b_a^{(h)}\log(\lambda^{2/3}) + b_a^{(T)}\log(\lambda^{2/3})=0 \
{\rm mod}\, k_a \ee
where $T$ stands for the triplet $SU(5)$ partners of the Higgses.
Then by absorbing various $SU(5)$ symmetric pieces in the bare
coupling, we can rewrite the formula as
\ba\label{TotalOneLoopV2} {16 \pi^2\over g_a^2(\mu)}  =  {16 \pi^2
\,k_a \over \tilde g^2}\!\!\!\!\!\! & &\!\!\! +\,
b_a^{MSSM}\log{M_{GUT}^2\over \mu^2} + b_a^{(T)} \log{M_{GUT}^2\over
M_{KK}^2} +\delta_a^g + \delta_a^{\bf 10} + \delta_a^{\overline{\bf
5\!}\,}\eol\ea
where
\be M_{GUT} = \lambda^{1/3} M_{KK} \ee
and $b_a^{(T)} = (2/3,0,1)$. In other words, comparing with the
analogous formula for conventional four-dimensional models, it looks
like unification gets modified by the appearance of an effective
pair of Higgs triplets with masses $M_T^{eff} = M_{KK}$ below the
GUT scale, as well as some small corrections $\delta_a$ from the GUT
breaking sector.

 One may contrast the situation in $F$-theory models with
earlier work on compactifications of $M$-theory on ALE-fibered
manifolds of $G_2$ holonomy \cite{Friedmann:2002ty}. In that case
charged matter is localized at points on the $G_2$ manifold and so
does not contribute a tower of KK modes to the one-loop running, and
there was no parametric separation between the KK scale and the GUT
scale. The situation in $F$-theory is a bit more complicated.

Recently the moduli dependence of the effective couplings has been
considered from the point of view of $4d$ supergravity, where the
NSVZ-KL formula relates holomorphic couplings to physical couplings
in the effective four-dimensional supergravity theory
\cite{Kaplunovsky:1994fg,Kaplunovsky:1995jw}:
\be {1\over g_a^2(\mu)}\ =\ {\rm Re}(f_a) + b_a \log
\left({\Lambda\over \mu}\right) + \ \log g_a^2 - c_a \hat {\cal K} -
\log \det Z \ee
In particular, it was argued that this formula predicts a new
unification scale \cite{Conlon:2009qa}. Namely after substituting
$\Lambda \sim M_{Pl}$, $f_a \sim f$, $\hat {\cal K} = - 2 \log {\cal
V}$ and $Z \sim {\cal V}^{-2/3}$, we can recombine the terms and
find an effective shift $M_{Pl} \to M_{Pl} {\cal V}^{-1/3}$.
Furthermore, this shift was argued to be due to tadpole cancellation
from a microscopic point of view, i.e. the appearance of new states
at this scale which make some divergent loop integrals finite.

We find the claims of \cite{Conlon:2009qa} somewhat surprising
because the effective $4d$ supergravity does not know about such
microscopic details of the compactification. To illustrate this,
suppose we broke the GUT group through discrete Wilson lines
instead of fluxes. Some toy models of this type were constructed
in \cite{Donagi:2008ca} (they have some exotics, but that is
besides the point here). The divergent terms only depend on the
$F_Y$ field strength, so in the case of discrete Wilson line
breaking all divergent terms must be $SU(5)$ symmetric and cannot
affect the unification scale. Thus in this case the unification
scale must be the KK scale $M_{KK} = 1/R_S$, as we also find from
the formulae derived above. Yet the argument in
\cite{Conlon:2009qa} is insensitive to such microscopic details.
Also the idea that tadpoles can only be cancelled when $M_{Pl}$ is
finite is not quite true, see the discussion in
\cite{Buican:2006sn}. It can already happen on the second order
neighbourhood of the brane. We believe this indicates that the
appearance of the new `winding' scale purely from a generic
supergravity argument is likely not physical.

\newsubsection{Toy models}

We would like to be able to get an idea of the size of the heavy
threshold corrections, and their effect on unification. Before we
start, we state some general formulae. With some simple algebra, the
relation between the GUT scale parameters and the parameters
measured at low energies can be expressed as \cite{Dienes:1995sv}
\ba \sin^2 \theta_W (M_Z) &=& {3\over 8}\left[1-(b_1 - {5\over 3}
b_2){a\over 4\pi} \log(M_{GUT}^2/ M_Z^2)\right] +\delta^{sin} \eol
\alpha_3^{-1}(M_Z) &=& {3\over 8}\left[{1\over a} -(b_1+b_2-{8\over
3}b_3){1\over 4\pi}\log (M_{GUT}^2/ M_Z^2)\right]
+\delta^{\alpha_3}\eol \ea
where
\ba\label{Phenodelta} \delta^{sin} &=& -{3\over 8}{a\over
4\pi}(\delta_1 - {5\over 3}\delta_2) \eol \delta^{\alpha_3} &=&
-{3\over 8}{1\over 4\pi}(\delta_1 + \delta_2 -{8\over 3}\delta_3)
\ea
and $a^{-1} = \alpha^{-1}_{em}(M_Z) =127.9 \pm 0.1$.

These equations will be used in the following way. We use the
experimental values of $\alpha_{em}(M_Z)$ and $\sin^2 \theta_W(M_Z)$
to predict $M_{GUT}$. Then we run them back down to predict
$\alpha_3(M_Z)$. Further the apparent GUT scale
$\overline{M}_{GUT}$, defined as the scale where $\alpha_1^{-1}$ and
$\alpha_2^{-1}$ meet by extrapolating from low energy data, is given
by
\be \overline{M}_{GUT}^2\ =\ M_{GUT}^2 \exp {\delta_1 - {5\over 3}
\delta_2\over b_1-{5\over 3} b_2}\ =\ M_{KK}^2 \lambda^{2/3}\exp
{\delta_1 - {5\over 3} \delta_2\over b_1-{5\over 3} b_2}  \ee
The heavy threshold corrections affect the relation between the
value of the KK scale and the apparent GUT scale, but not the value
of the apparent GUT scale because this is defined by extrapolating
low energy data.

The experimental values in the $\overline{\rm MS}$ scheme are given
by
\be \sin^2 \theta_W(M_Z) = 0.2312 \pm 0.0002, \qquad
\alpha_3^{}(M_Z) = 0.1176 \pm 0.002. \ee
Then without stringy threshold corrections, $M_Z \sim 91.2 \,{\rm
GeV}$, and using $\alpha^{-1}_{em}(M_Z)$ and $\sin^2 \theta_W(M_Z)$
as input, we find
\be M_{GUT} =1.8\,\times 10^{16}\, {\rm GeV}, \quad
\alpha_{GUT}^{-1} \sim 24.4 , \qquad \alpha_3(M_Z) = 0.115 \ee
We see that although $\alpha_3^{-1}$ comes out slightly too large,
the one loop running of the MSSM gives a remarkably accurate
prediction.

However we should really include various corrections due to two loop
running, scheme conversion, and light SUSY thresholds. The effects
can be summarized \cite{PDG,Alciati:2005ur,Raby:2008gh} by adding a
correction $\delta^{2-loop}_3/4\pi \sim -0.82$ and
$\delta^{light}_3/4\pi \sim -0.12$. Including these effects, instead
we find $\alpha_3(M_Z) \sim 0.129$ which is a much larger
discrepancy. Turning it around and using the experimental value of
$\alpha_3(M_Z)$ as input, we would get $\alpha_3(\overline{M}_{GUT})
= \alpha_{GUT}(1+\epsilon_3)$ where $\epsilon_3 \sim -4\%$. In order
to fix this, from equation (\ref{Phenodelta}) we see that we effectively
need a positive contribution of the
heavy thresholds of the order $\delta^{\alpha_3} \sim +0.94$.

Our result for the leading corrections is given in equation
(\ref{TotalOneLoopV1}) or equivalently equation
(\ref{TotalOneLoopV2}). We see that amongst these, there is one
universal contribution that we have written as the contribution of
an effective pair of triplets $b_a^{(T)}\log M_{GUT}^2/M_{KK}^2$.
Now it is well-known in the context of four-dimensional $SU(5)$
models that lowering the masses of the Higgs triplets improves
agreement with measured deviations from unification, so this looks
like good news. However in four-dimensional $SU(5)$ models, this is
also one of the main sources of tension, as lowering the triplet
masses also enhances dimension five proton decay
\cite{Goto:1998qg,Murayama:2001ur}. In our KK models we have an
analogous issue because the intermediate states are tied to the KK
scale, and dimension five proton decay operators need to be
sufficiently suppressed. Moreover we cannot choose the KK scale
freely as it is tied to the ratio of the GUT and Planck scales.

Let us illustrate this with some numerics. Suppose that
$M_{GUT}/M_{KK}\sim 10^2$. Using the above formulae, and including
the two-loop and light SUSY threshold corrections, we find that the
unification scale gets lowered to $1.3 \times 10^{16}\, {\rm GeV}$
and the prediction for $\alpha_3(M_Z)$ is $0.115$, so we have
essentially restored the successful prediction. If on the other hand
we have $M_{GUT}/M_{KK}\sim 10^{0.5}$, then the unification scale is
$1.7 \times 10^{16}\, {\rm GeV}$ and the predicted value for
$\alpha_3(M_Z)$ is $0.126$. In this case, the remaining heavy
threshold corrections have to make up for the difference.

In the remainder we would like to discuss the model dependent
threshold corrections. We will compute the determinants for some toy
models, the special cases of line bundles $\cO(n,-n)$ on ${\bf P}^1
\times {\bf P}^1$ and line bundles on genus one matter curves.

In order to compute the torsion for $\cO(n,-n)$, we first need the
torsion for the line bundle $\cO(k)_{{\bf P}^1}$. The eigenvalues we
need are well-known -- they are just the energy eigenvalues for a
charged particle moving on a sphere in a magnetic field. For the
sphere, the eigenvalues of the Laplacian on zero forms are given by
$l(l+1)$ with multiplicity $2l+1$. The eigenvalues of $\Delta_k$ are
given by $l(l+|k+1|)$ with multiplicity $2l+|k+1|$. Thus we need
\be
\zeta_k(s) = \sum_l {2l+|k+1|\over l^s (l+|k+1|)^s}
\ee
This was evaluated in \cite{Hortacsu:1979fg,Weisberger:1986qd,K1},
with the result
\ba \zeta'_k(0)&=& 4\,\zeta'_R(-1) -\half (k+1)^2 +
\sum_{m=1}^{k+1}(2m-|k+1|)\log m \ea
where $\zeta'_R(-1) \sim -0.165421$. Note that this is symmetric
under $k \to -2-k$. The torsion is then given by $\left.{\bf
T}_{\cO(k)}\right|_{R\to 1} =-\half \zeta'_k(0)$.

Now we can use this to compute ${\bf K}_Y$. We use the product
formula for Ray-Singer torsion \cite{RaySinger,K2}
\be  {\bf T}(M_1\times M_2,E_1\times E_2) = \chi(M_1,E_1)\, {\bf
T}(M_2,E_2) +
 {\bf T}(M_1,E_1)\,  \chi(M_2,E_2) \ee
Since $ \chi({\bf P}^1,\cO(n)) = n+1 $ we have
\ba
 {\bf T}_{\cO(n,m)} &=& (n+1) {\bf T}_{\cO(m)} + (m+1) {\bf T}_{\cO(n)}
\ea
We take the radii of the two ${\bf P}^1$'s to be equal, since we
imagine an embedding of ${\bf P}^1 \times {\bf P}^1$ in $B_3$ such
that the difference between the two rulings becomes homologically
trivial. Now we can take $L^{5/6} = \cO(n,-n),n\not = 0$, so that
\ba \delta_a^g &=& 2\,\,b_a^{5/6}\,\left. ({\bf T}_{5/6} - {\bf
T}_0)\right|_{R=1} \eol &=&  2\,\,b_a^{5/6}\,\left. ( (n+1){\bf
T}_{\cO(-n)} +(1-n) {\bf T}_{\cO(n)}-2 {\bf
T}_{\cO(0)})\right|_{R=1} \ea
with $b_a^{5/6} = (50/3,6,4)$. In order to have no massless triplet
gauge bosons we need $n=1$, but the model makes sense for any $n$.
Using (\ref{Phenodelta}), for $n=1$ we get
\ba \delta^{g,sin} &=& - {5a\over 4\pi}\left. ( 2{\bf T}_{\cO(-1)}
-2 {\bf T}_{\cO(0)})\right|_{R=1}= {5 a \over 8\pi} \eol
\delta^{g,\alpha_3} &=& - {9\over 4\pi}\left. ( 2{\bf T}_{\cO(-1)}
-2 {\bf T}_{\cO(0)})\right|_{R=1}= {9\over 8\pi} \ea
Thus we see in this toy example that the corrections work in
opposite directions: for $n=1$ the sign of $\delta^{g,sin}$ is
positive so the KK scale is increased, decreasing
$\alpha_3^{-1}(M_Z)$, but the sign of sign of $\delta^{g,\alpha_3}$
serves to reduce the discrepancy. Plugging in mathematica, the net
effect however is a slight increase in the discrepancy.

Now let us examine contributions from the matter curves. We discuss
first $\Sigma_{\bf 10}$ but the case of $\Sigma_{\overline{\bf
5\!}\,}$ is completely analogous. Since the net GUT breaking flux
through each matter curve must be zero, the matter curve cannot be a
${\bf P}^1$. Thus the first non-trivial case corresponds to
$\Sigma_{\bf 10}$ having genus one. Let us recall some results about
Ray-Singer torsion for $T^2$, which we take to be flat with modular
parameter $\tau$. For flat line bundles we have \cite{RaySinger}
\be {\bf T}_{L_z} = \log |e^{\pi i v^2\tau}\vartheta_1(z,\tau)
\eta(\tau)^{-1}| \ee
where $z = u-\tau v$ specifies a point on the dual $T^2$, and $L_z$
is the corresponding line bundle on $T^2$. When $z\to 0$, the theta
function has a zero and the torsion is minus infinity. This is
because one of the eigenvalues of the Laplacian goes to zero in this
limit. Thus for $z=0$ (i.e. for the trivial line bundle) we have to
remove this zero and the torsion is instead given by
\be {\bf T}_{\cO} = \lim_{z\to 0}{\bf T}_{L_z} - \half \log
\lambda_0 = \lim_{z\to 0}{\bf T}_{L_z} - \half \log {4\pi^2
|z|^2\over {\rm Im}(\tau)^2} \ee
Now we consider
\be L^{1/6}|_{\Sigma_{\bf 10}} \otimes M_{\bf 10} \otimes
K_{\Sigma_{\bf 10}}^{1/2} = \cO, \quad  \qquad L^{5/6}|_{\Sigma_\bt}
=L_z \ee
Using (\ref{Phenodelta}), this leads to the following corrections
\ba\label{Sigma10correction} \delta^{{\bf 10},sin} &=&   -{3\over 8}
{a_{em}\over 4\pi}({28\over 3} {\bf T}_{\cO} - {28\over 3}{\bf
T}_{L_z}) \eol \delta^{{\bf 10},\alpha_3} &=& -{3\over 8}{1\over
4\pi}(4{\bf T}_{\cO} - 4 {\bf T}_{L_z} ) \ea
where we used that ${\bf T}_{L_z} = {\bf T}_{L_{-z}}$ and all the
torsions should be evaluated at $R \to 1$. Evaluating in mathematica
for generic $\tau$ and $z$, the corrections work in opposite
directions, and the net effect is again a slight increase in the
discrepancy. For generic values of the moduli, the eigenvalues are
of order the compactification scale and the corrections they give
upon integrating out are small. For special values of the
moduli, in this example $z\to 0$, an eigenvalue may become
parametrically lighter than the KK scale and give large corrections.
In our example we can easily find small values of $z$ which repair
the discrepancy, eg. for $\tau = I$ and $z=0.1-0.1 I$ (and including
the two-loop and light SUSY thresholds, but none of the other heavy
thresholds) we find $\alpha_3(M_Z)\sim 0.1157$. However this correction
then comes mainly from parametrically light charged states which presumably
enhance proton decay significantly.

%
\begin{figure}[th]
\begin{center}
            \scalebox{.7}{
               \includegraphics[width=\textwidth]{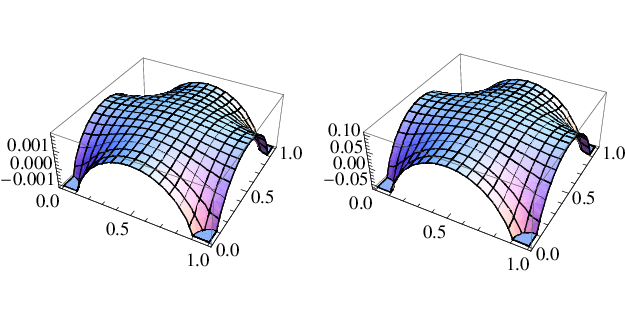}
               }
\end{center}
\vspace{-.5cm}
\begin{center}
\parbox{14cm}{\caption{ \it A plot of the threshold corrections
(\ref{Sigma10correction}) localized on $\Sigma_\bt$, here assumed to
be a square torus. On the left the value of $\delta^{{\bf 10},sin}$
is indicated along the $z$-axis, on the right the $z$-axis measures
the value of $\delta^{{\bf 10},\alpha_3}$. The remaining axes
correspond to the bundle modulus $z = x - \tau y$ with $0\leq x,y
\leq 1$. \label{Sigma10Thresh}}}
\end{center}
\end{figure} 

Similarly we can consider a toy model for the threshold corrections
coming from $\Sigma_{\overline{\bf 5\!}\,}$. Typically this matter
curve has genus larger than that of $\Sigma_{\bf 10}$, however just
to get an idea for the size of the corrections let us assume that
$\Sigma_{\overline{\bf 5\!}\,}$ has genus one and
\be
L^{-1/2}|_{\Sigma_{\overline{\bf 5\!}\,}} \otimes M_{\overline{\bf 5\!}\,} \otimes
K_{\Sigma_{\overline{\bf 5\!}\,}}^{1/2} = \cO \quad \qquad
L^{5/6}|_{\Sigma_\bfb} = L_z
\ee
Then we find
\ba \delta^{{\overline{\bf 5\!}\,},sin} &=&  -{3\over 8} {a_{em}\over
4\pi}({4\over 3} {\bf T}_{\cO} - {4\over 3} {\bf T}_{L_z}) \eol
\delta^{{\overline{\bf 5\!}\,},\alpha_3} &=& -{3\over 8}{1\over 4\pi}(-4{\bf T}_{\cO} + 4 {\bf T}_{L_z} )
\ea
Again, it is dangerous to compare these contributions to those from
$\Sigma_{\bf 10}$, since in a given model the curves and line
bundles on them cannot be identical. At any rate as long as ${\bf
T}_{\cO} - {\bf T}_{L_z}$ is positive, which according to the
results for (\ref{Sigma10correction}) is true for a large range of
$z$, the effect of both corrections is to reduce the discrepancy.

\newsubsection{Conclusions}

Since this section was rather long, we would here like to summarize
our conclusions. We argued that the full expression for the leading
corrections to the gauge kinetic terms in a local model is given by
\be\label{TCorrection} {16 \pi^2\over g_a^2(\mu)} \ =\  {16 \pi^2
\,k_a \over g^2} + b_a \log{M_{GUT}^2\over \mu^2} + b_a^{(T)} \log
{M_{GUT}^2\over M_{KK}^2} + \delta_a^g + \delta_a^{\bf 10} +
\delta_a^{\overline{\bf 5\!}\,}  +\delta_a^{(other)} \ee
The correction $\delta_a^{(other)}$ is due to two-loop running and
light SUSY thresholds. The correction  $b_a^{(T)} \log {M_{GUT}^2/
M_{KK}^2}$ is suggestively written as a colour triplet correction
and has the same effect as lowering the Higgs triplet masses in
conventional $4d$ GUTs, although it had a rather different origin
as a one-loop divergence in $F$-theory. It improves the agreement
between the predicted and the measured value of $\alpha_3(M_Z)$,
which as we reviewed differs slightly from the naive value
expected form unification. However to get complete agreement
requires lowering $M_{KK}$ by two orders of magnitude. This is
hard to obtain even if we assume dimension five proton decay can
be sufficiently suppressed, as the ratio $M_{KK}/M_{GUT}$ is tied
to the Planck scale and can't be very large. Thus the remaining
discrepancy must come from the correction terms $\delta_a^g$,
$\delta_a^{\bf 10}$ and $\delta_a^{\bfb}$. We showed that they can
be expressed in terms of Ray-Singer torsion which depends only on
complex structure moduli. In our toy models, these one loop
corrections could either increase or decrease $\alpha_3(M_Z)$, but
if the values obtained in figure \ref{Sigma10Thresh} are typical,
they seem a bit too small to make up the difference (recall we
need $\delta_{\rm heavy}^{\alpha_3} \sim +0.94$). Hence we
conclude that the threshold corrections generally go in the right
direction, but precise agreement with experimentally measured
deviations from unification probably implies extra structure
beyond what is assumed here.

There are some interesting differences with $4d$ models. We saw that
the KK scale is lowered compared to the GUT scale. This could affect
some of the phenomenology because masses of KK modes and certain
stabilized moduli are set by $M_{KK}$ or some power thereof, rather
than $M_{GUT}$. Analogous to four-dimensional models with an extra
pair of Higgs triplets and $M_T < M_{GUT}$, this lower scale
improves the agreement with measured deviations from unification
(through the $b_a^{(T)} \log {M_{GUT}^2/ M_{KK}^2}$ term) and
enhances dimension five proton decay if present. Different from
four-dimensional models, the effective scale $M_T \sim M_{KK}$ is
tied to the ratio of the GUT and Planck scales, and dimension six
proton decay is also set by this scale. This will be discussed in
more detail in the next section.

However in practice this effect is rather small. The two scales
differ by a factor
\be M_{GUT}/M_{KK} = \lambda^{1/3} \ee
Although $\lambda$ depends on the UV completion, generically we
should expect $\lambda = \Lambda/M_{KK} \sim R_{B_3} R_{S}
m_{10}^2$. At tree level, we have $m_{10} \sim
M_{KK}\alpha_{GUT}^{-1/4}$ and hence
\be \lambda\sim M_{Pl}^{1/3}M_{KK}^{-1/3} \alpha_{GUT}^{-1/6} \sim
M_{Pl}^{1/3}M_{KK}^{-1/3} \ee
since $\alpha_{GUT}^{-1/6}\sim 25^{1/6} \sim 1$. Then we have the
relation
\be M_{KK} \sim M_{GUT}(M_{GUT}/M_{Pl})^{1/8}.\ee
Here $M_{Pl}$ is the reduced Planck mass, $M_{Pl} \sim 2.4 \,\times
10^{18}\,{\rm GeV}$. With $M_{GUT}/M_{Pl}\sim 10^{-2}$,  the KK
scale ends up less than half an order of magnitude below the GUT
scale.

While we are at it, we believe that a conceptually cleaner approach
would be to define a notion of Ray-Singer torsion directly for Higgs
bundles. That is, we should really consider the Laplacian for $\bar
D = \delb_A + \Phi$ and then define the `Higgs bundle torsion' as in
equation (\ref{RSDefinition}), but with the modified Laplacian. Then
in (\ref{TCorrection}) we would get a single correction
$\delta^{Higgs}_a$. This should reduce to $\delta_a^g
+ \delta_a^{\bf 10} + \delta_a^{\overline{\bf 5\!}\,} $, because the
modes become sharply localized on the matter curve when the brane
intersection angles are large, and this is controlled by a K\"ahler
modulus. Recall that the torsion is essentially independent of
K\"ahler moduli. (See however \cite{Akerblom:2007np}).

It is not completely clear how we could go about computing such
`Higgs bundle torsion' directly. However apart from
a divergent term it should be independent of the K\"ahler moduli,
so we can use heterotic/$F$-theory duality to compute it. On the
heterotic side, the threshold corrections are computed by
the torsion of the $E_8$ bundle \cite{Kaplunovsky:1992vs,Bershadsky:1993cx}.
Using results of \cite{DW7}, we expect that
in a suitable limit we get the local $F$-theory contribution.

One may also apply our techniques to study threshold corrections for
perturbative IIb GUT models. Here one has two expansion parameters
and so one should distinguish between $\alpha'$ and $g_s$
corrections. The former were investigated in \cite{Blumenhagen:2008aw}.
The one-loop contribution would also give
rise to holomorphic Ray-Singer torsion.

\newpage

\newsection{Proton decay}
\seclabel{ProtonDecay}

In this section we discuss some constraints on $F$-theory models due
to observational constraints on proton decay\footnote{For recent
reviews of some of the issues in proton decay, see eg.
\cite{Nath:2006ut,Raby:2002wc}.}. We will denote the ${\overline{\bf
5\!}\,}$ modes consisting of $(L,d^c)$ by ${\overline{\bf 5\!}\,}_m$
($m$ for `matter) and the ${\overline{\bf 5\!}\,}$  mode consisting
of $(H_d,T_d)$ by ${\overline{\bf 5\!}\,}_h$. Of course we should
engineer our model so that after breaking the GUT group there are no
massless colour triplets. Massless chiral matter is assumed to live
on 7-brane intersections rather than in the bulk of the 7-brane.
Some issues in models with bulk matter have been clarified recently
and can now also be discussed \cite{Donagi:2011jy,DW6}.

\newsubsection{Fibered root systems and their monodromies}

In order to be fairly concrete, we first recall some aspects of
local geometries for 7-branes and the description of the matter
curves in terms of spectral covers
\cite{DonagiSpectral,Friedman:1997yq,Curio:1998bva}. Some of the
statements here are not quite rigorous. A more rigorous description
can be found in \cite{Donagi:2009ra}.

\begin{figure}[t]
\begin{center}
            \scalebox{.45}{
               \includegraphics[width=\textwidth]{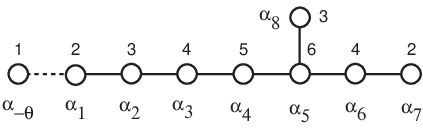}
               }
\end{center}
\vspace{-.5cm} \caption{ \it The extended $E_8$ Dynkin diagram and
Dynkin indices.}\label{E8Dynkin}
\end{figure}

We focus on an $E_8$ ALE (or $dP_8$) fibration over our surface
$S_2$, unfolded to an $SU(5)$ singularity. The two-cycles of the ALE
are labelled by the roots of $E_8$, in such a way that the
intersection matrix of the simple roots agrees with the Cartan
matrix of $E_8$. We will use the labeling shown in figure 
\ref{E8Dynkin}. We pick a maximal subgroup $SU(5)_H \times
SU(5)_{GUT}$ where
\be \{\alpha_{-\theta}, \alpha_1,\alpha_2 , \alpha_3\} \ee
are the creation operators for the adjoint representation of
$SU(5)_H$, and similarly
\be \{\alpha_5, \alpha_6,\alpha_7, \alpha_8\} \ee
are the creation operators for the adjoint representation of
$SU(5)_{GUT}$. Here $\alpha_{-\theta}$ is the negative of the
highest root, so that
\be \alpha_{-\theta} = -\sum d_i \alpha_i \ee
where $d_i$ are the Dynkin indices. We keep the cycles $\{\alpha_5,
\ldots, \alpha_8\}$ zero size, so they correspond to the exceptional
cycles of the $SU(5)$ singularity, but give a finite size to
$\{\alpha_{-\theta}, \ldots , \alpha_4\}$. There is a remnant gauge
symmetry, namely the Weyl group of $SU(5)_H$, which may be used to
glue the latter set of cycles together across local patches. This
Weyl group, denoted $\mathscr{W}_{A_4}$, is generated by the
reflections
\be \{W_{\alpha_{-\theta}},W_{\alpha_1} ,\ldots,W_{\alpha_3}\} \ee
and plays the role of the structure group breaking $E_8$ to
$SU(5)_{GUT}$. As we vary the ALE over $S_2$, the roots undergo
monodromies, and define a non-trivial local system (i.e. a
flat bundle) over $S_2$.

To parametrize the sizes of the cycles, we introduce the dual basis
$\omega_i$ of two-forms on the ALE such that $\int_{\alpha_j}
\omega_i= \delta_{ij}$, and further the generator $\omega_\theta$
satisfying $\omega_\theta + d_i\omega_i=0$. Then on each local patch
of $S_2$, the fibration is described by
\be\label{OmegaExpansion}  \Omega(t)\ =\ \Omega_0 + t_i \wedge
\omega_i + ({\rm higher\ order}) \ee
where $t_i$ is a (2,0) form on the patch and $t_i\equiv 0$ for
$i=5,\ldots,8$ in order to enforce the $SU(5)$ singularity.

Although in a local patch on $S$ we can make an expansion
(\ref{OmegaExpansion}), since globally the $t_i$ and $\omega_i$
undergo monodromies by the $\mathscr{W}_{A_4}$ Weyl group, it is not
very convenient to work with them. A more convenient way to specify
such a fibration is by working with the invariant polynomials. Let
us first define
\be \lambda_1 = t_4, \quad \lambda_2 = t_4 + t_3, \quad \ldots,
\quad \lambda_5 = t_4 + \ldots + t_{-\theta}
 \ee
Then $\mathscr{W}_{A_4}$ acts as the symmetric group on the
$\lambda_i$. Then to specify the fibration, we need to specify
precisely five global holomorphic sections
$\{a_0,a_2,a_3,a_4,a_5\}$. In our local patch they are given by
\be a_n/a_0 = \sigma_n(\lambda_i), \qquad n=2,\ldots,5 \ee
where $\sigma_n$ is the $n$th symmetric polynomial. The combination
$\sigma_1(\lambda_i)$ is zero automatically due to the constraint
$t_{-\theta} = - \sum d_i t_i$. The $a_i/a_0, i=2,..,5$ are the
Casimirs of the VEV of the adjoint field $\Phi^{2,0}$ of the
eight-dimensional gauge theory. It is not hard to see that the Chern
classes of $\{a_0,a_2,a_3,a_4,a_5\}$ are given by
\cite{Bershadsky:1996nh,Friedman:1997yq}
\be [a_i=0]\  =\ (6-i) c_1 - t \ee
where $c_1 = c_1(TS_2) = -c_1(K_S)$ and $-t = c_1(N_S)$.

Now let us write the Calabi-Yau four-fold more explicitly in terms
of the $a_i$. We introduce variables $\{u,z,x,y\}$ of degrees
$\{1,1,2,3\}$. Then we may write the sections as a single expression
\be p_5 = a_0\,z^5 + a_2\, xz^3 + a_3\, yz^2 + a_4 \,x^2z + a_5\, xy
\ee
The $dP_8$ fibration over $S$ is now simply given by the following
equation:
\be 0 = y^2 + x^3 + f xz^4 + gz^6 + u\,p_5(x,y) \ee
To get an affine space, we can set $u\to 1$. Moreover the terms
$fxz^4$ and $gz^6$ are subleading and irrelevant for our purposes
here, so we may set them to zero. Then we arrive at the equation of
a deformed $E_8$ singularity:
\be 0 = y^2 + x^3 + a_0 z^5 + a_2 xz^3 + a_3 yz^2 + a_4 x^2 z + a_5
xy \ee
It is frequently desirable to keep the $dP_8$, but this is very
similar and most of the comments in this section will go through.

The above data may also be represented by a spectral cover. The
$dP_8$ fibration is equivalent to the spectral cover defined by
$p_5=0$ in the elliptically fibered Calabi-Yau $y^2 = x^3 + fx +
g$. In the limit that $f,g \to 0$, the $T^2$ fibration degenerates
to a ${\bf P}^1$-fibration over $S_2$, and $z' = y/x$ is
identified with a coordinate on the ${\bf P}^1$. Then $p_5=0$
yields a spectral cover in the ${\bf P}^1$-fibration which is
equivalent to the (compactified) $E_8$ ALE fibration. The spectral
line bundle gets mapped to $G$-flux on $Y_4$. The locus where
$\lambda_i=\lambda_j$ on the spectral cover for some $i\not = j$
is called the ramification locus (or ramification divisor, because
it is complex codimension one). The spectral cover is smooth at
this locus, but the vertical derivative vanishes. The projection
of the ramification locus on $S$ is the branch locus. As we circle
around this locus, the cycles undergo the monodromies by the Weyl
group that we described above.

The adjoint representation of $E_8$ decomposes into the adjoint of
$SU(5)_H\times SU(5)_{GUT}$ plus some remaining representations,
which can be read off from the Dynkin diagram:
\be {\bf 248}\  =\ ({\bf 24},{\bf 1}) + ({\bf 1},{\bf 24}) +
(\bfv,\bt) + (\bfb,\btb) + (\bt,\bfb) + (\btb,\bfv) \ee
The $E_8$ root $\alpha_4$ is a weight for the $({\bf 5},\bt)$
representation of $SU(5)_H \times SU(5)_{GUT}$ (since it extends the
$SU(5)_H$ root system to $SU(6)$ and the $SU(5)_{GUT}$ root system
to $SO(10)$), and acting with the remaining creation operators give
the remaining weights of this representation. Similarly $-\alpha_4$
gives the $(\bfb,\btb)$, $2\alpha_4 + \alpha_3+ 2\alpha_5 + \alpha_6
+ \alpha_8$ gives the $(\bt,\bfb)$ and
$-2\alpha_4-\alpha_3-2\alpha_5-\alpha_6-\alpha_8$ gives the
$(\btb,\bfv)$. For more details on the decomposition of the $E_8$
roots under $SU(5)_H \times SU(5)_{GUT}$, see appendix
\ref{E8roots}.

Over a generic point on $S_2$, only $\{\alpha_5,\ldots,\alpha_8\}$
have zero size, and wrapping membranes (or rather $(p,q)$ strings)
on these cycles gives the eight-dimensional $SU(5)$ vector
multiplet. However over a complex codimension one locus on $S_2$, an
extra cycle may shrink to zero, and wrapping a membrane on it gives
rise to a hypermultiplet. When $a_5=0$, we have $\lambda_i \to 0$
for some $i$. Suppose we have $\lambda_1 \to 0$. Then the cycle
$\alpha_4$ shrinks to zero, and the singularity type is enhanced
from $SU(5)_{GUT}$ to $SO(10)$, as we can easily read of from the
Dynkin diagram in figure \ref{E8Dynkin}. In fact as we can see from
appendix \ref{E8roots}, when $\lambda_1 \to 0$ there are actually
twenty cycles shrinking to zero size simultaneously, precisely
fitting in the $\bt \oplus \btb$ of $SU(5)_{GUT}$. Thus the extra
massless states fit in a hypermultiplet in the $\bt$ of
$SU(5)_{GUT}$, and this locus this corresponds to the matter curve
$\Sigma_\bt$:
\be \Sigma_{\bt} = \{a_5=0 \} \ee

On the other hand, suppose $\lambda_i + \lambda_j \to 0$ for some
$i\not = j$. Eg. suppose we have $\lambda_1 + \lambda_2 \to 0$. Then
the cycle $\alpha_3 + 2\alpha_4 + 2\alpha_5 +\alpha_6 +\alpha_8$
shrinks to zero size, and computing the extended Dynkin diagram, we
see that the singularity type is enhanced from $SU(5)_{GUT}$ to
$SU(6)$. In fact according to appendix \ref{E8roots}, when
$\lambda_1 + \lambda_2 \to 0$ there are actually ten cycles
shrinking to zero size simultaneously, precisely fitting in the
$\bfv \oplus \bfb$ of $SU(5)_{GUT}$. Thus the extra vanishing cycles
form a hypermultiplet in the $\bfv$ of $SU(5)_{GUT}$, and this locus
corresponds to the matter curve $\Sigma_\bfv$.

We can deduce an explicit equation for $\Sigma_\bfv$ as follows.
When $\lambda_i + \lambda_j \to 0$ for $i\not = j$, it means that
$p_5$ has solutions which are interchanged under $y\to -y$.
Equivalently if we write
\be p_5 = (a_0 + a_2 x + a_4 x^2) + y (a_3 + a_5 x) = P(x) + y Q(x)
\ee
then $P(x)$ and $Q(x)$ should have a common zero. This is measured
by the vanishing of the resultant, so when
\be  \Sigma_\bfv = \{R_{8c_1-3t} = a_0 a_5^2 -a_2a_3a_5+a_4
a_3^2=0\} \ee
we get the matter curve $\Sigma_\bfv$.

%
\begin{figure}[t]
\begin{center}
            \scalebox{.5}{
               \includegraphics[width=\textwidth]{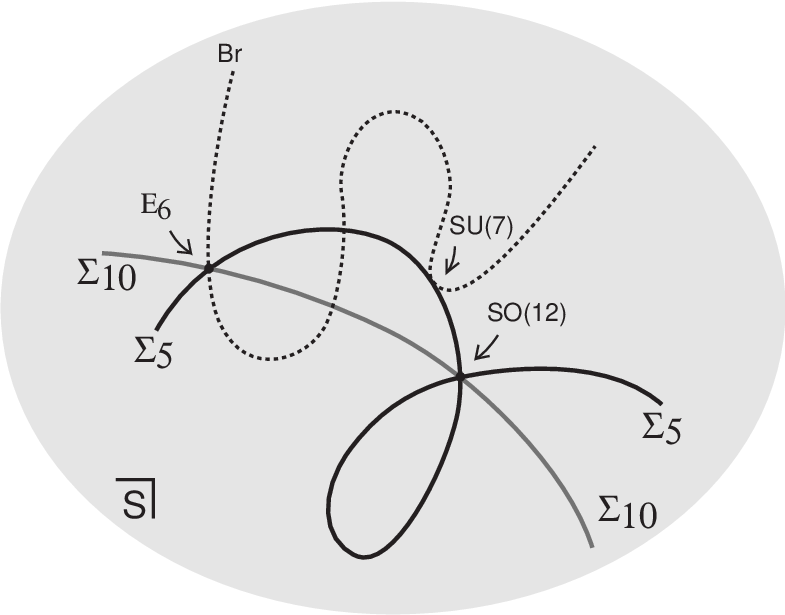}
               }
\end{center}
\vspace{-.5cm}
\begin{center}
\parbox{14cm}{\caption{ \it Schematic picture of the matter curves
and branch locus on $S$, and their intersections, for generic values
of the complex structure moduli. \label{Enhancements}}}
\end{center}
\end{figure} 

Now recall that the Yukawa couplings are given by the triple overlap
of wave functions:
\be\label{Yukawa} \int_S {\rm Tr}(A^{0,1} \wedge A^{0,1} \wedge
\Phi^{2,0}) \ee
where by the trace we just mean that one should pick a gauge
singlet. This coupling is well-defined on Dolbeault cohomology
classes. Since our wave functions have representatives that are
localized on the matter curves, the Yukawa couplings get
contributions only from the intersection of $\Sigma_\bt$ and
$\Sigma_\bfb$. When these curves intersect, we have $a_5=0$ and
$R=0$ so we must also have $a_3=0$ or $a_4=0$. If $a_5=a_4=0$, then
up to Weyl reflections we can say that the cycles $\alpha_4$ and
$\alpha_3$ shrink to zero and the singularity type is enhanced to
$E_6$. If we wrap a membrane (or really ($p,q$) string from the
perspective of the IIb space-time) on the cycles $\alpha_4, \alpha_4
+ \alpha_3 + 2\alpha_5 + \alpha_6 + \alpha_8$ and
$-2\alpha_4-\alpha_3-2\alpha_5 -\alpha_6-\alpha_8$, giving us states
in the $\bt,\bt$ and ${\bf 5}$ of $SU(5)_{GUT}$ respectively, then
at the $E_6$ intersection point there is a relation in homology
\be\label{homrelation} (\alpha_4) +  (\alpha_4 + \alpha_3+ 2\alpha_5
+\alpha_6 + \alpha_8) +
(-2\alpha_4-\alpha_3-2\alpha_5-\alpha_6-\alpha_8) = 0 \ee
Pictorially, this means that the three membranes wrapping these
three cycles may be combined into a topologically trivial
configuration. From the perspective of the IIb space-time, it means
that on this intersection the ends of the three open $(p,q)$ strings
can be combined into a gauge singlet.

In terms of group theory, the statement (\ref{homrelation}) is
equivalent to the fact that if we decompose the ${\bf 78}$ of $E_6$
under $SU(5)_{GUT} \times U(1)_a \times U(1)_b$, which yields
$\bt_{1,1} + \bt_{-1,1} + \bfb_{0,2} + c.c$ plus additional neutral
matter, then the ${\rm Tr}({\bf 78}^3)$ contains a contribution $\bt
\cdot \bt \cdot \bfv$. Thus at such an intersection we get a
contribution to the up type Yukawa coupling, inherited by KK
reduction from the cubic coupling (\ref{Yukawa}) in the
eight-dimensional gauge theory:
\be \{ a_5 = a_4 = 0\} \quad \to \quad \bt\cdot \bt \cdot {\bf 5}
\ee

Now consider the other type of intersection point, where
$a_5=a_3=0$. Here we have (up to Weyl reflections) $\alpha_4=0$ and
$\alpha_9=0$, where we define
\be \alpha_9 = \alpha_2 + 2\alpha_3 + 2\alpha_4 + 2 \alpha_5 +
\alpha_6 + \alpha_8 \ee
Note that $\alpha_9$ is a weight for the $(\bt,\bfb)$ of $SU(5)_H
\times SU(5)_{GUT}$, with size $\lambda_2 + \lambda_3$, and by
acting with raising and lowering operators we can get the remaining
states in this representation. Computing the inner products, we see
that the Dynkin diagram of $\{\alpha_4, ..,\alpha_9\}$ is that of
$SO(12)$, so the singularity type is enhanced to $SO(12)$. Now note
that $\alpha_4 + \alpha_9$ is still a root of $SO(12)$, so we can
make gauge invariant combinations of the type $\bt\cdot
\bfb\cdot\bfb$, yielding the down type Yukawas:
\be \{ a_5 = a_3 = 0\} \quad \to \quad \bt\cdot \bfb\cdot\bfb \ee
The curve $R=0$ clearly has a double point singularity at this
intersection point. As pointed out in \cite{Hayashi:2008ba}, if we
keep track of the vanishing cycles along each branch, we see that
the charged open strings, and hence the hypermultiplet in the
$\bfv$, actually naturally propagate on the normalization of the
matter curve, i.e. they propagate on the resolution obtained by
replacing each double point by two distinct points.

This accounts for all the intersections of $\Sigma_\bfv$ and
$\Sigma_\bt$. However there is another type of singularity
enhancement that occurs generically at codimension two on $S$ but
does not correspond to the intersection points above. Let us define
the root
\be \tilde{\alpha}_8 = \alpha_3 + 2 \alpha_4 + 3 \alpha_5 + 2
\alpha_6 + \alpha_7 + 2 \alpha_8 \ee
This is one of the weights of the $(\bt,\bfb)$ representation of
$SU(5)_H \times SU(5)_{GUT}$, with size $\lambda_1 + \lambda_2$. Now
suppose that $\tilde{\alpha}_8$ and $\alpha_9$ vanish at the same
time. The Dynkin diagram generated by
$\{\tilde{\alpha}_8,\alpha_8,\alpha_5,\alpha_6,\alpha_7,\alpha_9 \}$
is that of $SU(7)$. Now let us apply to $\alpha_9$ the raising
operator corresponding to the highest root of $SU(5)_{GUT}$,
yielding:
\be \tilde{\alpha}_9 = \alpha_2 + 2 \alpha_3 + 2 \alpha_4 + 3
\alpha_5 + 2 \alpha_6+ \alpha_7 + 2\alpha_8 \ee
The difference between these two roots is
\be\label{551Yukawa} \tilde{\alpha}_8 - \tilde{\alpha}_9 = -\alpha_2
- \alpha_3 \ee
with size $\lambda_1 - \lambda_3$. Hence we see that $\alpha_2 +
\alpha_3$ is also a vanishing two-cycle, but if we wrap a membrane
on it we get a hypermultiplet $N$ which is neutral under
$SU(5)_{GUT}$. This corresponds to one of the non-normalizable
complex structure moduli describing deformations of the 7-branes.
Indeed $\alpha_2 + \alpha_3$ is a root for $SU(5)_H$, so we see that
$N$ transforms in the adjoint of $SU(5)_H$. In the heterotic string,
it would be identified with a vector bundle modulus. Note that
complex structure moduli like $N$ are not localized along any matter
curve -- they are supported in the bulk.

To arrange such an $SU(7)$ point, we need (up to Weyl reflections)
that $\lambda_1 + \lambda_2 = \lambda_2 + \lambda_3=0$, but
$\lambda_1,\lambda_3 \not = 0$, because that would impose three
conditions and so will not happen generically at codimension two.
Thus this corresponds to some intersection of the branch locus with
$\Sigma_\bfv$. But not all intersections of the branch locus with
$\Sigma_\bfv$ are $SU(7)$ points, so how do we distinguish them
geometrically?

Let us enumerate the possible intersections between $\Sigma_\bfv$
and the branch locus. One such type of intersection corresponds to
$\lambda_i = \lambda_j=0$ for $i\not = j$. As we saw above, these
are the $E_6$ enhancement points where the top quark Yukawa
couplings are localized, and $\Sigma_\bfv$ also intersects
$\Sigma_\bt$ there. We could also have $\lambda_1+\lambda_2=0$
intersecting with the locus where $\lambda_i-\lambda_j=0$, but both
$i$ and $j$ are not equal to one or two. This is generically a
transverse intersection and there is no singularity enhancement at
the intersection; lifting to the spectral cover, the ramification
locus does not even intersect the zero section here. Even when the
intersection between $\Sigma_\bfv$ and the branch locus is
accidentally not transversal, we can make it transversal by a small
complex structure deformation. Finally we could have the $SU(7)$
points where $\lambda_1 + \lambda_2 = 0$ and
$\lambda_i-\lambda_j=0$, and precisely one of $i$ or $j$ is equal to
one or two (they shouldn't both be equal, as this would give us back
the $E_6$ intersection points). A simple local model near such an
enhancement point is given by the two-fold spectral cover
\be\label{2foldcover} z_1 + z_2 s + s^2 = 0 \ee
where $z_1,z_2$ are local coordinates on $S$, and $s$ is a local
coordinate on the canonical bundle of $S$. In this model,
$\Sigma_\bfv$ is given by $z_1=0$ and the branch locus is given by
$z_2^2 - 4 z_1=0$. Although both curves are smooth, we see that the
intersection is not transversal: the branch locus becomes tangential
to $\Sigma_\bfv$ at the intersection $z_1=z_2=0$, where the
singularity is enhanced to $SU(7)$. The non-transversality is
required by the analytic structure and cannot be removed by a small
deformation (as this would just amount to a redefinition of $z_1$
and $z_2$). Hence such intersection points can be distinguished
geometrically.

From the homology relation (\ref{551Yukawa}), we see that we can
make a gauge invariant coupling:
\be \{R = Br = 0, \ {\rm det}(R',Br')=0 \} \quad \to\quad \bfv \cdot
\bfb\cdot {\bf 1} \ee
where $Br=0$ denotes the branch locus of the spectral
cover.\footnote{ One may also write an explicit equation for the
branch locus: it corresponds to the discriminant, here given by the
resultant of $p_5$ and its
vertical derivative. Eg. for the two-fold cover in
(\ref{2foldcover}) we get ${\rm Res}(z_1 + z_2 s + s^2, z_2 + 2
s)=4z_1 - z_2^2$.} Of course getting a massless $\bfv/\bfb$ pair
localized on a connected component of $\Sigma_\bfv$ generally
requires some fine-tuning of the complex structure moduli, and such
pairs are easily lifted. Turning on an expectation value for $N$
yields a mass term for the hypermultiplet. This corresponds to
deforming the branch locus and $\Sigma_\bfv$ on $S$.

The $SU(7)$ enhancement points are on different footing from the
$E_6$ and $SO(12)$ points; they are generally smooth points of
$\Sigma_\bfv$ and it seems unlikely that the $\bfv \cdot \bfb\cdot
{\bf 1} $ couplings are localized there. Further, one may also get
$\bt \cdot \btb \cdot {\bf 1}$ couplings, and it does not appear to
require singularity enhancements.

\newsubsection{Extra $U(1)$'s and their $D$-terms}
\subseclabel{ExtraU(1)s}

In the $SU(5)$ GUT models $R$-parity is not guaranteed. In order to
prevent couplings of the type ${\bf 10}_m \cdot {\overline{\bf
5\!}\,}_m \cdot {\overline{\bf 5\!}\,}_m$, we have to somehow be
able to distinguish between ${\overline{\bf 5\!}\,}_h$ and
${\overline{\bf 5\!}\,}_m$ zero modes. This can be achieved by
localizing the ${\overline{\bf 5\!}\,}_h$ and ${\overline{\bf
5\!}\,}_m$ on different matter curves, charging them differently
under an extra gauge symmetry, or both.

The idea of using $U(1)$ symmetries has long been discussed in the
heterotic literature, and is implicit in the construction of bundles
by extension. As already noted in \cite{Tatar:2006dc} the discussion
can be largely translated to $F$-theory. In the discussion of the
previous section we assumed that the spectral cover is smooth and
irreducible, in which case the holonomy group of the spectral cover
breaks $E_8$ to $SU(5)_{GUT}$. However if the spectral cover
degenerates, under certain conditions one may get additional gauge
symmetries.

The five-fold spectral cover can be degenerated in various ways.
When the spectral cover becomes reducible, for instance if the
sections $a_i$ are such that we have a 3+2 factorization
\be p_5 = (b_0 + b_2 x)(d_0 + d_2 x + d_3 y) \ee
with $b_i,d_j$ generic, then the structure group of the spectral
cover is reduced from $\mathscr{W}_{A_4}$ to
$\mathscr{W}_{A_1}\times \mathscr{W}_{A_2}$. In order to
understand the low energy gauge group, we also need to know what
happens with the gauge field on the intersection of the reducible
pieces. A rank one sheaf on the reducible surface consists of a
pair of line bundles on the reducible components, together with a
gluing morphism on the intersection
\cite{Donagi:2010pd,Donagi:2011jy}. If the gluing morphism
vanishes identically, then there is an extra $U(1)_X$ symmetry
that commutes with the structure group; in fact in this example
there may be additional massless $W$-bosons, and the GUT group is
enhanced to $SU(6)$, but this is typically broken to $SU(5)_{GUT}
\times U(1)_X$ through $U(1)_X$ flux. If the elliptic fibration
admits a second section, then we can also break to $SU(5)_{GUT}
\times U(1)_X$ by using an abelian Higgs field. At the
intersection of the two reducible pieces of the spectral cover,
one finds chiral fields charged under $U(1)_X$, whose internal
zero modes correspond to the gluing morphism mentioned above. If
the $D$-terms for $U(1)_X$ can be satisfied so that these charged
moduli have zero VEVs, then we have a candidate for an extra
$U(1)_X$ symmetry.

The $U(1)_X$ could still be lifted by other effects. In section 2 we
saw that the $U(1)_X$ also gets a contribution to its mass from its
couplings to axions, if the matrices
\be
\Pi^X_M \ = \ \int_{Y_4} \beta_M \wedge \omega^X \wedge G
\ee
are non-zero. It is also possible that $\omega^X$ may not be
extended globally. Thus the fate of the extra $U(1)$ can not be
completely understood from the spectral cover for the local model
alone.

In the supergravity approximation, we expect the Fayet-Iliopoulos
parameters to be proportional to $G\wedge J\sim F \wedge J$. From
the analysis of Becker and Becker we know that if we are working
within $11d$ supergravity on a smooth resolved Calabi-Yau, then we
must have $J \wedge G = 0$, but this need not be true in the
$F$-theory limit, as their may be new light fields entering the
$D$-term potential. To understand the dependence of the masses on
the moduli, we need to write the $D$-term potential more precisely.
We now describe how to do this in the four-fold picture. A brief
summary of the conclusions here was already included in
\cite{Donagi:2010pd}, in relation to $M5$-instantons.

To find the $D$-terms in a supersymmetric manner, it is
convenient to dualize the Green-Schwarz couplings. The K\"ahler
potential for the K\"ahler moduli is
\be\label{TKaehlerPotential} {\cal K}\ =\ M_{Pl}^2\, \hat{\cal K}\ =\ -2\,M_{Pl}^2 \, \log\,
{1\over 6} m_{10}^6 \int_{B_3} J^3 \ee
The K\"ahler moduli are defined as
\be T_a\ =\  \half m_{10}^4\int_{D_a} J \wedge J -i\int_{D_a} C_4\ee
Then we may rewrite
\be {\cal K}(T) \sim -\half \,M_{Pl}^2\,\log\, {8\over 6} ({\rm
Re}\,T_a )({\rm Re}\,T_b )({\rm Re}\,T_c )d^{abc} \ee
where $d_{abc} = D_a\cap D_b\cap D_c$, and indices are raised and
lowered using the metric $\hat{\cal K}_{ab}$.

In the presence of 7-branes, the RR field $C_{(4)}$ obtains an
anomalous transformation law:
\be\label{C4anom} \delta C_{(4)} \sim \sum_m {\rm Tr}(\Lambda_m\,
F_m)\wedge \delta^{(2)}(D_m) - {\rm Tr}(\Theta\, R)\wedge
\delta^{(2)}(D_m) \ee
This transformation law is determined by anomaly cancellation. We
can try to write this in terms of four-fold data as follows:
$C_{(4)}$ gets promoted to $C_{(6)}$ with two indices on the
elliptic fiber, the $F_m$ get lifted to $G$-flux, and the
$\Lambda_m$ get lifted to a harmonic two-form $\Lambda =\Lambda_m
\omega^m$. Then we are motivated to write the variation under gauge
transformations as
\be\label{C4anomV2} \delta C_{(6)}\ \sim\ G \wedge \Lambda +
I_6(R,\Theta) \ee
One can derive this more directly from the $M$-theory perspective,
which is the proper way to do it. Due to the Chern-Simons term in
$11d$ supergravity, one finds that
\be d^2 C_6 + \half G \wedge G + I_8(R) = 0 \ee
By descent we then arrive at (\ref{C4anomV2}). Now the K\"ahler
moduli are not invariant under a gauge transformation of $U(1)_X$:
\be\label{gaugedisom} {\rm Im}(T_a)\ \to\ {\rm Im}(T_a) + q_a^X
\lambda_X \ee
Therefore the K\"ahler potential for the moduli is modified to
\be \hat{\cal K}(T_a + T_a^\dagger )\ \to\ \hat{\cal K}(T_a +
T_a^\dagger - q_a^X V_X) \ee
so that under a gauge transformation, the action is invariant. This
K\"ahler potential now includes the dualized Green-Schwarz
couplings, which are of the form $q_a^X\,A_\mu^X\, \del^\mu {\rm
Im}(T_a)$, in a manifestly supersymmetric manner. The
Fayet-Iliopoulos term is given as the coefficient of $\int d^4
\theta\, V_X$, i.e.
\ba\label{SUGRAFI} \xi^X\ =\ M_{Pl}^2 \left.{\del \hat{\cal
K}\over \del V_X}\right|_{V=0} &\sim & M_{Pl}^2 { q_a^X ({\rm
Re}\,T_b )({\rm Re}\,T_c )d^{abc} \over ({\rm Re}\,T_a )({\rm
Re}\,T_b )({\rm Re}\,T_c ) d^{abc} } \ea
One can also understand this as the moment map associated to the
gauged isometry (\ref{gaugedisom}). The $D$-term potential is given
by
\be\label{DPotential}
V_D\ =\ \half g_{X}^2 \left(\xi^X - \sum q^X_\phi
|\phi|^2\right)^2 \ee
where $\phi$ are recombination moduli or matter fields with charge
$q^X_\phi$ under $U(1)_X$. More generally if there are multiple
$U(1)$'s, there could be mixing and we should replace $g_X^2$ by the
inverse of the holomorphic gauge kinetic function ${\rm Re}(f)$. It
follows that we only need to determine the precise coupling $q_a^X$.
This can be read off from (\ref{C4anom}) or (\ref{C4anomV2}) and the
definition of the K\"ahler moduli. From (\ref{C4anomV2}) we read off
that
\be\label{qaX} q_a^X\ \sim\ \int_{Y_4} G \wedge \omega^X \wedge
P(\pi^*_{Y_4} D_a) \ee
The symbol $P$ is meant to represent the Poincar\'e dual. Defining
parameters $t^i$ by $J = t^i P(D_i)$, we claim that $({\rm
Re}\,T_j)({\rm Re}\, T_k) d^{ijk} \sim {\rm vol}(B_3)^3 \, t^k$.
Substituting in (\ref{SUGRAFI}), we find that
\be \xi^X\ \sim\ m_{10}^4 {\int_{Y_4} G \wedge \omega^X \wedge J }
\ee
where we used $M_{Pl}^2 \sim m_{10}^8 {\rm vol}(B_3)$. This is
expected, because it is related by supersymmetry to the matrices
$\Pi_M^X$ discussed earlier which describe the coupling between
$U(1)_X$ gauge fields and RR axions. We would like to use this in
a regime which is far form the $11d$ supergravity limit, where we
derived this expression (and where the equations of motion yield
${\sf G} \wedge J = 0$). The non-renormalization theorem for
Fayet-Iliopoulos terms gives us some confidence that it can be
extrapolated to $F$-theory. Furthermore, the kinetic terms of
extra $U(1)$'s typically scale as
\be {1\over g_X^2}\ \sim\ \int_{Y_4} \omega_X \wedge * \omega_X\
\sim\ m_{10}^{3}{\rm vol}(B_3)^{1/2} \ee
Here we used the relation between $M$- and $F$-theory to convert
the volume of $Y_4$ measured in eleven-dimensional Planck units to
the volume of $B_3$ measured in ten-dimensional Planck units. Note
that such gauge fields are generally not localized on the cycle
wrapped by the GUT brane, so the scaling is not set by $1/R_S$. In
the non-compact limit they freeze out, which further shows that
their fate depends on the UV completion.

Note that the charged fields $\phi$ in (\ref{DPotential}) whose
$D$-terms could possibly compensate for a non-zero
Fayet-Iliopoulos term are obtained by quantizing wrapped
$M2$-branes. They are very heavy and have been integrated out in
$11d$ supergravity, and cannot be condensed. This is consistent
with the fact that ${\sf G}\wedge J=0$ in $11d$ supergravity.
Extrapolating to $F$-theory by varying K\"ahler moduli, these
states becomes light and should be included in the effective
action. Even so, the effective action describes the linearized
deformations, so the Fayet-Iliopoulos parameter must be small in
this regime, so that all the masses of the fields in the effective
action are parametrically below the KK scale.

In particular, using the above expressions for generic moduli
dependence and non-zero Fayet-Iliopoulos term we find that the
masses squared scale like $g_X^2 \xi^X \sim 1/R_{B_3}$, which
exceeds the KK scale. Even with the scaling $g_X^2 \sim {\rm
vol}(B_3)^{-2/3}$ seen in type IIb (where such $U(1)$ gauge
symmetries can be cleanly localized on four-cycles), the masses
are not suppressed compared to the KK scale. Under these
circumstances a consistent effective Lagrangian with a $U(1)_X$
symmetry requires the $\xi^X$ to vanish identically, or else we
are expanding around the wrong background. This applies for
example to models with only a single K\"ahler modulus. However
when there are multiple K\"ahler moduli one may be able to tune
some ratio to get a parametrically small VEV, although one still
has to ensure that all effective cycles of $B_3$ are large
compared to the $10d$ Planck scale when we do this tuning.

Since we extrapolated expressions from $11d$ supergravity, one may
wonder if we should trust this at all. Let us look at possible
corrections to $\xi^X$. The non-renormalization theorem for the
Fayet-Iliopoulos term says that there could be a perturbative
correction at one loop. In the context of the heterotic string, it
is well-known that for an anomalous $U(1)_X$ there could be a
one-loop contribution that is quadratically divergent. In the
context of perturbative type II, the evidence seems to indicate that
such a correction is absent if tadpoles are cancelled (by
open/closed duality). Furthermore, under heterotic/$F$-theory
duality it appears that the heterotic one-loop contribution gets
mapped to a tree level contribution (see appendix C of
\cite{Donagi:2008ca}). It also matches qualitatively with the Higgs
bundle picture \cite{Donagi:2011jy}, though a precise match requires
a more detailed analysis of the degeneration limit involved.

Even if extra $U(1)$ symmetries are unbroken perturbatively, they
may still be violated by non-perturbative effects. Consider the
action of a $D3$-instanton wrapping a four-cycle $D$:
\be e^{-vol(D) + i\int_D C_{(4)}} \ee
From (\ref{C4anom}) we see that the instanton action shifts under
$\Lambda_m$ when either (1) the worldvolumes coincide ($D_m=D$) and
$\int_D F_m \wedge c_1(K) \not = 0$, or (2) the worldvolumes
intersect ($D \cap D_m = \Sigma_m$) and $\int_{\Sigma_m} F_m \not =
0$. (Again this implicitly uses a weakly coupled IIb picture and is
slightly imprecise, it is better to use (\ref{C4anomV2}) and obtain
$q_D^X$ as in (\ref{qaX}).) In these cases, the $D3$-instanton can
generate superpotential terms which violate the $U(1)$ symmetry, by
an amount equal and opposite to that of the instanton action. For
instance in the case of an extra $U(1)_{B-L}$ symmetry, we might
generate terms of the form
\be\label{InstCouplings} {\mathscr L}\quad \supset \quad d^2\theta\,
(Q D L + L E L + DUD )\,f(m)\,e^{-V_{D3} + i \int_{D3} C^{(4)}_{RR}}
\ee
where $f(m)$ is a one-loop determinant depending on complex
structure moduli, but not on K\"ahler moduli. Such effects are
exponentially suppressed and so we have a good approximate
symmetry. To understand how instanton induced corrections to
holomorphic couplings of charged fields such as
(\ref{InstCouplings}) get generated requires a more in depth
analysis of the chiral two-form on the $M5$-brane, in particular
of certain correlation functions of this field. This is discussed
in \cite{Progress}. See also \cite{Blumenhagen:2010ja}.

As an aside, $F$-theory has fewer anomalous $U(1)$'s than
perturbative IIb, and thus many more instantons contributing to the
superpotential \cite{Progress}. In particular, it has been argued
that in type IIb, a $D3$ instanton wrapped on the Standard Model
cycle (or intersecting it) cannot contribute to the superpotential
\cite{Blumenhagen:2007sm}. This is because $C_4$ has an anomalous
transformation under the $U(1) \subset U(5)$, proportional to the
number of generations. To compensate for this, one has to consider
derivatives of the superpotential, which effectively puts
charged fields in front of the exponentiated instanton action. In
$F$-theory by contrast, there is no such $U(1)$ with a mass small
compared to the KK scale, and so the effective action contains only
the $SU(5)$ gauge fields and does not respect this $U(1)$ symmetry.
Therefore such an instanton can contribute to the superpotential in
$F$-theory \cite{Progress}. Lifting of instanton zero modes for
finite coupling is also familiar in field theory%
\footnote{We would like to thank Chris Beasley for emphasizing this.}. %
A similar story also holds in IIa: when we lift to $M$-theory, all
anomalous $U(1)$'s have masses at or above the KK scale
\cite{Pantev:2009de,Witten:2001uq}.

Given a five-fold spectral cover, there's only a limited number of
ways in which we can degenerate the cover to gain additional $U(1)$
symmetries. The possible degenerations and the resulting $U(1)_X$
symmetries were discussed in \cite{Tatar:2006dc}. The simplest
possibilities are a $4+1$-split, which leads to an extra
$U(1)_{B-L}$ symmetry, and a $3+2$-split, which leads to an
$SU(6)_{GUT}$ group (typically broken to $SU(5)_{GUT} \times U(1)_X$).

 The extra $U(1)$ symmetries could be used for various
phenomenological purposes, such as for preventing rapid proton
decay, or for generating flavour structure, as already advocated
in the context of heterotic models. In one scenario, the $U(1)$
selection rules are maintained in perturbation theory. Then the
instanton action $e^{-vol}$ plays the role of the Froggatt-Nielsen
field. A second scenario is to explicitly break the $U(1)$
symmetry, but preserve some of its selection rules through the
holomorphic zero mechanism.

\newsubsection{Proton decay operators of dimension four and five}

We would like to have the benefits of some of the possible extra
$U(1)$ symmetries, without some of their drawbacks. We propose to do
this as follows. When we split the spectral cover to gain an extra
$U(1)$, we also break up the matter curves, and multiplets with
different $U(1)_X$ charges get localized in different ways. The
perturbative vanishing of superpotential couplings violating these
symmetries now comes about because overlap integrals vanish due to
the localization of the wave functions. The crucial ingredient here
however is localization properties of the wave functions of matter
fields, rather than the $U(1)_X$ symmetry, so we can imagine
splitting up the matter curves without splitting up the spectral
cover itself. Localization would still forbid suitable couplings,
but not lead to a preserved $U(1)_X$ symmetry that forbids too many
couplings. Similar ideas have been advocated in the phenomenology
literature on extra dimensions for a while, although one of the
difference with that literature is that due to the holomorphic
nature of superpotential calculations, the effect of localization of
the wave functions is much more pronounced here.

Let us consider an $SU(5)_{GUT}$ model. Clearly for a generic choice
of sections, the matter curves will be irreducible. Hence if we want
$\Sigma_\bfb$ to be reducible, we must require $R$ to factorize into
two global holomorphic sections, $R_{8c_1-3t} =
b_{8c_1-3t-q}\,b_{q}$ for some $q \in H^2(S_2)$. Note this does not
necessarily mean that $\Sigma_\bt$ also factorizes. We will take
$\Sigma_{\bfb_m}$ to correspond to $b_{8c_1-3t-q}=0$, and we will
further factorise $b_q=b_u \, b_d$ so that $b_u=0$ and $b_d=0$
correspond to $\Sigma_{{\bf 5}_h}$ and $\Sigma_{\bfb_h}$
respectively (it will be clear later why this is needed).

 Therefore one solution to
eliminate dimension four proton decay is as follows: we must
factorize $R$ as above, and in order to eliminate the
 ${\bf 10}_m \cdot {\overline{\bf
5\!}\,}_m \cdot {\overline{\bf 5\!}\,}_m$ Yukawa coupling we must
ensure that $a_5 = b_{8c_1-3t-q}=0$ implies that also $b_d=0$. This
is just saying that all down type couplings correspond to ${\bf
10}_m \cdot {\overline{\bf 5\!}\,}_m \cdot {\overline{\bf 5\!}\,}_h$
Yukawa couplings. Or if we are content with eliminating the down
type Yukawas at tree level, we could simply require that
$\Sigma_{{\bf 10}_m}$ and $\Sigma_{{\overline{\bf 5\!}\,}_m}$ do not
intersect at all. Such a configuration still allows for up type
Yukawa couplings. The latter option actually seems somewhat
attractive since only the top quark Yukawa coupling is
experimentally observed to be of order one.

Now we come to the dimension five operators, i.e. the quartic terms
in the superpotential. Let us first make some general remarks about
quartic (and higher order) terms in the superpotential. A
straightforward dimensional reduction of the eight-dimensional
Yang-Mills theory only yields cubic interaction vertices in the
superpotential (\ref{Yukawa}). So in order to get a quartic
interaction in the low energy effective theory valid below the KK
scale, we have to take two cubic vertices and connect them by a
propagator of a mode with mass at least the KK scale. Starting with
the eight-dimensional Yang-Mills theory, the only modes that can
appear in the intermediate channel are the KK modes of the
eight-dimensional theory and the matter hypermultiplets. Corrections
to the eight-dimensional Yang-Mills theory from more massive modes
could give further contributions, but would also be relatively
suppressed, and probably decouple in the $M_{Pl} \to \infty$ limit.
Summing the propagators of all the massive KK modes yields the
Green's function $G_{\delb}$ of the $\delb$ operator. Thus a
four-point coupling in the superpotential is recovered by computing
an amplitude of the form
\be \int_S \vev{\psi_1 \wedge \psi_2\, G_{\delb}\, \psi_3 \wedge
\psi_4 + \psi_2 \wedge \psi_3\, G_{\delb}\, \psi_4 \wedge \psi_1}
\ee
where $\psi \sim A^{0,1} + \Phi^{2,0}$ is the Dolbeault cohomology
class paired with a four-dimensional massless chiral field. In other
words, superpotential terms are computed only from the fermionic
part of the eight-dimensional gauge theory, i.e. from the
holomorphic Chern-Simons theory \cite{Witten:1992fb}. The quartic
terms in the superpotential correspond to a (length three) Massey
product. Since they express topological information, they can also
be computed in alternative ways that are manifestly independent of
the K\"ahler moduli \cite{Merkulov}. However in the following it
will be useful to keep the picture with intermediate propagating KK
modes in mind.

In four-dimensional models, the dimension five operators
$d^2\theta\, QQQL$ and $d^2\theta\, UDUE$ leading to proton decay
can be generated by exchange of massive higgsino triplets. If
present, they lead to the decay $p \to K^+ \bar{\nu}$ with a
lifetime that is hard to reconcile with experiment, at least in
the more minimal versions of supersymmetric GUTs. In $F$-theory
models there is a natural solution to the doublet-triplet
splitting problem, which basically eliminates the triplet partners
of the Higgses. However there are still massive KK modes with the
same quantum numbers (charged in the representation $({\bf
1},\overline{\bf 3\!}\,)_{1/3}$ and its conjugate) supported on
the matter curves. Their exchange would lead to the same dimension
five operators suppressed only by a single power of $M_{KK}$,
which is slightly below $M_{GUT}$, and thus leads to essentially
the same problems as in four-dimensional models. So we would like
to suppress the leading contributions to these operators. (There
are of course other scenarios than we discuss; for example in
split supersymmetry, dimension five proton decay is also
suppressed, due to large masses for the squarks).

Massive fields in the $({\bf 1},\overline{\bf 3\!}\,)_{1/3}$
representation naturally propagate on any of the matter curves which
support a ${\bf 5}$ or $\overline{\bf 5\!}\,$, i.e. curves which
support $(L,d^c)$, the Higgses and possibly messenger fields for
supersymmetry breaking. This means such dimension five operators are
generically present and we have to work to avoid them. If we want
${\bf 10}_m\cdot  {\overline{\bf 5\!}\,}_m \cdot {\overline{\bf
5\!}\,}_h$ down type Yukawa couplings then there will be
interactions of the form
\be Q L T_d + U D T_d \ee
where $T_d$ is a massive KK triplet supported on
$\Sigma_{{\overline{\bf 5\!}\,}_h}$. Now further since we have ${\bf
10}_m \cdot {\bf 10}_m \cdot {{\bf 5\!}\,}_h$ up type Yukawa
couplings, there will also be couplings of the form
\be Q Q T_u + U E T_u\ee
where $T_u$ is a massive KK triplet supported on $\Sigma_{{{\bf
5\!}\,}_h}$.

Thus our options seem to be (1) arrange the matter curves so that a
mass term of the form $m T_u T_d$ is forbidden, (2) engineer some
$U(1)$ symmetry so that the dimension five operators are forbidden,
or (3) arrange the matter curves so that the down type Yukawas, and
hence also $QLT_d + UDT_d$ are classically forbidden. The third
scenario was already discussed in the context of eliminating the
dimension four operators, however on its own it now seems less
appealing. Whatever quantum effect would end up generating the down
type Yukawas of the right order of magnitude would presumably also
end up generating $Q Q T_u + U E T_u$ with coefficients that are too
large, unless we also eliminate $m T_u T_d$ mass terms. Hence we
concentrate on the first two scenarios.

For the first scenario, we must make sure that either $\Sigma_{{{\bf
5\!}\,}_h}$ and $\Sigma_{{\overline{\bf 5\!}\,}_h}$ do not
intersect, or if they do it must be a type of intersection that does
not allow for a ${\bf 5}\cdot{\overline{\bf 5\!}\,}$ mass term. This
is the reason why in the above we required the Higgs curve to
factorize as $b_q = b_u b_d$, and as we see we actually need a
slightly stronger condition on the allowed intersections. It is also
the same condition that would eliminate the classical $\mu$ term, so
we see that these two issues are connected. Note though that whether
or not $\Sigma_{\bfb_m}$ and $\Sigma_{\bfb_{h}}$ intersect, we may
still get Giudice-Masiero operators of type $H_uH_dS^\dagger$ where
$S$ is a complex structure modulus, because this is a term in the
K\"ahler potential rather than a term in the superpotential. If the
curves do not intersect, the overlap between the gaussian tails of
the wave functions gives a suppression of order $\exp
\alpha_G^{-1/2}\sim e^{-5}$, since $\alpha_G^{-1/4}$ is the typical
distance between non-intersecting matter curves.

Perhaps surprisingly, the equation of the matter curve
$\Sigma_\bfv$ naturally allows for factorizations satisfying all
these requirements. For instance we can take
\be\label{FactorSigma10}
 \Sigma_{\bfv_{h_u}} = \{ h = 0\}, \qquad a_4 =  \alpha_4 h, \qquad a_2 = \alpha_2 h,
 \qquad a_0 = \alpha_0 h.
 \ee
Then we factorize
\be \begin{array}{lccc}
 0 = a_0 a_5^2 - a_2 a_3 a_5 + a_3^2 a_4 =&
\underbrace{h} & \underbrace{(\beta_1 a_5 + \alpha_4 a_3}) &
\underbrace{(\beta_2 a_5 + a_3)} \eol
 & \Sigma_{\bfv_{h_u}} & \Sigma_{\bfb_{h_d}} & \Sigma_{\bfb_m}
\end{array}
\ee
where $\alpha_0 = \beta_1 \beta_2$, $\alpha_2 = -\beta_1
-\alpha_4\beta_2$. If we can localize the MSSM fields on the
indicated curves, then there is no dimension four proton decay
because we eliminated the double points on each irreducible piece
of $\Sigma_\bfv$. (More precisely, we also have to make sure that
the spectral sheaf decomposes, i.e. there is no gluing morphism).
If we take $\alpha_4=1$ and arrange all three pieces of
$\Sigma_\bfv$ to be non-intersecting save for the down type Yukawa
intersections $a_3 = a_5 = 0$, then there is also no dimension
five proton decay and the mu-term is absent. This happens despite
the absence any extra $U(1)$ symmetries, because the spectral
cover subject to these conditions is generally still irreducible.
Moreover if the curves are non-intersecting apart from the
required Yukawa couplings, then there are no infinitesimal
deformations that could recombine the curves. Therefore the
absence of the undesired operators is not dependent on moduli
stabilization or fine tuning.%
\footnote{It has recently been suggested that such a scenario for
eliminating proton decay is ruled out due to subleading
corrections to the localized contributions in the overlap
integrals, so that we cannot make the couplings vanish exactly.
This is only partially true. The localization of Yukawa couplings
for massless chiral fields is an exact statement due to its nature
as a product on Dolbeault cohomology \cite{Donagi:2008ca}; there
are no subleading corrections except for non-perturbative ones. On
the other hand, the overlap with KK modes is not sharply localized
as KK modes are not Dolbeault cohomology classes. Thus for
dimension five proton decay we can only argue that it is
suppressed. The superpotential of course is still exact, and we
should understand this suppression as having to do with the
K\"ahler potential.}

One should be careful with the fluxes however. In order to get
only an $H_u$ on $\Sigma_{H_u}$ and an $H_d$ on $\Sigma_{H_d}$, we
want a non-zero hypercharge flux through these components. As we
discussed earlier, the hypercharge flux is always orthogonal to
homology classes dual to linear combinations of $c_1$ and $t$, so
the equation for $\Sigma_{H_u}$ cannot simply be a polynomial in
the $a_i$. If we appropriately factor the $a_i$ (restricted to
$S$), then in principle we can evade this constraint. It is not
hard to see that with the above expressions, the hypercharge flux
through  $\Sigma_{H_u}$ and $\Sigma_{H_d}$ must be equal and
opposite. In the above factorization this requires $\alpha_4\not =
1$, and thus reintroduces dimension five proton decay through
massive triplets propagating on $\Sigma_{H_d}$, unless we also
ensure that there are no simultaneous solutions to $a_5=0$ and
$\alpha_4=0$.

If one insists on extra $U(1)$ symmetries, then what are the
possibilities for preventing dimension five proton decay? Embedding
in $SO(10)$ does not really help because the ${\bf 16}^4$ contains a
singlet. One could try to hope embedding in $E_6$ (and breaking to
$SU(5)_{GUT}$ by turning on a $U(1)\times U(1)$ bundle) will work,
because the ${\bf 27}^4$ does not contain a singlet. It has been argued
that the extra $U(1)$ symmetry is not sufficient
to prevent all dimension five operators, which can still be
generated through triplet exchange \cite{Pati:1996fn}. It seems that
by further embedding in $E_7$ or $E_8$, sufficient $U(1)$ symmetries
are available. We still need to explain how such
symmetries are broken since they should not survive at low energies.
We explained some options for breaking extra $U(1)$'s in subsection
\ref{ExtraU(1)s}.

We would like to mention that an additional interesting possibility
is to break the $U(1)$ symmetries through certain subleading corrections
to the local model. To engineer this is we use a
partially factorized `local' spectral cover, so as to get accidental
$U(1)$ symmetries that forbid dimension four and/or dimension five
proton decay in the local model, but embed it in a global model in
which such symmetries are broken. The breaking of the $U(1)$
symmetries is then due to the subleading terms to the local ALE
fibration, and for generic global UV completions of a local model we
would expect such breaking to occur, because generic subleading
terms would typically destroy any factorizability properties of the
leading terms. Said differently, an exact extra $U(1)$ symmetry
requires a global harmonic two-form with one index on the base and
one on the fiber, but such forms are hard to come by as the elliptic
fibration generically has $I_1$ singularities and the discriminant
locus is irreducible away from the GUT brane. This is yet another
manifestation of the principle that the fate of $U(1)$ symmetries is
not determined locally but depends on the UV completion, as we saw
earlier for $U(1)_Y$.

From the four-dimensional perspective this means that the $U(1)$
symmetries are only broken by operators suppressed by a high scale
which is not native to the local model. There has been recent
progress on defining a stable degeneration limit of global models
in which a local model splits off from the rest of the Calabi-Yau
\cite{DW7}. One could use this technology to implement such an
idea.

Another possible solution to the mu-problem and doublet/triplet
splitting problem was recently described in \cite{DW6}. The idea
is to put the Higgs fields in the bulk of the $7$-brane, rather
than on a matter curve as discussed above. These models use a
further $U(1)$ symmetry to eliminate dimension four proton decay.

Even if we suppressed the classical dimension four and five
operators, they may still be generated non-perturbatively through
$D3$-instanton effects. Such effects are presumably too small to be
observed.

The issue of dimension four and five proton decay has been much
studied since this paper appeared, and we believe the last word has
not been said on it.

\newsubsection{Operators of dimension six}

Assuming we have engineered the dimension four and five operators
leading to proton decay to be sufficiently suppressed, we now turn
to the dimension six operators (the four-fermi terms). These can be
mediated by massive gauge bosons in the representation $({\bf
2},{\bf 3})_{-5/6}$. The internal wave functions of these particles
are scalar fields supported in the bulk of the 7-brane.\footnote{To
be fair, as seen in section \ref{OneLoopKK} the internal wave
functions for the longitudinal modes of these gauge bosons are exact
one-forms on $S_2$. As we consider couplings to conserved currents,
these modes are not relevant.} We will follow the formulation of
\cite{Friedmann:2002ty}, making adjustments for the $F$-theory
setting.

In four-dimensional $SU(5)$ GUTs, dimension six proton
decay\footnote{In flipped $SU(5)$ or $SO(10)$ models there are
additional dimension six operators mediated by heavy gauge bosons in
the $({\bf 2},{\bf 3})_{-1/3}$ representation. Also, dimension six
operators mediated by exchange of scalars in the $({\bf
1},\overline{\bf 3\!}\,)_{1/3}$ are presumed absent due to whatever
mechanism eliminated the dimensions five proton decay operators.}
comes from exchange of massive gauge bosons in the $({\bf 2},{\bf
3})_{-5/6}$, resulting in the matrix element
\be\label{4dDim6Decay} {\mathscr M}\sim  g_{GUT}^2 {J_\mu
\tilde{J}^\mu(0)\over M^2} \ee
with $J^\mu \sim \overline{\Psi}_{\bf 10} \gamma^\mu \Psi_{\bf 10} +
\overline{\Psi}_\bfb \gamma^\mu \Psi_{\bfb}$. In $F$-theory, there
is a whole tower of massive gauge bosons with these quantum numbers.
They are KK modes of the eight-dimensional $SU(5)$ gauge field,
whose internal wave functions are eigen-functions of the scalar
Laplacian on $S$ (the 7-brane worldvolume). Thus we can immediately
write down the analogous expression:
\be {\mathscr M} \sim g_8^2 \int_{S\times S} d^4 z_1\, d^4 z_2\,
{\bf j}^\mu(0,z_1)\, \tilde{\bf j}_\mu(0,z_2)\, G(z_1,z_2) \ee
Here the currents are constructed from the wave functions of the
fermionic zero modes, ${\bf j}^{\,\mu}(x,z) \sim
\overline{\Psi}\gamma^\mu \Psi(x,z) $, with $x$ denoting coordinates
on ${\bf R}^4$ and $z$ denoting coordinates on $S$, and $G(z_1,z_2)$
is the Green's function for the scalar Laplacian on $S$, for scalar
fields valued in the representation $({\bf 2},{\bf 3})_{-5/6}$. The
fermionic zero modes are of the form
\be \Psi(x,z) = \sum_{i}\chi_i(x) \otimes \psi_i(z) \ee
where $\chi_i$ is a four-dimensional spinor and $\psi_i(z)$ is an
internal fermionic zero mode on the 7-brane worldvolume. The fermion
zero modes, as discussed in detail in \cite{Donagi:2008ca},
correspond to harmonic $(0,q)$ forms on $S$. In order to extract the
physical amplitudes, they must be properly normalized, i.e. $\int_S
d^4 z\, \psi_i^\dagger \psi_i = 1$. Then we can pull out the
four-dimensional part and write the matrix element as\footnote{We
are being a little crude and use the label $i$ only to indicate the
gauge indices; in principle there are also flavour indices.}
\be {\mathscr M} \sim g_8^2\  \sum_{i,j} {J_{i,\mu}
\tilde{J}_j^\mu(0)}\  \int_{S\times S} d^4 z_1\, d^4 z_2\,
\psi_i^\dagger \psi_i(z_1)\, \tilde{\psi}_j^\dagger
\tilde{\psi}_j(z_2)\, G(z_1,z_2) \ee
where $J_i^\mu = \overline{\chi}_i \gamma^\mu \chi_i$. We can
formally write the Green's function as
\be G(z_1,z_2) \sim \sum_k {\phi_k^*(z_1) \phi_k(z_2)\over M_k^2}
\ee
where the $\phi_k(z)$ are the normalized eigenmodes of the Laplacian
for $({\bf 2},{\bf 3})_{5/6}$-valued scalars. This representation
makes the analogy with the four-dimensional expression
(\ref{4dDim6Decay}) more clear.

If the integral is finite (i.e. if the sum over KK modes is
convergent), then we will end up getting an expression of the same
general form as (\ref{4dDim6Decay}), with $M$ replaced by the KK
scale. Therefore we would like to investigate the UV behaviour of
the amplitude.

Let us first consider the ${\bf 10}^4$ amplitude. Then both $\psi$
and $\tilde \psi$ are localized on $\Sigma_{\bf 10}$, and both $z_1$
and $z_2$ are points on $\Sigma_{\bf 10}$. The Green's function
diverges as
\be G(z_1,z_2) \sim {1\over d(z_1,z_2)^2 } \ee
as $z_1 \to z_2$. Here $d(z_1,z_2)$ is the geodesic distance on $S$.
In order to estimate the integral, we first change variables from
$\{z_1,z_2\}$ to $\{z_1-z_2,z_1+z_2\}$. To make sense of this on a
Riemannian manifold, we use the exponential map to set up a local
coordinate system. This cannot be done globally, but for large $d$
the integral is effectively cut off and the only scale that can
enter the large $d$ integration is $R_{KK}$. Since we are interested
in the UV behaviour, we should concentrate on $z_1$ close to $z_2$.
The change of variable introduces a Jacobian. To leading order for
$z_1$ close to $z_2$, the Jacobian should not depend on the metric
and we can use the flat space expressions. For our purposes we can
also approximate the densities $\psi_i^\dagger \psi_i$ as constant
of order $1/R^2$ on the matter curve, and vanishing off the matter
curve. Then the integral over $z_1 + z_2$ gives the volume of
$\Sigma_\bt$ which is of order $R^2$. The integral over $z_1 - z_2$
gives
\be \int_{S\times S} d^4 z_1\, d^4 z_2\, \psi_i^\dagger
\psi_i(z_1)\, \tilde{\psi}_j^\dagger \tilde{\psi}_j(z_2)\,
G(z_1,z_2) \sim {1\over R^2} \int_\epsilon r dr \,{1\over r^2} \ee
where $r = |z_1 - z_2|$. The integral is cut-off at $r\sim R$ on the
upper limit due to the fact that $\Sigma_{\bt}$ is compact. However
we are interested in the lower limit, and here it diverges
logarithmically. To evaluate it, we need to go beyond the
eight-dimensional field theory approximation. Any subleading
corrections to our approximations though lead to integrals that are
convergent as $r\to 0$, and can be evaluated without knowing
anything about the UV completion.

The divergence indicates sensitivity to UV physics and so will be
regularized by the non-perturbative completion of $F$-theory. Thus
we expect that the divergence will be effectively cut off by
including new modes with wave-lengths of order the ten-dimensional
Planck length. Therefore we `regularize' the integral by cutting off
the integration at $r \sim 1/m_{10}$, where $m_{10}$ is the
ten-dimensional Planck scale. Then our estimate for the size of the
amplitude becomes
\be {\mathscr M} \sim {g_8^2 \  M^2_{KK}\log(M_{KK}/m_{10})}\
{J_{\bt,\mu} \tilde{J}_{\bt}^\mu(0)} \ee

Even if we would have kept track of all the numerical factors and
been able to evaluate the integrals, we do not know the exact
coefficient due to lack of knowledge of the proper regularization of
the Green's function. The dependence on the cut-off scale is very
mild however. It is interesting to compare the parametric dependence
of this expression with the analogous one for four-dimensional GUTs.
Using our earlier expressions,  $g_8^2 M_{KK}^2  \sim
\alpha_{GUT}/M_{KK}^2$, so we can write this as\footnote{This
corrects an error in v1. The logarithmic dependence was pointed out in
\cite{Wijnholt:2008db,TWv3}.}
\be  {\mathscr M} \sim {\alpha_{GUT}^{}\log(\alpha_{GUT}^{-1})}\
{J_{\bt,\mu} \tilde{J}_{\bt}^\mu(0) \over M_{KK}^2} \ee
There are two important qualitative differences here compared to the
analogous expression for conventional four-dimensional GUTs. First,
the amplitude is proportional to $M_{KK}^{-2}$ rather than
$M_{GUT}^{-2}$. This is expected because the modes mediating this
process are KK modes. Secondly, there is a logarithmic enhancement
as $\alpha_{GUT}\to 0$ compared to the expression for
four-dimensional GUTs. This came about because the amplitude was
divergent and sensitive to physics at the ten-dimensional Planck
scale. In practice we have $\alpha_{GUT}^{-1} \sim 25$ and the
enhancement is not very large. Also it is probably hard to get any
universal results about the numerical values of the integrals. (According
to a recent conjecture however, we can numerically approximate the
hermitian metric on the Higgs bundle \cite{Donagi:2011jy}.
If correct, it means that in principle we could systematically
approximate the integrals, although we would still not know
the correct value of the cut-off).

For the $\bt^2 \bfb^2$ amplitude, we need to distinguish two cases.
The easiest is if $\Sigma_{\bt}$ and $\Sigma_\bfb$ do not intersect,
a possibility we considered above in order to suppress dimension
four and dimension five baryon number violating operators. Then
their separation depends on complex structure moduli but is at most
order $R\sim V_S^{1/4}$ (scale of $S$), and the least we can get is
something of the form
\be\label{Dim6RDecay} {\mathscr M} \sim {g_8^2   V_S^{-1/2}}
{J_{\bt,\mu} \tilde{J}_{\bfb}^\mu(0)} \sim {\alpha_{GUT}}
{J_{\bt,\mu} \tilde{J}_{\bfb}^\mu(0) \over M_{KK}^2} \ee
Modulo the appearance of $M_{KK}$ instead of $M_{GUT}$, this has the
same parametric dependence as the four-dimensional expression.

If we allow $\Sigma_{\bt}$ and $\Sigma_\bfb$ to come closer or if
they intersect at a point, we need to be more careful. Although at
the down type Yukawa intersection the curve $\Sigma_\bfv$ has a
double point singularity, we will interpret this as two branches of
distinct components of $\Sigma_\bfv$ intersecting there, one
component supporting matter and the other supporting the Higgs. As
we have seen, other intersections lead to dimension four proton
decay. Then $\Sigma_{\bfb_m}$ and $\Sigma_\bt$ intersect each other
transversely there, and we may take $\Sigma_\bt$ to be defined by
$z_2=0$ locally, and $\Sigma_{\bfb_m}$ by $z_2 = \alpha z_1$ for
some constant $\alpha$.

Let us use $w_1$ as the local coordinate along $\Sigma_\bt$ with
$z_1(w_1) = w_1$, and $w_2$ be a local coordinate along
$\Sigma_{\bfb_m}$, such that $z_1(w_2) =w_2$ and $z_2(w_2) = \alpha
w_2$. Let us further assume that the wave functions of the matter
fields are non-vanishing at the intersection. This may not
necessarily be the case, but the integrals only get more convergent
when the wave-functions have such vanishing behaviour, so we may
restrict to the case when they do not vanish. Then we get the
following integral:
\be g_8^2 \int d^2 w_1 d^2 w_2\, {\bf j}_\bt(w_1) G(w_1,w_2) {\bf
j}_\bfb(w_2) \sim g_8^2 R^{-4}\int d^2 w_1 d^2 w_2 {1\over
|w_1-w_2|^2 + |\alpha w_2|^2} \ee
Let us change variables from $(w_1,w_2)$ to $(w_{12},w_2)$, where
$w_{12} = w_1 - w_2$. The Jacobian for this change of variable is
simply the identity. Therefore the integral becomes
\be g_8^2 R^{-4}\int d^2 w_{12} d^2 w_2 {1\over |w_{12}|^2 + |\alpha
w_2|^2} \ee
By a further change of variables we get
\be g_8^2 R^{-4}|\alpha|^{-2} \int_\epsilon d^4 x {1\over |x|^2}
\sim g_8^2 R^{-4}|\alpha|^{-2}\int_\epsilon r^3 dr {1\over r^2} \ee
which converges at short distances. Therefore this amplitude can be
calculated in the effective eight-dimensional gauge theory, we do
not need to know anything about the UV completion to compute this
amplitude. Just as in (\ref{Dim6RDecay}) the parametric dependence
is the same as conventional four-dimensional models
(\ref{4dDim6Decay}), except with $M$ being $M_{KK}$ rather than
$M_{GUT}$. The only exception could be if for some reason the
intersection is not transverse and the angle $\alpha$ is zero, which
could yield logarithmic behaviour, but this is certainly not
generic.

\newsubsection{Conclusions}

In this section we investigated the issue of proton decay in
$F$-theory GUTs. In generic models there is nothing to prevent fast
proton decay, so there must be some extra structure. We discussed
some options for eliminating significant proton decay through
dimension four and five operators, and simultaneously solve the
classical mu-problem, by factorizing the matter curves in certain
ways and ensuring that we have the right flux through each
irreducible component, or by using extra $U(1)$ symmetries which are
broken by the subleading corrections to the local ALE fibration in a
global model. Although it appears possible to satisfy experimental
constraints, ideally one would like to connect this with other
issues in flavour physics, and further research is desirable.

The dimension six operators receive contributions from massive KK
gauge bosons. Their wavefunctions are supported in the bulk of the
7-brane, and so such contributions can likely not be suppressed. We
have found two qualitative differences with conventional
four-dimensional models: there is a parametric difference between
the $p \to \pi^0 e^+_L$ and $p\to \pi^0 e^+_R$, and further the
scale differs parametrically from the GUT scale. Although there is
uncertainly about the numerical prefactors, these qualitative
differences are universal and could potentially serve as smoking
guns for $F$-theoretic GUT models.

Processes coming from the ${\bf 10} \cdot {\bf 10}$ OPE may be
enhanced with respect to those coming from the $\bt \cdot \bfb$ OPE,
due to a parametric enhancement of dimension six proton decay by a
factor $\log (\alpha_{GUT}^{-1})$. Whether this would translate to
an actual enhancement in the real world also depends on the
numerical factors, which are hard to calculate, but at least it
constitutes a qualitative difference with four-dimensional models.
Such a qualitative difference is not uncommon in Kaluza-Klein
models, but the precise form depends on the dimensionality: in
$M$-theory/IIA models of unification, an $\alpha_{GUT}^{-1/3}$
enhancement in the $\bt\cdot \bt$ OPE was found compared to
four-dimensional GUTs \cite{Friedmann:2002ty,Klebanov:2003my}.

The second effect comes from the KK scale being lowered compared
to the GUT scale, which is parametric when the GUT group is broken
by $U(1)_Y$ flux. The origin of this effect was explained in
section \ref{KKThresholds}. If indeed the KK scale is lowered even
slightly, we get a rather large enhancement of proton decay since
$|\mathscr{M}|^{-2}$ scales as $M_{KK}^4$. This is a surprising
difference with conventional four-dimensional models, where the
dimension 6 operators are mediated by $X$ and $Y$ bosons which
have masses of order $M_{GUT}$. This situation is also different
from the $M$-theory unification models; in \cite{Friedmann:2002ty}
it was found that after taking into account one-loop threshold
corrections in $M$-theory, there is no parametric separation
between $M_{KK}$ and $M_{GUT}$.

The proton life-time is of the form
\be \tau_p \sim |\mathscr{M}|^{-2} m_p^{-5} \ee
In four-dimensional GUTs, with $M_{GUT} \sim 3\times 10^{16}\,{\rm
GeV}$, we get \cite{Raby:2002wc} $\tau(p\to \pi^0e^+) \sim 10^{35
\pm 1}$ yrs. This is out of range for current experiments, which
provide a bound $\tau(p\to \pi^0 e^+)
> 5\times 10^{33}$ yrs. The numerical pre-factor in $F$-theory
models depends on the details of the geometry of the configuration,
so it seems hard to make a universal statement. If, hypothetically,
the numerical factors are similar to those of four-dimensional
models, then with $\alpha_{GUT} \sim 1/25$ the lifetime for $p\to
\pi^0 e^+_L$ could get reduced by a factor of 10 compared to $p\to
\pi^0 e^+_R$ due to the first effect. Due to the second effect, for
$\Lambda \sim R_{B_3}m_{10}^2$ the lifetime through dimension six
decays may be brought down by a factor of $10^{1.5}$, bringing it to
the verge of detection.

\vspace{1cm} \noindent {\it Acknowledgements:}

MW would like to thank K. K\"ohler for correspondence on Ray-Singer
torsion; J.~Conlon for discussions on the unification scale in
string models; and I. Adam, Y. Oz and S. Theisen for discussion
related to this project. MW would also like to thank UPenn, LPTHE Jussieu,
CUHK, HK University, CERN, Cambridge University, and the Simons workshop at
SUNYSB for hospitality while this work was in progress and the opportunity to present
some of these results. The research of MW was
supported in part by a Curie fellowship under contract number
MRTN-CT-2004-512294. R.D. is partially supported by NSF grant DMS
0612992 and NSF Research and Training Grant DMS 0636606.

\vspace{2cm}

\newpage

\appendix

\renewcommand{\newsection}[1]{
\addtocounter{section}{1} \setcounter{equation}{0}
\setcounter{subsection}{0} \addcontentsline{toc}{section}{\protect
\numberline{\Alph{section}}{{\rm #1}}} \vglue .6cm \pagebreak[3]
\noindent{\bf Appendix {\Alph{section}}:
#1}\nopagebreak[4]\par\vskip .3cm}

\newsection{GUT breaking fluxes constructed through the cylinder map}
\seclabel{HetGfluxes}

In this appendix we would like to give a more detailed discussion of
quantization constraints on GUT breaking fluxes in heterotic and
local $F$-theory models. We will see how to construct models without
exotic matter, and we will learn that this requires the
Noether-Lefschetz fluxes of \cite{Donagi:2009ra}.

We consider an elliptically fibered Calabi-Yau three-fold $Z$ with a
$U(5) \times U(1)$ bundle $V \oplus K^{-1}$. We require that ${\rm
det}\, V = K$ so that the structure group embeds in $SU(6) \subset
E_8$. In other words, we consider a rank six bundle with structure
group $U(5)$. We further assume our bundle $V$ can be constructed
using spectral covers. Such bundles sometimes have the larger
structure group $SU(5)\times U(1)$ rather than $U(5) =
SU(5)\times_{Z_5} U(1)$, which can cause trouble with exotic matter
as we will see below. Then the spectral data consist of a degree six
spectral cover and a rank one spectral sheaf on it. The six sheets
may be labelled by the following roots of $E_8$:
\be\label{SU6Cover}
\begin{array}{lcl}
  \sigma_1=\alpha_5 & \qquad & \sigma_4=\alpha_5+\alpha_4 + \alpha_3 + \alpha_2 \\
  \sigma_2=\alpha_5+\alpha_4 &  & \sigma_5=\alpha_5+\alpha_4 + \alpha_3 + \alpha_2 + \alpha_1 \\
 \sigma_3= \alpha_5+\alpha_4 +
\alpha_3 &  & \sigma_6=\alpha_5+\alpha_4 + \alpha_3 + \alpha_2 +
\alpha_1 + \alpha_{-\theta}
\end{array} \ee
The degree six spectral cover decomposes into an irreducible
five-fold spectral cover $\pi_{C_5}: C_5 \to S$ with sheets
$\{\sigma_2, \ldots, \sigma_6\}$ and a degree one piece which is
just $\sigma_1 =\sigma_{B_2}$, so $C_6 = C_5 \cup \sigma_1$. Since
we allow for a $U(5)$ structure group, we have a wider set of
choices for the spectral line bundle $L_5$ on $C_5$:
\be c_1(L_5) = -\half c_1(C) + \half \pi_{C_5}^* c_1(B_2) + \gamma
\ee
with $\Sigma = C \cap \sigma_{B_2}$. For $SU(5)$ bundles, one would
also require that $\pi_{C_5*}\gamma =0$. Here instead we turn on a
line bundle $\zeta^{-1}$ on $\sigma_1$ so that
\be c_1(\zeta) = c_1(\pi_{C_5 *}\gamma) = c_1(\det(V)|_{B_2})\in
H^2(B_2). \ee

In fact this is still not the most general construction; we may
further twist $V$ by a line bundle $Q$ with $c_1(Q) = q p_Z^*
\Sigma$. Then we find \cite{Blumenhagen:2006wj}
\be c_1(V\otimes Q) = n q p_Z^* \Sigma + p_Z^* c_1(\zeta) \ee
which determines the line bundle $K$ we have to turn on to
compensate. Turning on $q\not = 0$ corresponds to making $\sigma_1$
different form the zero section $\sigma_{B_2}$, i.e. changing the
four-fold rather than turning on $G$-flux. In terms of the
eight-dimensional gauge theory, it corresponds to breaking the gauge
group by turning on a non-zero profile for an abelian Higgs field.
For more discussion of the latter as a mechanism for breaking the
GUT group, see \cite{Donagi:2009ra}. In the following however we
keep $q=0$.

In order to break the GUT group without massless lepto-quarks, we
want $c_1(\zeta)$ to be a primitive class with $c_1(\zeta)\cdot
c_1(B_2) = 0$, $c_1(\zeta)\cdot t = 0$ and $c_1(\zeta)^2 = -2$.
Furthermore, for $L_5$ to exist the first Chern class of $L_5$ must
be an integer class on $C_5$. As a simple Ansatz, we could try to
take $\gamma$ of the form
\be\label{SpectralAnsatz} \gamma = \lambda \gamma_u + {1\over 5}\,
\pi_{C_5}^* c_1(\zeta) \ee
where $\pi_{C_5 *}\gamma_u=0$. Now we claim that if $\gamma_u$ is
given by the universal flux:
\be\label{uniflux} \gamma_u = 5 [\Sigma_\bt] - p^*(\eta - 5 c_1) \ee
then these conditions are not compatible, as we now show. First of
all, note that any non-integrality in ${1\over 5}\, \pi_{C_5}^*
c_1(\zeta)$ cannot be cancelled by $\half r + \lambda \gamma_u$
unless $c_1(\zeta)$ is a multiple of $\eta - 5 c_1$. But this is
impossible on a del Pezzo surface as $\eta - 5 c_1$ is effective, hence
$c_1\cdot (\eta- 5c_1) >0$,  but we had
$c_1(B_2)\cdot c_1(\zeta) = 0$. (One may strengthen this and also rule
out other cases like $dP_9$). Therefore
$p^*c_1(\zeta)$ should be divisible by $5$. But then
$p^*c_1(\zeta)^2$ must be divisible by $25$. However
\be \pi_{C_5}^*c_1(\zeta) \cdot \pi_{C_5}^*c_1(\zeta)=
c_1(\zeta)\cdot \pi_{C_5*}\pi_{C_5}^*c_1(\zeta) = -10 \ee
which does not contain any squares, so $\pi_{C_5}^*c_1(\zeta)$ is in
fact primitive. We conclude that there is a problem with exotic
matter for GUT breaking by $U(1)$ fluxes in the heterotic string, if
we only use this special flux. This is the heterotic version of the
puzzle we discussed in section \ref{RefinedGUTBreaking}: the $U(5)$
bundle cannot be the product of an $SU(5)$ bundle and a $U(1)$
bundle if we want to avoid exotic matter. In particular,
previously proposed models along these lines turn out to have exotic matter.

We will now show that exotic matter can be avoided if we use more
general fluxes, namely the Noether-Lefschetz fluxes discovered in
\cite{Donagi:2009ra}. Consider a flux of the form
\be c_1(L_5)\ =\ \half r + \lambda \gamma_u + \alpha\  =\ \half r +
\tilde \gamma + {1\over 5} \pi_{C_5}^*c_1(\zeta) \ee
where $c_1(\zeta) = \pi_{C_5*}\alpha$, $r =- c_1(C_5) +  \pi_{C_5}^*
c_1(B_2)$. In the last equality, we have artificially split the flux
into a flux $\tilde \gamma$ that commutes with $SU(5)_{GUT}$:
\be \pi_{C_5*}\tilde \gamma = \lambda\, \pi_{C_5*} \gamma_u +
\pi_{C_5*}(\alpha - {1\over 5} \pi_{C_5}^*\pi_{C_5*}\alpha)=0 \ee
and a flux ${1\over 5} \pi_{C_5}^*c_1(\zeta)$ that is responsible
for breaking $SU(5)_{GUT}$ to the Standard Model. However ${1\over
5} \pi_{C_5}^*c_1(\zeta)$ is generally not an integer class. So we
cannot simply turn off the GUT breaking flux if we are to satisfy
the quantization constraints. This would result in an inconsistent
model.

With the above definitions, $c_1(L_5)$ is naturally an integral
class when $\lambda-\half$ is an integer and $\alpha$ is an integral
class. So we only need to construct a $(1,1)$ class $\alpha$ such
that $\pi_{C_5*}\alpha \cdot c_1=0$, $\pi_{C_5*}\alpha \cdot t=0$
and $(\pi_{C_5*}\alpha)^2 = -2$. This can certainly be done, for
instance as in \cite{Donagi:2009ra}, but it requires that we
adjust/stabilize some of the moduli. Otherwise the only available
fluxes are the class of the matter curve and pull-backs of classes
in $B_2$, which leads to the universal flux (\ref{uniflux}) and thus
to exotic matter. Hence we conclude that absence of light scalar and
vector lepto-quarks is perfectly consistent with the quantization
conditions, provided we go slightly outside of the usual framework
of (\ref{SpectralAnsatz}) and make use of all available fluxes.

If we use the construction in \cite{Donagi:2009ra} then we have
$\int_{\Sigma_\bt} \alpha = 0$ and turning on $\alpha$ does not
change the net chirality. So we immediately get some toy models with
three generations and no exotic matter by using the explicit
examples in section 4 of \cite{Donagi:2008ca} and turning on
$\alpha$. The main downside of these toy models is that for generic
complex structure moduli, they have no light Higgses (i.e the model
suffers from the $\mu$-problem), and if we would tune to get such
Higgses, the model would typically suffer from fast proton decay and
the Yukawa couplings would not be hierarchical. Some extra structure
is needed to explain the mu-problem, the stability of the proton and
flavour.

Let us be a little more explicit about the resulting $G$-fluxes in
$F$-theory. Again we focus on the case of $SU(5)_{GUT}$ models
broken to the Standard Model by hypercharge flux. Then this is a
special case of an $Sl(6)$ cover, which splits as $5+1$. All the
other components of the $E_8$ cover can be expressed in terms of
this fundamental cover. We use the following notation, similar to
appendix C of \cite{Donagi:2008ca}:
\ba \pi:  Y_4 \to B_3   & &  {\rm elliptic\ fibration\ } \eol
\sigma_{B_3}: B_3 \to Y_4 & &  {\rm the\ section\ } \eol
 \rho: B_3 \to B_2   & &  P^1 {\rm \ fibration} \eol
 Z \subset Y_4 & & \pi^{-1} {\rm \ of\ a\ section\ of\ }\rho. \eol
p: Y_4 \to B_2 & & dP_9 {\rm \ fibration.} \eol
\pi_C: C_6 \to B_2 & &
{\rm the\ heterotic\  spectral\  cover} \eol p_R: R \to C_6 & & {\rm
the\ ``cylinder",\ or\ union\ of\ lines\ in\ the\ } dP_8{\rm 's}
\eol
        & & {\rm (i.e.\ sections\ of\ } dP_9{\rm 's, \ disjoint\ from\ } \sigma {\rm )\ parametrized}\eol
        & & {\rm  by\ points\ of\ } C. \eol
j: (C_6 = R \cap Z) \subset R & &  {\rm  the\ inclusion\ ``at\
infinity"} \eol
i: R \hookrightarrow Y & & {\rm the\ natural\
inclusion.} \eol
{\bf Pr}: H^i(Y_4) \to H^i(Y_4) & & {\rm orthogonal\
projection\ on\ } [\pi^*H^i(B_3)]^\perp \eol \ea
We further consider the split $C_6 = C_5 \cup C_1$, where
$C_1$ coincides with the zero section $B_2$.
In $F$-theory, the $G$-flux dual to the bundle $V\oplus
\det(V)^{-1}$ is then
\ba {\sf G}\ &=&  {\bf Pr}[ i_{R_5*} p_{R_5}^* \gamma - i_{R_1*} p_{R_1}^*
c_1(\zeta)] \eol[2mm] &=& i_{R_5*} p_{R_5}^* \gamma - i_{R_1*} p_{R_1}^*
c_1(\zeta) - n_\gamma [dP_9] \ea
Here we used $\gamma = c_1(L_5) - \half r$ to represent the flux of $L_5$.
We recall that $c_1(\zeta) = - \pi_{C_5 *} \gamma$. We compute
that
\be n_\gamma = \gamma \cdot_{C_5} \Sigma - c_1(\zeta)\cdot_{B_2}
c_1(K_{B_2}) \ee
Since we are assuming that $c_1(\zeta)$ is orthogonal to $K_S$,
the last part simply drops out. The map $p_R$ associates lines in
$dP_8$ (or sections of $dP_9$) with points on $T^2$ and is defined
using the embedding $SU(6) \in E_8$ discussed in equation
(\ref{SU6Cover}). Note that in the notation of section
\ref{RefinedGUTBreaking}, essentially we have a fractional line
bundle $L^{1/6} = \zeta^{1/5}$ on each sheet of the irreducible
five-fold cover $C_5$, and $L^{-5/6} = \zeta^{-1}$ on the sixth
sheet.

It remains to say something about the $D$-terms. There are two
cases to be addressed. First, it is possible that our abelian
gauge symmetry does not couple to RR axions for topological
reasons, as we imposed for hypercharge. In this case, by
supersymmetry there is also no corresponding Fayet-Iliopoulos
term. Such configurations are naturally polystable, with slope
zero.

Now suppose that we do have such a coupling. We take a general
K\"ahler class of the form
\be
J = t_1 p^*J_{B_2} + t_2 J_\infty
\ee
where $J_{B_2}$ is a class in $H^2(B_2)$, and $J_\infty$ is the
Poincar\'e dual of $B_2$ in $B_3$ (where we use the embedding at
infinity here -- this differs from the embedding at the location
of the singular locus only by a class in $H^2(B_2)$). we need both
$t_1$ and $t_2$ large in Planck units. The small angle limit,
where we can trust the $8d$ gauge theory description, corresponds
to $t_1 >> t_2$. We are interested in the Fayet-Iliopoulos term
\be
\xi\ \sim\ m_{10}^4\int_{ Y_4} G \wedge J \wedge \omega^Y
\ee
The intersection of $G$ with $J_\infty$ would generally be
non-zero, but in the present case we know that $\omega_Y$ is
supported at the zero section (see equation
(\ref{HypDualExpression})), whereas $J_\infty$ is localized at the
infinity section, and hence the intersection vanishes. The
Fayet-Iliopoulos parameter then only depends only on the
intersection of $G$ with $p^*J_{B_2}$, which is proportional to
$J_{B_2} \cdot_{B_2} c_1(\zeta)$. In particular, it vanishes if
$J_{B_2} \cdot c_1(\zeta)=0$. For more general $U(1)$'s, the
Fayet-Iliopoulos parameter can depend on the extension to the
global model.

\newsection{Metric anomaly for holomorphic torsion}
\seclabel{MetricAnomaly}

Let $X$ denote a complex manifold of dimension $d=2n$ with metric
$g$, and let $V$ be a holomorphic bundle on $X$ with Hermitean
metric $h$. Let ${\bf T}(X,V)$ denote the Ray-singer torsion for
$V$. The torsion is not invariant under general changes of the
metrics $g$ and $h$. Consider a one parameter family of metrics
$g_t,h_t$. The variational formula of \cite{BGS3} is as follows:

\ba {1\over 2\pi}{\del\over \del t}{\bf T}(X,V) &=& {1\over
2\pi}\half \sum (-1)^q {\rm Tr}(\alpha_t P_{0,q}^t) \eol & &- \half
\int_X \left.{\del\over \del s}\right|_{s=0}\left[ {\bf Td}{1\over
2\pi}(iR + s g^{-1}\del_t g) \,{\bf ch}{1\over 2\pi}(iF +
sh^{-1}\del_t h) \right]_{n+1} \eol \ea
Here $\alpha_t = *_t^{-1}\del_t *_t$, the operator $P^t_{0,q}$
projects on the zero modes of the indicated degree, and the
subscript $n+1$ on the second line means that one should take the
piece of the expression of degree $n+1$.

In order to get the scale dependence, we consider rescaling the
metric as $g_{\mu\nu} \to tg_{\mu\nu}$. That is, we only need the
following special case: $h^{-1}\del h=0, g^{-1}\del g = {1\over t}
I_{2n}$ ($1/t$ times the $2n\times 2n$ identity matrix). We claim
that
\be *_t\,\omega_{0,q} \propto t^{d/2-q} \ee
Here $t^{d/2}$ arises from the factor of $\sqrt{g}$ in $*$, and
$t^{-q}$ arises because we need $q$ factors of $g^{\mu\nu}$ to
contract the indices of $\omega_{0,q}$ with the $\epsilon$ tensor.
Therefore
\be\label{Zeroanomaly}
\sum (-1)^q {\rm Tr}(\alpha_t P_{0,q}^t)
={1\over t} \sum (-1)^q h^q (d/2-q) ={1\over t} ({d\over 2}\chi(V) +
\sum (-1)^{q+1}q h^q) \ee
where $h^q$ is the number of zero modes in $\Omega^{0,q}(V)$.

For our purposes we need two special cases. We use the splitting
principle in order to simplify the calculation. Recall that
\be Td(x) = {x\over 1-e^{-x}} = 1 + \half x + {1\over 12 } x^2 -
{1\over 720} x^4 + \ldots \ee
We also implicitly assume that $V$ has rank one. The calculation is
easily adjusted for higher rank.

 \noindent \underline{ For a curve:}
\ba \left.{\del\over \del s}\right|_{s=0} \left[ {\bf Td}(T) {\bf ch}(V) \right]_2 &=&
 {1\over 2\pi t} {\del\over \del s}|_{s=0}\left[ (1+\half(x+s)+{1\over 12}(x+s)^2) (1+c_1 + ch_2)
 \right]_2 \eol
&=& {1\over 2\pi t} \left.{\del\over \del s}\right|_{s=0}\left[ ch_2 + \half (x+s)c_1 + {1\over 12}(x+s)^2
 \right] \eol
 &=& {1\over 2\pi t} (\half c_1(V) + {1\over 6}c_1(T)) \eol
 \ea
In the above equation, in order to avoid having to keep writing
$s/2\pi t$, we redefined $s \to 2\pi t\, s$ which of course gives
the same result. Combining with (\ref{Zeroanomaly}), we get
\be 2\,{\bf T}(X,V) \sim \left[(h^0(X,V) - \int_X (\half c_1(V) +
{1\over 6}c_1(T))\right] \log t \ee
One may check this expression is invariant under Serre duality,
${\bf T}(X,V) = {\bf T}(X,K\otimes V^*)$.

\noindent
\underline{ For a surface:}
\ba \left.{\del\over \del s}\right|_{s=0}\!\! \left[{\bf Td}(T) {\bf ch}(V) \right]_3 \!\!&\!=\!&\!\!
{1\over 2\pi t}  \left.{\del\over \del s}\right|_{s=0}\!\!
\left[\prod_{i=1,2}\!\left( (1+\half(x_i+s)+{1\over 12}(x_i+s)^2\right) \sum_j ch_j(V)
 \right]_3 \eol
 \!&\!=\!&\! {1\over 2\pi t} (ch_2(V) + {5\over 12}c_1(V)c_1(T) + {1\over 24}c_1(T)^2+{1\over 12}c_2(T) )\eol
\ea
Combining with (\ref{Zeroanomaly}), we get
\ba 2\, {\bf T}(X,V) &\sim& \left[ \sum (-1)^{q}(2-q) h^q(X,V)
\right. \eol & & \qquad \qquad\left.
 - \int_X (ch_2(V) + {5\over 12}c_1(V)c_1(T) + {1\over 24}c_1(T)^2 +{1\over 12} c_2(T))\right] \log t
\eol
\ea
Again one may check this expression behaves appropriately under
Serre duality, in this case ${\bf T}(X,V)=-{\bf T}(X,K \otimes
V^*)$. This is a useful check on some of the coefficients.

\newsection{Roots of $E_8$}
\seclabel{E8roots}

 In this brief section we would like to explicitly
write out the roots of $E_8$ and their decomposition under the
$SU(5)_H\times SU(5)_{GUT}$ subgroup of $E_8$. We use the labelling
of the $E_8$ Dynkin diagram shown earlier in this paper, and define
$\alpha_{-\theta}$ to be the negative of the highest root:
\be \alpha_{-\theta} = -2 \alpha_1
-3\alpha_2-4\alpha_3-5\alpha_4-6\alpha_5-4\alpha_6-2\alpha_7-3\alpha_8
\ee
The adjoint representation is $248$-dimensional, consisting of $8$
Cartan generators, $120$ positive and $120$ negative roots. To make
the table less cluttered, will list only $120$ of the roots; the
remaining roots are the negative of those that are written. For the
case of the $({\bf 1},{\bf 24}) $ and $({\bf 24},{\bf 1}) $
representations, we further have to add the Cartan generators in
order to recover the full representation. The roots are divided up
as follows:

\noindent
$({\bf 1},{\bf 24}) $:
 \be
 \qquad\qquad
\left\{
\begin{array}{c}
 \alpha_5,\ \alpha_6,\ \alpha_7,\ \alpha_8 \\
\alpha_5 + \alpha_6,\ \alpha_6 + \alpha_7,\ \alpha_5 + \alpha_8,\ \alpha_5 + \alpha_6 + \alpha_8, \\
 \ \alpha_5 + \alpha_6 + \alpha_7 ,\  \alpha_5 + \alpha_6 + \alpha_7 + \alpha_8.\\
\end{array}
\right\}
\ee

\noindent
$({\bf 24},{\bf 1}) $:
\be \qquad \qquad
\left\{
\begin{array}{c}
\ \alpha_{-\theta},\ \alpha_1,\ \alpha_2,\ \alpha_3,\\
\ \alpha_{-\theta} + \alpha_1,
\  \alpha_1 + \alpha_2,\  \alpha_2 + \alpha_3,
\ \\
\
\alpha_{-\theta} + \alpha_1+ \alpha_2, \\
\ \alpha_1 + \alpha_2+\alpha_3 \\
\  \alpha_{-\theta} + \alpha_1+ \alpha_2+ \alpha_3.\\
\end{array}
\right\}
\ee

\noindent
$({\bf 5},{\bf 10})$:
\be  \qquad
\left\{
\begin{array}{c}
\alpha_4   \\
\alpha_3 + \alpha_4           \\
\alpha_2 + \alpha_3 + \alpha_4       \\
\alpha_1+   \alpha_2 + \alpha_3 + \alpha_4         \\
\alpha_{-\theta}+\alpha_1+   \alpha_2 + \alpha_3 + \alpha_4           \\
\end{array}
\right\}
+
\left\{
\begin{array}{c}
0 \\
\alpha_5 \\
\alpha_5 +\alpha_6\\
\alpha_5 + \alpha_8 \\
\alpha_5 + \alpha_6 + \alpha_7\\
 \alpha_5 + \alpha_6 + \alpha_8 \\
  \alpha_5 + \alpha_6 + \alpha_7 + \alpha_8 \\
   2 \alpha_5 + \alpha_6  + \alpha_8 \\
   2 \alpha_5 + \alpha_6 +\alpha_7 + \alpha_8 \\
   2 \alpha_5 + 2 \alpha_6 +\alpha_7 + \alpha_8\\
\end{array}
\right\}
\ee

\noindent
$(\bt,\bfb)$:
\be \qquad\qquad
\left\{
\begin{array}{c}
\alpha_3 + 2 \alpha_4   \\
\alpha_2 + \alpha_3 + 2 \alpha_4 \\
 \alpha_2 + 2\alpha_3 + 2 \alpha_4    \\
\alpha_1 + \alpha_2 + \alpha_3 + 2 \alpha_4                    \\
\alpha_1 + \alpha_2 + 2\alpha_3 + 2 \alpha_4                       \\
\alpha_1 + 2\alpha_2 + 2\alpha_3 + 2 \alpha_4    \\
\alpha_{-\theta} + \alpha_1 + \alpha_2 + \alpha_3 + 2\alpha_4   \\
\alpha_{-\theta} + \alpha_1 + \alpha_2 + 2\alpha_3 + 2\alpha_4   \\
\alpha_{-\theta} + \alpha_1 + 2\alpha_2 + 2\alpha_3 + 2\alpha_4   \\
\alpha_{-\theta} + 2\alpha_1 + 2\alpha_2 + 2\alpha_3 + 2\alpha_4   \\
\end{array}
\right\}
+
\left\{
\begin{array}{c}
2 \alpha_5 + \alpha_6 + \alpha_8 \\
2 \alpha_5 +\alpha_6 + \alpha_7 +
                        \alpha_8 \\
2 \alpha_5 + 2 \alpha_6 + \alpha_7 +
                        \alpha_8 \\
3 \alpha_5 + 2 \alpha_6 + \alpha_7 +
                        \alpha_8 \\
3 \alpha_5 + 2 \alpha_6 + \alpha_7 +
                        2 \alpha_8 \\
\end{array}
\right\}
\ee

\end{document}